\documentclass[useAMS,usenatbib]{mn2e}
\usepackage{graphicx}

\newcommand{\br}{{\bf r}}
\newcommand{\vv}{{\bf v}}
\newcommand{\bF}{{\bf F}}
\newcommand{\vU}{{\bf U}}
\newcommand{\vF}{{\bf F}}
\newcommand{\vH}{{\bf H}}
\newcommand{\vG}{{\bf G}}
\newcommand{\vS}{{\bf S}}

\newcommand{\emma}{{\texttt{EMMA }}}

\newcommand{\dom}[1]{#1}
\begin{document}

\title[EMMA]{EMMA: an AMR cosmological simulation code with radiative transfer}

\author[D. Aubert et al. ]{Dominique Aubert$^{1}$\thanks{E-mail:
dominique.aubert@astro.unistra.fr}, Nicolas Deparis$^{1}$ and Pierre Ocvirk$^{1}$\\
$^{1}$Observatoire Astronomique de Strasbourg, CNRS UMR 7550, Universite de Strasbourg, Strasbourg,France}

\maketitle

\begin{abstract}
\dom{\texttt{EMMA} is a cosmological simulation code aimed at investigating the reionization epoch. It handles simultaneously collisionless  and gas dynamics, as well as radiative transfer physics using a moment-based description with the M1 approximation. Field quantities are stored and computed on an adaptive 3D mesh and the spatial resolution can be dynamically modified based on physically-motivated criteria. }

\dom{Physical processes can be coupled at all spatial and temporal scales. We also introduce a new and optional approximation to handle radiation~: the light is transported at the resolution of the non-refined grid and only once the dynamics have been fully updated, whereas thermo-chemical processes are still tracked on the refined elements. Such an approximation reduces the overheads induced by the treatment of radiation physics. A suite of standard tests are presented and passed by \texttt{EMMA}, providing a validation for its future use in studies of the reionization epoch.}

\dom{The code is parallel and is able to use graphics processing units (GPUs) to accelerate hydrodynamics and radiative transfer calculations. Depending on the optimizations and the compilers used to generate the CPU reference, global GPU acceleration factors between x3.9 and x16.9 can be obtained.  Vectorization and transfer operations currently prevent better GPU performances and we expect that future optimizations and hardware evolution will lead to greater accelerations.}
\end{abstract}


\section{Introduction}
Starting in the 70s, numerical simulations have been part of the astrophysicists tool kit to investigate regimes where non-linearities, strong coupling between different physics  and multi-scale processes are dominant. The study of structure formation in a cosmological context is an archetypal example of such intricacies and has thus driven the rise and the growth of the so-called 'cosmological simulations' for decades now. These simulations have successively included collisionless dynamics and hydrodynamics, and implements routinely sub-resolution physics such as star formation or feedback from supernovae to model the galaxy formation process.

Recently, the challenges encountered in galaxy formation theory and the near advent of new observatories capable of investigating the Universe at redshifts $z>6$ (such as JWST or SKA) produced an interest toward the study of the reionization epoch. Interesting in its own sake, as a great cosmic transition produced by the first sources and capable of reionizing the Universe, this epoch will be studied \textit{in-situ} and will provide insights on the initial stages of the structure formation process (see e.g. reviews by  \citealt{BAR01,PRI12}). Reionization is also thought to provide keys to understand the current state of galaxies, with the rise of a UV background capable of suppressing star formation in light objects (see e.g. \citealt{GNE00,HOE06,FIN11,WIS12}). Consequently, numerical simulations of the reionization started to appear 15 years ago to assess this process and prepare the advent of observational constraints (see \citealt{TRA11} for a review).

The common feature of such simulations is the inclusion of radiative transfer physics, to model the impact of the UV photons emitted by the first sources. Quite demanding in terms of computing resources, radiative transfer is often included as a post-processing step on  outputs of simulations to reduce this cost~: absorbents are static and ad-hoc recipes are included to model the negative feedback of radiation on sources (see e.g.\citealt{CIA03,ILI06,MCQ07,MEL06,ZAH07,BAE10,CHA12,OCV13,PAA13,OCV14,ZAW14}). Nevertheless, radiative hydrodynamics codes started to be implemented (like recently e.g \citealt{FIN11,ROS13,PAW15}). They cope with the reduced time scales (and therefore increased CPU cost) produced by radiation physics using massive parallelism (see e.g. \citealt{TRA07,NOR15}), methodological solutions (such as implicit solvers like e.g. \citealt{GON07,NOR15}) or specific physical regimes (such as a reduced speed of light like e.g. \citealt{GNE14,ROS13}). Whatever will be the solution, radiation will surely become an additional standard component of future cosmological simulation codes. This evolution is greatly illustrated by the number of participating codes to comparison projects such as \citet{ILI06} and \citet{ILI09}.

In this context, a new cosmological simulation code is presented here, \texttt{EMMA}\footnote{Electromagnetisme et Mecanique sur Maille Adaptative}, that includes collisionless physics, hydrodynamics and radiative transfer. It builds upon the experience gathered with previous codes such as \texttt{ATON} \citep{AUB08,AUB10} or a particle-mesh N-Body only  code \citep{AUB09}. Its ambition is to be able to tackle structure formation experiments with a special emphasis on the influence of radiation during the reionization epoch. \texttt{EMMA} is an adaptive mesh refinement code, parallel and with GPU-driven acceleration on a selection of its sub-modules, mainly the physics engines. As such, it shares a number of features with \texttt{ATON} and extend its field of application to coupled radiative hydrodynamics and to high resolution through mesh refinement.

The full methodology of \texttt{EMMA} is first described:  details on how data is managed, how the physics is solved and how parallelisation is implemented are presented. Then we present a selection of validations tests, from pure dark matter experiments to more complex situations with coupled physics. Finally we discuss the performances and the potential future developments of \texttt{EMMA}.

\section{Methodology}
\label{s:metho}
\subsection{Adaptive Mesh implementation}
\texttt{EMMA} relies on a grid-based description of the physical quantities and implements adaptive mesh refinement (AMR) techniques to increase dynamically its spatial resolution. The latter is typically of the order of a few cells, for all the different physics presented here, whether it is set by the smoothing of the gravitational force or the smearing of hydrodynamical shocks or ionization fronts. AMR permits to increase this resolution with a moderate cost in terms of memory consumption by providing finer meshes only at selected locations.

Several AMR implementation exists and \texttt{EMMA} adopts a Fully Threaded Tree (FTT, \citet{KHO98}) description of the data, sharing this core feature with other codes such as \texttt{ART} \citep{KRA97} or \texttt{RAMSES} \citep{TEY02}. In this framework, a fundamental grid divides a 3D space (usually cubic but not necessarily) in $2^{3\ell_c}$ cells where $\ell_c$ designates its refinement level and where the cells are arranged in a Cartesian manner. For instance $\ell_c=7$ corresponds to a fundamental $128^3$ grid where each direction is sampled along 128 points.  \dom{Hereafter this fundamental coarsest grid will be referred as the \textit{base} grid and $\ell_c$ is the \textit{base} level.}

These base level cells are the roots of oct-trees that fully describe the refined geometry of space. If a cell is refined, it points towards an octal structure (or \textit{oct}) that in turns points toward eight additional cells belonging to the $\ell_c+1$ level.
This hierarchy can be recursively maintained until a satisfying resolution (i.e. refinement level) is achieved. Conversely, base level cells can also be grouped in octs, that belong to coarser cells of level $\ell_c-1$. Recursively, this process can be repeated until a single $\ell=0$ cell is obtained. This specific cell is therefore the root of the global AMR structure that samples the 3D space where the numerical experiments take place.

In practice, the data is organized using octs as the fundamental structures where a level $\ell$ oct $O^\ell$ contains the following informations:
\begin{itemize}
\item the level $\ell$,
\item 8 cells $c^\ell_i$ with $i\in [0,7]$,
\item a pointer toward its parent cell $c^{\ell-1}$,
\item 6 pointers toward the 6 Cartesian neighbors cells of $c^{\ell-1}$,
\item 2 pointers toward the next and the previous octs in memory that belong to the same level $\ell$. Note that the arrangement of octs in memory is not related to spatial distribution of the octs in the computational domain. With this feature, octs are stored as a \textit{doubly linked list}.
\end{itemize}
A cell $c^\ell_i$, contains the following informations:
\begin{itemize}
\item its index $i\in [0,7]$,
\item the physical data at this location (density, potential, pressure, ionized fraction, etc...),
\item a pointer toward an oct $O^{\ell+1}$ if it is refined,
\end{itemize}

Octs are mainly used for data exploration and management. They are the mandatory intermediates in order to probe the neighbors of a a given cell. They are also the intermediate structure to scan all the cells at a given $\ell$ (i.e. at a given spatial resolution) as they store the pointers toward the previous and next member of the set of all the $O^\ell$. This scanning operations are performed through the doubly linked list.

Cells are mainly used to store physical data. They also store a pointer toward an oct if they are refined but that is all the relational information that they possess. Any query on a neighbor cell must be passed through their parent oct. This description reduces greatly (by a factor 8 typically) the number of relational pointers that would be required if each cell possessed its own set of neighbors: such an organization takes advantage of the fact that all the cells of an oct share a significant fraction of neighbors, at the cost of some operations to retrieve them via the octs.

Since geometrical relations between neighbors are explicit, the mesh can be arbitrarily refined without any constraint on the geometrical strategy of this refinement. Also there is no simple mapping between the position of an oct in memory and its geometrical location.  As a consequence, the fully threaded tree can be extremely versatile and follow closely any physical feature that requires high resolution at the cost of storing additional information. 

In refined grids, resolution jumps require special care. In particular these interface are source of errors and inaccuracies that must be controlled as much as possible. One standard requirement that must be enforced in the current description of AMR data is the fact that resolution jumps cannot be greater than unity for two neighbor cells. Hence the six neighbor pointers of an oct necessarily exist. However it also puts constraints on the way cells are refined and the procedure takes place as follow, in that order:
\begin{enumerate}
\item an oct containing a cell marked for refinement or already refined must be maintained to ensure that resolution jumps are smaller than two. Its parent cell is therefore marked for refinement.
\item if a cell is marked for refinement, so must be its 26 neighbors. It ensures a smooth transition between low and high resolution regions by creating a buffer of high resolution cells at the interface and also prevents large resolution jumps.
\item if a cell satisfies some user-defined criterion, it is marked for refinement. This criterion can be physically motivated (based on physical values or gradients) or based on location (for zoom simulation for instance).
\end{enumerate}
\dom{In practice, step (i) is applied to all the cells that belong to a given level in a single pass. Afterwards the step (ii) criterion is likewise applied to all the cells of that level, then step (iii), resulting in 3 successive passes on all the cells of a given level.  } The marking procedure can be repeated an arbitrarily number of times, especially to increase the thickness of the buffer regions at the transition between regions at 2 different resolutions. In practice, this procedure is applied twice, similarly to \texttt{RAMSES} or \texttt{ART}. It ensures that a cell that is refined on a physical basis (criterion 3 above) will be produced with an outer layer of 2 neighbor cells at the same resolution. 
Once this marking has been completed, cells can be refined by creating a new oct to be attached to it or coarsened by suppressing its child oct. When refined, all the relations must be computed while the physical quantities are straightforwardly injected from coarse values. 

\subsection{Solvers}

\texttt{EMMA} tracks the evolution of 3 `fluids', coupled to each other. First a collisionless fluid, sampled by particles in a standard particle-in-cell description. It is aimed at modeling the dynamics of stars and non-baryonic dark matter. Second, the code solves the Euler equations to describe the gas dynamics. \dom{We} chose to follow \texttt{RAMSES} by using a piecewise linear method 'a la' MUSCL-Hancock driven by HLLC Riemann solvers. Third, \texttt{EMMA} models the propagation of radiation using a moment description of the radiative transfer (RT hereafter)  equation with the M1 closure relation. Preliminary atomic processes are also included to describe the cooling and ionization that takes place on top of the adiabatic evolution of the gas. Finally, \emma deals with cosmological settings through the use of supercomoving coordinates, which are briefly discussed.

\subsubsection{Collisionless dynamics}
Collisionless dynamics are handled through a Monte-Carlo sampling of phase space using particles (see e.g. \citet{HOC81} for a reference).  Each particle consists in a structure of data that contains:
\begin{enumerate}
\item its fundamental properties: mass, 3D position ${\bf r}= x,y,z$ and velocity ${\bf v} =v_x,v_y,v_z$,
\item the level of the cell it belongs to,
\item 2 pointer towards the next and the previous particle that belong to the same cell.
\end{enumerate} 
Additionally, if collisionless dynamics is included, a non-refined cell contains a pointer toward the first particle it contains. If a cell is split, the particles of the coarse cell are passed to the newly created cells. If an oct is destroyed, all the particles are assigned to the parent cell. By means of linked lists, the code can therefore scan all the particles of a given cell.

Each particle is evolved in phase space thanks to the usual Newton's equations :
\begin{eqnarray}
\frac{d \bf v}{d t}&=&-\nabla \phi(\bf r),\\
\Delta \phi(\bf r)&=&4\pi G \rho(\bf r),
\end{eqnarray}
where $\rho(\bf r)$ stands for the 3D total matter density and $\phi(\bf r)$ is the associated gravitational potential. Updating the phase space coordinates of a particle therefore implies an evaluation of the density and the potential fields on the refined mesh.

The density is computed in all cells of the AMR grid using a standard cloud-in-cell (CIC) interpolation scheme. In practice it is performed level by level, according to the time stepping procedure described in section \ref{s:tstep}. If an unsplit cell $c^\ell_i$ contains particles, they are assigned to cells of the same level $\ell$ according to the CIC scheme. Furthermore if a neighbor cell is split, the contribution of the particles in its $\ell+1$ cells is also taken in account to compute the density of $c^\ell_i$. \dom{Conversely, if a neighbor cell is coarser and unsplit, its particles will also contribute to the density of $c^\ell_i$. In these two situations, the CIC extent of the particles will correspond to the level $\ell$ resolution, resulting in a CIC density equals to the one that would be obtained from a uniform grid at level $\ell$.} Finally if $c^\ell_i$ is split, its density is computed by averaging the 8 values of its child cells.

The density being available, the potential is obtained by solving the Poisson equation. In the case of \texttt{EMMA}, the solution is obtained by means of relaxation techniques. They allow for a great flexibility (notably compared to FFTs based approaches) in dealing with complex domain geometries, arbitrary boundary conditions and grids with multiple resolutions. One can use the simple Jacobi iteration formula where the estimation $p+1$ for the potential in the cell of Cartesian coordinates $(i,j,k)$ can be computed from a previous estimation $p$ via:
\begin{eqnarray}
\phi^{p+1}_{i,j,k}&=&\frac{\phi^{p}_{i+1}+\phi^{p}_{i-1}+\phi^{p}_{j+1}+\phi^{p}_{j-1}+\phi^{p}_{k+1}+\phi^{p}_{k-1}}{6}\\
&&+\frac{4\pi G\Delta x^2 \rho}{6}\label{e:jacobi}
\end{eqnarray}
The convergence rate of this process is slow, especially to achieve convergence on scales comparable to the total volume. As can be seen in eq. \ref{e:jacobi}, information can only be propagated from one cell to the next one, making such formula inefficient at determining the low frequency modes of the solution. It can be marginally accelerated by means of over-relaxation or by using the Gauss-Seidel iteration technique, analog to Jacobi but where new estimates of the potential $\phi^{p+1}$ are used on the RHS of equation \ref{e:jacobi} as soon as they are available instead of $\phi^p$.
 
Greater acceleration rates can be obtained by using a multigrid algorithm. The first stage is a `smoothing operation' where a simple relaxation formula such as eq. \ref{e:jacobi} is used to remove spurious high frequency features in the current estimation of the potential, i.e. in the regime of scales it is the most efficient. Then this estimate of $\phi^p$ is coarsened through a restriction operation and applied as a first guess for a new relaxation stage, which is more efficient at low resolution for two reasons : first the number of sampling points is reduced (typically by a factor of 8 in 3D) and second, the convergence of low frequency modes is increased as their effective scale is reduced relative to the influence radius of eq. \ref{e:jacobi}. Once this low resolution solution is converged, it is interpolated back (also called prolongation step) and after a few relaxation steps at full resolution a converged solution is obtained.


This two level-process (between $\ell$ and $\ell-1$)  can be recursively applied to $\ell-1$ and $\ell-2$ and so on. In \texttt{EMMA}, we typically proceed until $l=2$ (i.e. using a $4^3$  field) to achieve a $10^-3$ convergence of the solution with only $\sim 5$  iterations at full resolution. We take advantage of the AMR structure to store residuals and perform the calculations on coarse levels. Prolongation is performed via linear interpolation, whereas restriction is done by averaging on the 8 cells of an oct.
 
For fine levels with $\ell>\ell_c$, the potential is computed by simple relaxation, using a red-black Gauss-Seidel smoother, with boundary conditions provided by the $\ell-1$ cells that surround high resolution patches. However, the initial potential evaluation is obtained by interpolation of the coarse solution. Therefore, convergence is quickly obtained from this first value that is already close to the correct one. For \texttt{EMMA}, convergence at the $10^{-3}$ level are obtained after $~10$ iterations in a typical cosmological simulation. Fine cells at resolution jumps must compute the potential using high resolution values that do not exist on the coarse side of the jumps. In this case, the potential is interpolated at the requested location from its coarse value.

The potential being available, the \dom{acceleration} field ${\bf f}=-\nabla \phi$ is computed by simple derivation and is interpolated back at the particles positions again accordingly to the CIC scheme. If a particle lies close to the interface between two regions at different levels $\ell$ and $\ell-1$, the force field is interpolated from the coarsest $\ell-1$ level. As such, it implies that a particle must penetrate significantly within high resolution regions to be sensitive to their force field, otherwise they will be driven by a lower resolution description of the potential. 

Particles are advanced thanks to a mid-point scheme. The $p+1$ value of the position and the velocities are computed from their $p$ value as:
\begin{equation}
{\bf r}_{p+1}={\bf r}_{p}+{\bf v}_{p+1/2}\Delta t^p
\end{equation}
where the mid-point velocity is provided by
\begin{equation}
{\bf v}_{p+1/2}={\bf v}_{p}+{\bf f}_p \frac{\Delta t^p}{2}
\end{equation}
and finally corrected as:
\begin{equation}
{\bf v}_{p+1}={\bf v}_{p+1/2}+{\bf f}_{p+1} \frac{\Delta t^p}{2}
\end{equation}
 
\subsubsection{Hydrodynamics}
Hydrodynamics are solved through a Eulerian description of conserved fluid quantities, which obey the set of Euler equations (see e.g. \citet{TOR97}):
\begin{equation}
\frac{\partial \vU}{\partial t}+\frac{\partial \vF(\vU)}{\partial x}+\frac{\partial \vG(\vU)}{\partial y}+\frac{\partial \vH(\vU)}{\partial z}={\bf S},
\label{e:euler}
\end{equation}
where $\vU=(\rho,\rho u,\rho v,\rho w,E)$ is the array of conserved quantities, the density, the 3 components of momentum and the energy. $\vF$, $\vG$ and $\vH$ are the 3 flux functions with
\begin{eqnarray}
\vF(\vU)&=&(\rho u, \rho u^2+p, \rho uv, \rho uw, u(E+p))\\
\vG(\vU)&=&(\rho v, \rho uv, \rho v^2+p, \rho vw, v(E+p))\\
\vH(\vU)&=&(\rho w, \rho uw, \rho vw, \rho w^2+p, w(E+p))
\end{eqnarray}
In association with conserved quantities, this is usual to also consider the primitive quantities, namely the density $\rho$, the velocities $\vv=(u,v,w)$ and the pressure $p$. The total energy is given by 
\begin{equation}
E=\frac{\rho}{2}(\vv^2)+\frac{p}{\gamma-1},
\label{e:EP}
\end{equation}
with contribution of the kinetic and the internal energy. Here $\gamma$ is the usual adiabatic exponent, equal to $5/3$ for an ideal mono-atomic gas. The High-Mach flow regime is taken in account using the recipe described in \citet{RAS06}.

The set of Euler equation is solved in a split fashion, dealing first with the pure transport part with a null r.h.s in eq. \ref{e:euler} then updating the solution by adding the contribution of source terms. The transport update of $\vU^p$ into $\vU^{p+1}$ is done explicitly (in 1D here for simplicity) via:
\begin{equation}
\vU^{p+1}_i=\vU^{p}_i+\frac{\Delta t}{\Delta x}(\mathcal{F}^p_{i-1/2}-\mathcal{F}^p_{i+1/2}).
\label{e:eulersol}
\end{equation}
\dom{This solution requires the knowledge of intercell fluxes $\mathcal{F}^p_{i\pm1/2}$ at instant $p$ between cells $i$ and $i\pm1/2$, through the resolution of Riemann problems.} Here we use a MUSCL scheme coupled to an HLLC Riemann Solver (see e.g. \citet{TOR94,TOR97}). \dom{First, conserved quantities are linearly reconstructed at the cell boundaries (in 1D for simplicity)~:
\begin{equation}
\vU_i^{L/R}=\vU_i^p\pm\frac{\Delta_i}{2}.
\end{equation}
Here $L/R$ designates the left and right reconstructed states at the cell $i$ boundaries in $x=0$ and $x=\Delta x$, the cell center being in $\Delta x/2$. $\Delta_i$ is the slope vector of the conserved quantities and is computed using neighbor cells values and a MinMod limiter to ensure a  monotonic solution. Before solving the Riemann problem, the boundary extrapolated values are also evolved by a time $\Delta t/2$~: 
\begin{equation}
\bar \vU_i^{L/R}=\vU_i^{L/R}+\frac{\Delta t}{2\Delta x}\left[\vF(\vU^L_i)-\vF(\vU^R_i)\right].
\end{equation}
The Riemann problem at intercell positions are solved using these evolved states, for instance  $\mathcal{F}^p_{i+1/2}$ is obtained from states $\bar \vU_i^R$ and $\bar \vU^L_{i+1}$. The MUSCL scheme achieves second order accuracy \citep{TOR94,TOR97}. As discussed previously, when dealing with cells at the interface between 2 regions with different resolutions, coarse quantities are interpolated in a conservative manner at the locations of the 'virtual' fine cells. At such interfaces, the low resolution values are assumed to be constant in time and accuracy reduces to first order \citep{KHO98,TEY02}. } The multi-dimensionality of the problem is dealt through an unsplit approach with a flux contribution of the 3 directions included at once in the update.

The stability of the solution is obtained by enforcing the Courant condition on the time step
\begin{equation}
\Delta t= C \frac{\Delta x}{3V_h},
\end{equation}
where $V_h$ is an estimate of the largest wave speed present throughout the computational domain and $C<1$. We follow the simple suggestion of Toro with 
\begin{equation}
V_h=\mathrm{max}\{|\vv|_i+a_i\}
\label{e:couranthyd}
\end{equation}
where $a_i$ stands for the sound speed at location $i$. It usually overestimates the limiting velocity and therefore underestimates the corresponding time step, but provides a robust estimation.

The source term in Eq. \ref{e:euler}, models deviation to pure conservation. In our case $\vS$ holds the contribution of the gravitational force to momentum and energy variation and expresses the coupling of gravitation with hydrodynamics :
\begin{equation}
\vS=(0,-\rho{\bf \vec\nabla}\phi,-\rho \vv{\bf \vec \nabla}\phi),
\end{equation}
where $\phi$ is the gravitational potential. Its inclusion in the solution (eq. \ref{e:eulersol}) is done in a 'explicit' manner with
\begin{equation}
\vS=(0,-\rho^{p}{\bf \vec \nabla}\phi^{p},-\rho^{p} \vv^{p}{\bf \vec \nabla}\phi^{p}).
\end{equation}
This contribution modifies the conservative quantities after the pure transport update. However they are taken in account within the MUSCL scheme to compute the interpolated left/right states before solving the Riemann problems. Of course hydrodynamics is also coupled to radiative physics and thermo-chemistry via the internal energy (or equivalently the pressure) of the gas~: photo-heating and cooling act as source and sink terms of this quantity and are treated by the radiative transfer engine. Finally, the hydro-engine is able to handle passive scalars that are being advected with the fluid. Among such scalars, one can currently find the hydrogen neutral fraction or in forthcoming developments one can think of the metallicity.

\subsubsection{Radiative Transfer}
\label{s:rt}
Propagation of radiation is dealt with using a moment based description : light is described as fluid, where its phase space distribution is averaged on velocities to focus on spatial fields. Among this family of radiation description, \texttt{EMMA} relies on the M1 approximation \citep{LEV84,GON07,AUB08}. 

Taking the first two moments of the Liouville Equation leads to the conservation equations of the ionizing photons density $N_\nu(\br,t)$ and flux $\bF_\nu(\br,t)$, in each bin of frequency $\nu$:
\begin{eqnarray}
\frac{\partial N_\nu}{\partial t}+\frac{\partial \bF_\nu}{\partial \br}&=&S_\nu-\kappa_N N_\nu,\label{e:pdert1}\\
\frac{\partial \bF_\nu}{\partial t}+c^2\frac{\partial {\bf P_\nu}}{\partial \br}&=&-\kappa_F \bF_\nu.
\label{e:pdert2}
\end{eqnarray}
Here, $\bf P_\nu(\br,t)$ stands for the radiative pressure tensor. This system of equation being opened, a closure relation is required to be able to solve it. The M1 closure relation is given by:
\begin{eqnarray}
{\bf P_\nu}&=&{\bf D_\nu} N_\nu\\
{\bf D_\nu}&=& \frac{3\chi-1}{2} {\bf I} +\frac{1-\chi}{2} {\bf n}\times{\bf n}.
\end{eqnarray}
Here, $\bf D_\nu$ stands for the Eddington tensor and its expression is set by the value of the $\chi$ quantity that varies between $1/3$ for a diffusive regime (i.e. $F_\nu\ll cN_\nu$) and $1$ for a pure transport regime (i.e. $F_\nu\sim cN_\nu$). 

The quantities on the r.h.s. of the conservation equations are the source of photons $S_\nu(\br,t)$ (expressed in photons per unit time per unit volume) and the two absorption terms, driven by the absorption coefficients $\kappa_N$ and $\kappa_F$, considered equal in \texttt{EMMA} with $\kappa_N=c\sigma_\nu n_H$. This terms couple the hydrodynamics and the radiative transfer by means of gas absorption with $n_H$ being the density number of neutral gas and $\sigma_\nu$ the photo-ionization cross-section at frequency $\nu$.

In the case of multi-frequency transfer, photons are gathered in so-called 'groups' of frequencies and the flux and number densities of a given group satisfy the above conservative equations. In practice, Eqs \ref{e:pdert1} and \ref{e:pdert2} can be integrated between two frequencies:
\begin{eqnarray}
\frac{\partial N}{\partial t}+\frac{\partial \bF}{\partial \br}&=&S-c\sigma_N n_H N,\label{e:pdert1i}\\
\frac{\partial \bF}{\partial t}+c^2\frac{\partial {\bf P}}{\partial \br}&=&-c\sigma_N n_H \bF,
\label{e:pdert2i}
\end{eqnarray}
with 
\begin{eqnarray}
N&=&\int_{\nu_1}^{\nu_2} N_\nu d\nu\\
\bF&=&\int_{\nu_1}^{\nu_2} \bF_\nu d\nu\\
S&=&\int_{\nu_1}^{\nu_2} S_\nu d\nu\label{e:sgroup}\\
\sigma_N&=&\frac{1}{N}\int_{\nu_1}^{\nu_2} \sigma_\nu N_\nu d\nu.\label{e:siggroup}
\end{eqnarray}
Typical frequency groups have limits set by the ionization levels of hydrogen and helium (even though \texttt{EMMA} does not currently handle Helium chemistry) i.e. [13.6,24.6, 54.4] eV or chosen to represent broad classes of different types of radiation such as UV, X and hard X-rays.

In practice, the update of radiative quantities is a two stages process~: first a conservative transport is performed, then non-conservative contributions (source and sinks) are added within a subsequent thermo-chemical solver (see Sec. \ref{s:thermo}). \dom{The set of homogeneous (with zero r.h.s) coupled equation is solved for ${\bf U}=(N,{\bf F})$ using fluxes ${\tilde F}=({\bf F},c^2 {\bf P} )$  with a simple explicit finite difference scheme (in 1D for sake of simplicity):}
\begin{equation}
{\bf U}^{p+1}_i={\bf U}^{p}_i +\frac{\Delta t}{\Delta x}({\mathcal F}^p_{i-1/2}-{\mathcal F}^p_{i+1/2})
\label{e:uprt}
\end{equation}
that must satisfies the usual Courant Condition:
\begin{equation}
c\le\frac{\Delta x}{\Delta t}.
\end{equation}
Here ${\mathcal F}^p_{i-1/2}({\bf U})$ represents the flux function at instant $p$ measured at the interface between the cell $i$ and $i-1$. This intercell flux is obtained by solving a typical Riemann problem at this interface : in \texttt{EMMA}, this flux is given by the Lax-Friedrich Formula:
\begin{equation}
\mathcal{F}_{i+1/2}({\bf U})=\frac{\tilde{F}_i+\tilde{F}_{i+1}}{2}-c\frac{{\bf U}_{i+1}-{\bf U}_{i}}{2}.
\label{e:glf}
\end{equation}
As in \texttt{ATON}, an implementation of the less diffusive HLL flux is also on the way. At this stage, conservative transport is completed and radiative quantities (N,\bF) are in an intermediate state, waiting for the contribution of source and sinks to be taken in account, as explained in the next section.

Before, it should be noted that the Courant condition ensures the stability of the scheme but impose that the numerical 'sampling velocity' $\Delta x/\Delta t$ must be greater than the speed of light. The cost of simplicity provided by the explicit solver is therefore a very fine temporal sampling, greatly enhancing the CPU-cost of the radiative transfer. This stringent constrains on the time step imposed by the Courant condition can be circumvented in two ways. The first one is simply by taking advantage of hardware acceleration (as in \texttt{ATON}). In \texttt{EMMA} this route is explored with GPUs as accelerating devices and described in section \ref{s:para}. The second one is to reduce the speed of light (\citealt{GNE01,ROS13}), taking advantage of the fact that in a large number of situations, the effective propagation of radiative information is performed through ionization fronts that propagate at smaller pace. This option can also be set in \texttt{EMMA}, as done for instance in Sec. \ref{s:reionsimu}.

\subsubsection{Thermal and chemical processes}
\label{s:thermo}
The thermal and 'chemical' processes encompass the atomic physics that will affect hydro and radiative quantities. Currently \texttt{EMMA} only handles atomic hydrogen processes. They contribute to change the number density and flux of photons (source and sinks), the ionization state of the gas and finally its internal energy:
\begin{eqnarray}
\frac{dN}{dt}&=&S-c\sigma_N n_H N +(\alpha_A(T) -\alpha_B(T))x^2 n_0^2, \label{e:therm1}\\
\frac{d\bF}{dt}&=&-c\sigma_N n_H \bF,\label{e:therm2}\\
\frac{d n_H}{dt}&=&(\alpha_A(T) x^2 -\beta(T) x(1-x))n_0^2- c n_H \sigma_N  N,\label{e:therm3}\\
\frac{d e}{dt}&=&c n_H \Sigma_E N-\Lambda(n_0,x,T),\label{e:therm4}
\end{eqnarray}  
where $n_H=(1-x)\rho/m_p=(1-x)n_0$ is the number density of hydrogen atoms (with individual mass $m_p$),  $x$ being their ionized fraction and $n_0$ being the total number of protons (i.e. neutral + ionized hydrogen). \dom{The quantities $\alpha_{A/B}$ and $\beta$ are respectively the case A/B recombination and collisional ionization rates. Eqs. \ref{e:therm1} and \ref{e:therm2} describe the influence of source and sinks on the radiative quantities~: in particular the last term in Eq. \ref{e:therm1} refers to the recombining ionizing radiation. Note that the on-the-spot approximation can easily be applied by setting $\alpha_A=\alpha_B$ in Eqs.  \ref{e:therm1} and \ref{e:therm3}.} Eq. \ref{e:therm3} details the competing effects of recombination and ionizations on the number of neutral hydrogen atoms. Eq. \ref{e:therm4} encodes the evolution of the internal energy density of the gas $e=p/(\gamma-1)$ due to atomic cooling (given by the cooling rate $\Lambda$) and to the photo-heating above the ionization threshold $\mathcal{H}=c n_H N \Sigma_E$. Here $\Sigma_E=(\sigma_E \langle E\rangle -\sigma_N E_\mathrm{13.6})$ and the quantity $\sigma_E$ is the energy averaged cross-section over the group of frequencies of interest:
\begin{eqnarray}
\sigma_E&=&\frac{1}{N \langle E \rangle}\int_{\nu_1}^{\nu_2} \sigma_\nu N_\nu h\nu d\nu,\\
\langle E\rangle&=&\frac{1}{N}\int_{\nu_1}^{\nu_2} N_\nu h\nu d\nu
\end{eqnarray}
where the latter quantity $\langle E \rangle$ is the average photon energy in the same group.

 In the case of multi-frequency transfer, Eqs (\ref{e:therm1}-\ref{e:therm4}) are solved for each frequency interval  (i.e. group) $[\nu_i,\nu_{i+1}]$ where $i$ stands for a group label. Within each frequency group $i$ the corresponding photon density and fluxes $(N_i,\bF_i)$ satisfy :
 \begin{eqnarray}
\frac{dN_i}{dt}&=&S_i-c\sigma_{N,i} n_H N_i +(\alpha_A(T) -\alpha_B(T))\delta_{i,1}x^2 n_0^2 \label{e:therm1g}\\
\frac{d\bF_i}{dt}&=&-c\sigma_{N,i} n_H \bF_i.\label{e:therm2g}
 \end{eqnarray}
These equations depend on the group cross-section $\sigma_{N,i}$ and source function $S_i$ obtained using Eqs \ref{e:sgroup} and \ref{e:siggroup} on the $[\nu_i,\nu_{i+1}]$ interval. The number of radiative conservative updates therefore scales as the number of frequency groups. The Kronecker symbol $\delta_{i,1}$ in Eq. \ref{e:therm1g} implies that the recombining radiation only contributes to the first group of ionizing photons, the closest to the hydrogen ionization frequency (see e.g. \cite{ROS13}). Again, if the on-the-spot approximation is used, this contribution is set to zero for all the groups. The thermo-chemical equations Eqs. \ref{e:therm3} and \ref{e:therm4} are also modified and the following expressions must be replaced:
\begin{eqnarray}
c n_H \sigma_N  N &\rightarrow & c n_H \sum_{i=1}^\mathrm{Ngroups} \sigma_{N,i}  N_i,\label{e:g2}\\
c n_H \Sigma_E N&\rightarrow& c n_H\sum_{i=1}^\mathrm{Ngroups}  \Sigma_{E,i} N_i.\label{e:g3}
\end{eqnarray}
$\mathrm{Ngroups}$ stands for the total number of frequency intervals considered. Eqs. \ref{e:g2} and \ref{e:g3} depend on the groups cross-sections and photons densities and effectively couple photons of different frequencies that are otherwise transported without any interactions. 

In practice, we solve the equations in the same order as above by sub-cycling the dynamical step $\Delta t$ with a chemical time step $\Delta \tau$. During the latter, all quantities are updated from state $p$ to $p+1$. $\Delta \tau$ is first evaluated from the $p$ state of the internal energy, with $\delta \tau= e^p/(\mathcal{H}^p-\Lambda^p)$. Then all the quantities are updated in the following manner and in that order:
\begin{eqnarray}
N^{p+1}&=&\frac{N^{p}+(S+(\alpha_A(T) -\alpha_B(T))x^2 n_0^2)\Delta \tau}{1+c\sigma_N (1-x^p) n_0}\\
\bF^{p+1}&=&\frac{\bF^{p}}{1+c\sigma_N (1-x^p) n_0}\\
x^{p+1}&=&1-\frac{\alpha_A(T^p) x^{p2} n_0 \Delta \tau+ 1-x^p}{1+\Delta\tau(\beta(T^p) x^p n_0 + c\sigma_N N^{p+1})}\\
e^{p+1}&=&e^p\\
&&+\Delta \tau c(1-x^{p+1}) n_0 N^{p+1} \Sigma_E \\
&&-\Delta\tau \Lambda(n_0,x^{p+1}, T^p)).
\end{eqnarray}
In this sequence, new information on a given quantity is immediately used to compute a subsequent one. We follow \citet{ROS13} and enforce that the internal energy must at most vary by 10\% (relatively) during $\Delta \tau$, otherwise the set of equation is recomputed with a time step divided by 2 until this condition is satisfied. In all our experiments, this procedure has been found to be accurate and robust enough. \dom{All the rates required to describe the atomic processes such as the recombination, the collisional ionization, and the cooling are taken from the compilation of \citet{THE98}. Photo-ionization cross-sections are taken from \citet{HUI97}.}

\subsection{Time stepping}
\label{s:tstep}

\begin{figure} 
\includegraphics[width=1.\columnwidth]{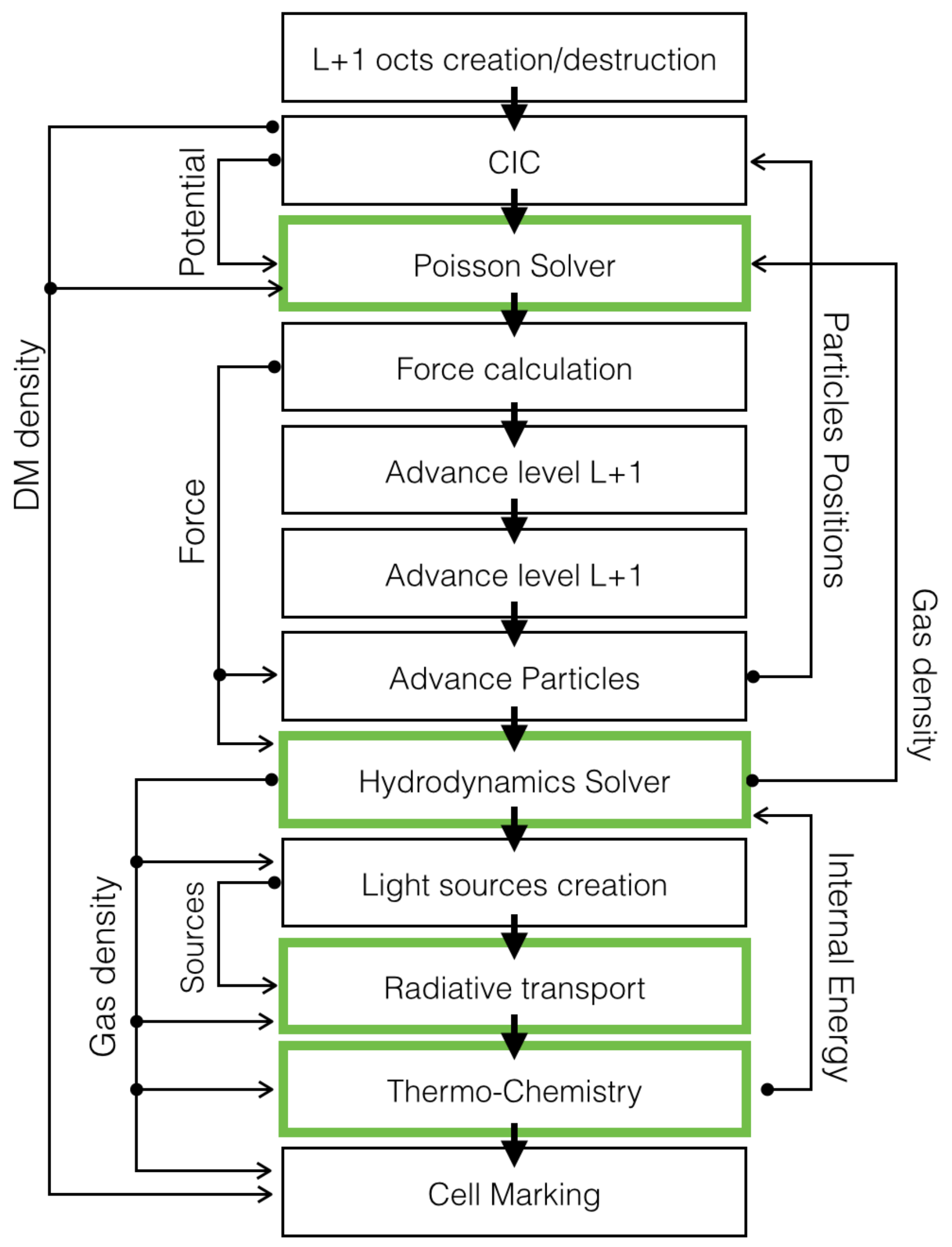}
\caption{Flow of sequence for a time step at a given level (from top to bottom). Side arrows describe the exchange of physical quantities between different modules to emphasize the most important couplings. Green boxes stand for modules that have been ported on GPU using CUDA. It is assumed that $\Delta t_\ell=2\Delta t_{\ell+1}$ and the $\ell+1$ level is advanced twice. \dom{This sequence is repeated recursively for all the finer levels.}}
\label{f:diag}
\end{figure}

The organization of time step is intimately constrained by the multi-level structure of the data. A single level time step is organized in quite an usual fashion and described in Fig. \ref{f:diag}. Coupling between the different physics occur at different levels, the most explicit ones being :
\begin{itemize}
\item between Gravity and Hydrodynamics through the total matter density and the gravitational force it creates,
\item between radiation and hydrodynamics through the density distribution, or the gas temperature.
\end{itemize}
On top of these explicit coupling, detailed in the next section, implicit ones also occur where e.g. a photo- evaporating halo could in principle affect the underlying dark matter distribution (in the same way supernovae feedback could affect it, even though they are not explicitly coupled, e.g. \citealt{PON12})
\begin{figure}
\includegraphics[width=\columnwidth]{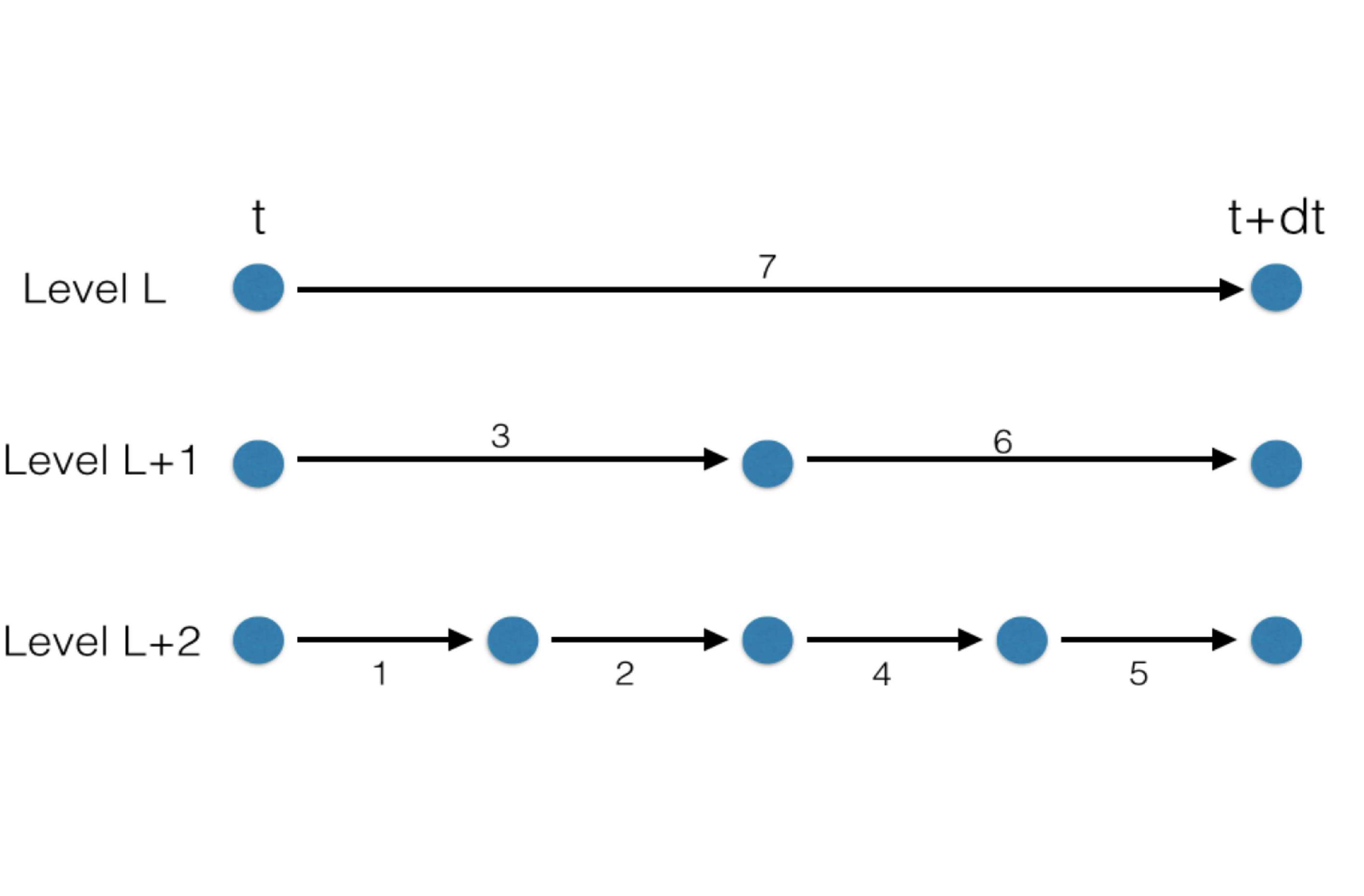}
\caption{The time-step hierarchy of 3 level of the AMR structure. The coarse level $\ell$ is advanced from $t$ to $t+dt$, which implies partial advance and temporary updates of fine levels $\ell+1$ and $\ell+2$ until they are synchronized with $\ell$. Arrows and numbers indicate the sequence of partial updates at different levels to perform this full update of the 3 nested levels.}
\label{f:tstep}
\end{figure}

When mesh refinement is enabled, coupling between levels must be taken into account with special care. In \texttt{EMMA}, updates are performed level by level, each level being updated with its own time step $\Delta t_\ell$. Typically, $\Delta t_{\ell}\sim\Delta t_{\ell+1}\times 2$ ($\ell$ corresponding to a level coarser than $\ell+1$). Advancing the solution on a level $\ell$ can be expressed as the following recursive expression:
\begin{equation}
A_\ell=R_{\ell+1} P_\ell A_{\ell+1} U_\ell M_{\ell},
\end{equation}
where $A_\ell$ stands for 'advancing the solution' at level $\ell$. $R_{\ell}$ is the refinement operator (i.e. creating/destroying $\ell+1$ octs from $\ell$ cells). $P_l$ and $U_l$ are the operators for the Poisson resolution and the update (i.e. moving particles, updating Eulerian fields) at level $\ell$. Finally, $M_\ell$ corresponds to the marking of the level $\ell$ cells for future refinements. The recursion is stopped at the maximal allowed level with $R_{\ell\mathrm{max}}=M_{\ell\mathrm{max}}=A_{\ell\mathrm{max}+1}=1$. 
For instance, a simulation with three resolution levels (e.g. $\ell=5,6,7$) will be fully updated on $\Delta t_5$ according to the following operations (Fig. \ref{f:tstep} details this step in a schematic manner) :
\begin{eqnarray}
A_5&=&R_5 P_5 \left[ R_6 P_6 \left[ A_7\right] U_6 M_6\right]\left[  R_6 P_6 \left[ A_7\right] U_6 M_6\right] U_5 M_5\nonumber\\
A_7&=&P_7 U_7 P_7 U_7
\end{eqnarray} 
where we assumed that $\Delta t_5=2 \Delta t_6= 4 \Delta t_7$. More generally, the time steps are constrained by :
\begin{equation}
\Delta t^1_\mathrm{\ell+1}+\Delta t^2_\mathrm{\ell+1}\le \Delta t_\ell.
\end{equation}
Once the level $\ell+1$ has been updated by $\Delta t^1_\mathrm{\ell+1}+\Delta t^2_\mathrm{\ell+1}$, the value $\Delta t_\ell$ is updated to synchronize the coarse level on the finer one.

\begin{figure}
\includegraphics[width=\columnwidth]{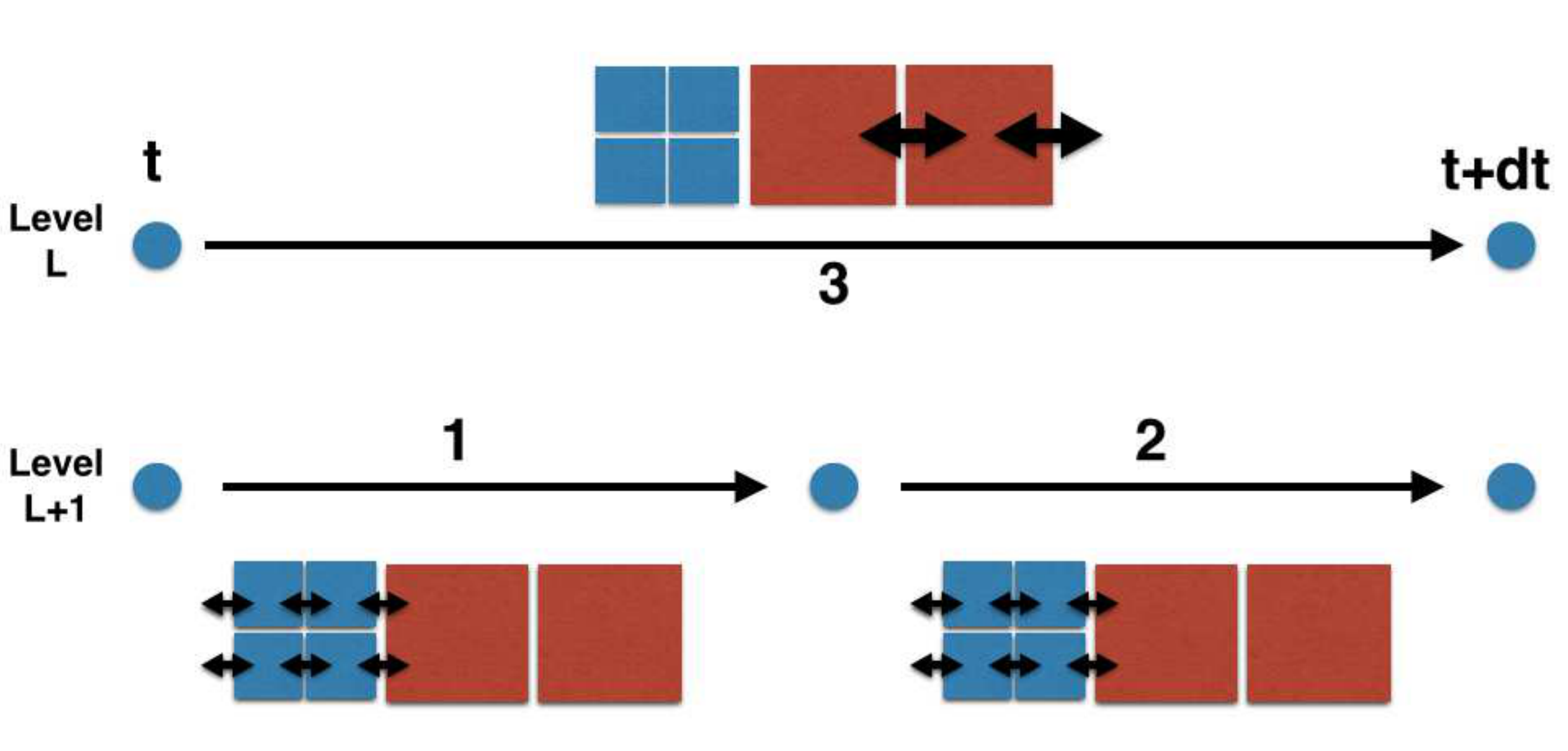}
\caption{Multilevel update of fluxes at the interface between two levels. To perform an update from $t$ to $t+dt$, fine level $\ell+1$ must perform 2 steps sub-cycled within a large step of the coarse level $\ell$. Updates are performed according to the 1,2,3 sequence~: note how the middle coarse cell is updated during the 3 steps with fine fluxes first and a final coarse flux,during the level $\ell$ update.}
\label{f:flux}
\end{figure}

As originally described by \citet{KHO98}, this nested hierarchy of time step has some strong implication on the flux update at interfaces between two levels. \dom{When conservative quantities are updated on level $\ell+1$, adjacent $\ell$ cells must take in account the fluxes produced during the subcycling of level $\ell+1$.} For instance, if a conservative quantity $U_\ell$ and $U_{\ell +1}$ in two adjacent cells must be updated on $\Delta t_\ell$, the following sequence must be obeyed (see also Fig. \ref{f:flux}):
\begin{enumerate}
\item $U_{\ell+1}^p$ is updated to $U_{\ell+1}^{p+1/2}$ using a flux $F(U_{\ell+1}^p,U_{\ell}^p)$. Meanwhile $U_{\ell}^{p+1}$ is temporarily updated using the same flux.
\item $U_{\ell+1}^{p+1/2}$ is updated to $U_{\ell+1}^{p+1}$ using a flux $F(U_{\ell+1}^{p+1/2},U_{\ell}^p)$. Again $U_{\ell}^{p+1}$ is temporarily updated using the same flux.
\item Finally, $U_{\ell}^{p+1}$ is fully updated taking in account the flux created by adjacent cells of level $\ell$.
\end{enumerate}
This partial update of coarse cells during fine time steps is one of the reason to enforce a maximal jump in resolution of unity between two adjacent cells. Larger jumps would create a complex hierarchy of partial update of coarse quantities, difficult to handle properly.

The amplitude of time steps is set by physical temporal scales that must be resolved to track properly their impact on the evolution of quantities. \texttt{EMMA} being a multi-physics code, a whole list of time scales are computed for all the cells of a given level and the smallest set its global time step:
\begin{itemize}
\item we follow \citet{TEY02}, and limit the rate of change of the cosmological expansion factor $\delta a(\Delta t_\mathrm{cosmo})/a <\epsilon$,
\item a particle cannot move on a scale larger than the size of a cell~: $\Delta t_\mathrm{pic}=\epsilon \Delta x_\ell/v_\mathrm{max}$,
\item the local dynamical time must be resolved~: $\Delta t_\mathrm{dyn}=\epsilon/\sqrt{G\rho}$,
\item the hydrodynamical Courant condition must be satisfied ~: $\Delta t_\mathrm{hyd}<\Delta x_\ell/V_h$ (see Eq. \ref{e:couranthyd})
\item If light sources are present, the radiation Courant condition must be satisfied ~: $\Delta t_\mathrm{rad}<\Delta x_\ell/c$. It ensures that light propagation is performed in a stable manner.
\end{itemize}
When full physics are included and effective, the most stringent condition is usually provided by $\Delta t_\mathrm{rad}$ and by orders of magnitude since usually $c\gg V_h$. This dominance can be reduced by setting a reduced speed of light. Furthermore as non-linearities increase (and even more when strong shocks such as induced by supernovae feedback will be included) the ratio $\Delta t_\mathrm{hyd}/\Delta t_\mathrm{rad}$ tends to decrease.
 
\section{Cosmological Setting}
\label{s:cosmo}
Cosmological experiments are implemented using the set of 'super-comoving' variables suggested by \citet{MAR98}. The transformation from physical to supercomoving variables are given by:
\begin{eqnarray}
\tilde \br&=& \frac{\br}{ar_*},\\
\tilde \vv&=&\frac{a\vv}{v_*},\\
\tilde \rho&=&\frac{\rho}{a^3\rho_*},\\
\tilde p&=&\frac{a^5p}{p_*},\\
\tilde \phi&=&\frac{a^2\phi}{\phi_*},\\
\tilde dt &=&\frac{dt}{a^2 t_*},\\
\tilde N &=& a^3 N r_*^3\\
\tilde \bF &=& a^4 r_*^2 t_* \bF,
\end{eqnarray}
where starred quantities stand for normalization units with $r_*=L$ (the box length), $t_*=2/(H_0 \sqrt{\Omega_m})$, $v_*=r_*/t_*$, $\rho_*=3H_0^2\Omega_m/(8\pi G)$ and $p_*=\rho_*v^2_*$. $H_0$ and $a(t)$ stand for the usual current Hubble parameter and the time-dependent expansion factor.

With this set of transformation, it can be shown that almost all the differential equations to be solved keep their standard expression for a $\gamma=5/3$ gas. The only notable exception is the Poisson equation which becomes:
\begin{equation}
\tilde \Delta \tilde \phi= 6 a \delta,
\end{equation}
where $\delta= \tilde \rho / \langle\tilde \rho\rangle -1$. Still, this equation remains typical of an elliptic equation that can be solved by all the methods already in place for the Newtonian field equation. Overall, the use of such a transformation greatly simplifies the implementation of cosmological settings in this kind of simulation code. 

\section{Parallelization and vectorization}
\label{s:para}
\texttt{EMMA} is a parallel code which includes two levels of multi-tasking. The first one is the standard multi-CPU mode, where the computational domain is distributed among several processes that communicate with each other via the MPI protocol. The second level of parallelism resides within an MPI-process where the local load is distributed among several threads of execution. In the case of \texttt{EMMA}, this local parallelisation is performed on GPUs but could in principle be extended to other modes of multi-threading such as local shared-memory parallelism among multiple CPU cores or other hardware accelerators. The second level of parallelism can be understood as a vectorization, where arrays of data are processed in parallels through the same set of instructions with minimal communications.

Bearing this two-levels parallelism in mind, \texttt{EMMA} has been designed to decouple as much as possible instructions that deal with the logistic of data from the ones that actually perform calculations. Logistics operations are defined as operating directly on the AMR tree~: e.g. cell marking and refinement, tree management and inter-process communications. These operations are handled by CPUs. Computing functions on the other hand expect arrays of data to be 'crunched', without any mention to tree-organized data or inter-process parallelism and return likewise array of results. The physics solvers belong to this second category and are meant to be processed by vector based-hardware, such as multi-core processors, GPUs or any other kind of co-processor. In between, a set of interface functions must be developed to perform gather/scatter operations from/to the AMR tree to/from the calculations arrays.
These aspects are developed in the next subsections.

\subsection{Distributed parallelism on multiple CPUs}
Distributed parallelism is handled through a space-filling curve domain decomposition. Such a curve provides a 1D mapping of a 3D grid by assigning a unique key to each oct as a function of its Cartesian position. The number $n_p$ of parallel processes being defined, the curve is split in $n_p$ successive parts with equal loads, thus assigning a set of octs to each process. The number $n_p$ can thus be arbitrary and the 1D mapping alleviates the need to deal with multiple boundaries along multiple directions. \texttt{EMMA} has been implemented with both a Peano-Hilbert space-filling curve and a slab-based key ordering for problems with unidirectional variations (such as the Shock Tube or the Zeldovich Pancake). Currently, the domain decomposition is performed at the \dom{level $\ell_c$ corresponding to the base resolution of the simulation~: all the $\ell_c$ octs are distributed among the processes, in such a way that each process possesses at least one such oct.}  All octs created from a level $\ell_c$ cell are assigned to the same process. At the current stage, \texttt{EMMA} does not perform any kind of load-balancing that could be obtained by sliding the limits of the 1D domains along the space-filling curve to optimize the distribution of work among the processes.

\begin{figure}
\includegraphics[width=\columnwidth]{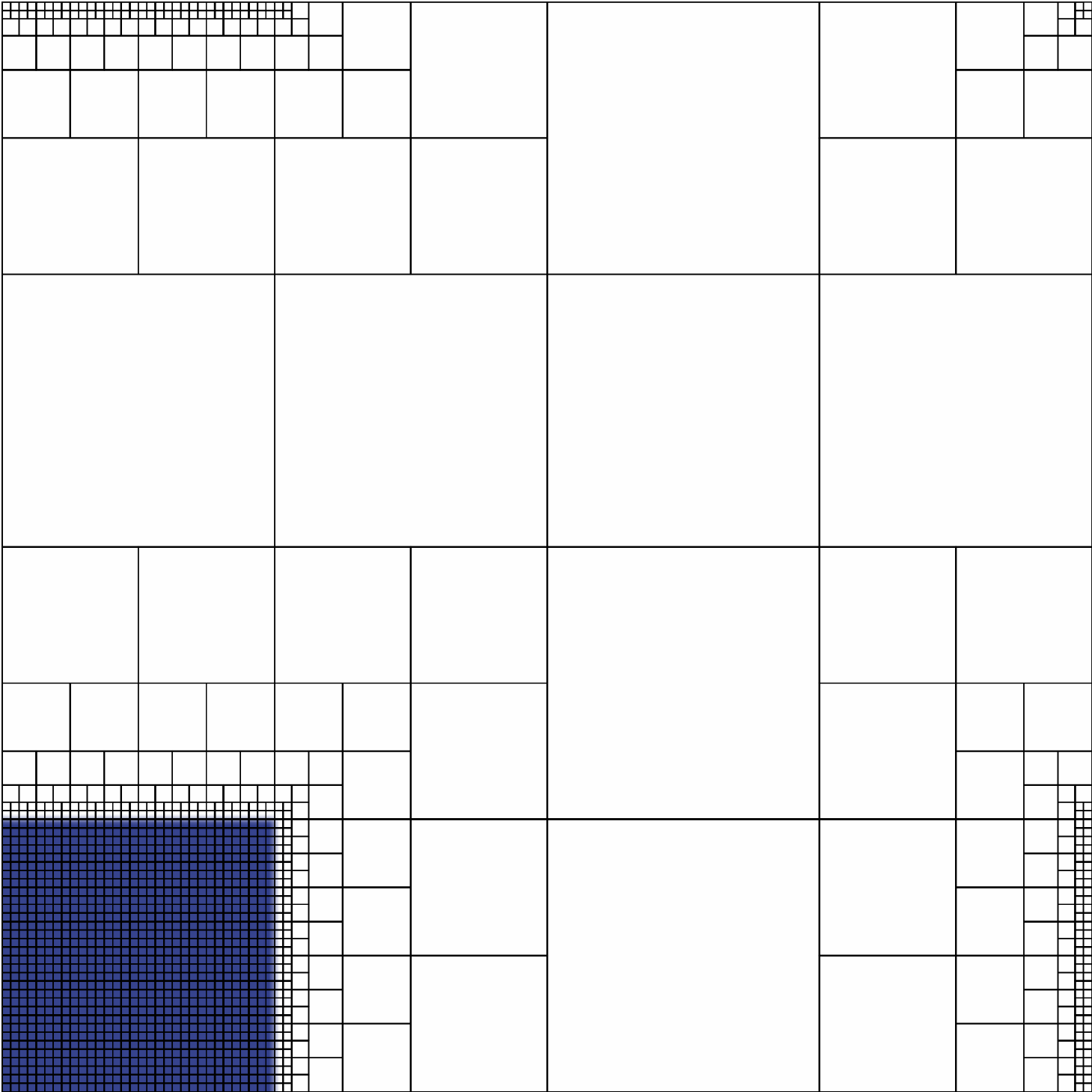}
\caption{An example of sub-domain seen by a processor in the essential tree paradigm. The processor has been assigned the lower left corner (shaded) but is aware of the whole computational domain. Distant regions are coarsened relative to close ones.}
\label{f:esstree}
\end{figure}
It should be emphasized that the AMR structure can only be fully exploited if it remains `consistent': no holes, no level jumps greater than 1 from one cell to another, pointers toward neighboring cells must exist, etc... Furthermore, a given process should be aware, at least partially, of the AMR tree structure of the neighbor processes. \texttt{EMMA} copes with these issues by employing the local essential tree decomposition \citep{WAR93,DUB96}~: each process, even though it has been assigned only a subset of the total volume, is aware of the whole hierarchy of nested octs but only at the levels relevant to its tasks (see Fig. \ref{f:esstree}). Neighbor cells directly in contact with its domain bring their whole tree from $\ell=1$ to $\ell_\mathrm{max}$, those being second order neighbors are one level coarser and so on. In practice, this local hierarchy is obtained at the initial building of the oct-tree: from the root $\ell=1$ oct, cells are refined down to the coarse level if they belong to the sub-volume assigned to the current processor or if they are direct neighbors of this sub-volume. It produces naturally a local tree for each processor, individually fully consistent and yet aware of the structure of the direct neighbors.

Communications are handled using the MPI-protocol and if a given process $P_0$ requires data present on other processes, it must be performed explicitly by specifying which octs should come from which other MPI process. This communication protocol in \texttt{EMMA} has been written in the following fashion:
\begin{itemize}
\item First, all processes $P_i$ build their own lists of neighbor MPI processes $\{P_j\}$ with $i\ne j$. 
\item For each member of $\{P_j\}$, a list of requests (i.e. of neighbor octs) is established by $P_i$ by storing their space-filling curve keys. This list of keys is sent from each $P_i$ to all its $\{P_j\}$~: $P_i$ acts as a client sending requests to neighbor servers. 
\item Likewise each $P_i$ receives a list of requests from the same sources~: $P_i$ acts as a server to neighbor clients.
\item Each client key is processed by $P_i$ through an hash table to relate the absolute key to a local pointer to an oct. The data is gathered and sent back to the clients $\{P_j\}$.
\item Meanwhile, the data from the servers $\{P_j\}$ is received and scattered back in the local tree by $P_i$.
\end{itemize}
This set of instructions is performed level by level and called by the Poisson, the hydrodynamics and the radiation solvers to update border cells that belong to  other processes and that have been remotely modified: the flux of information can be considered as outside-in. However there are situation where the information flux is inside-out: a process performed locally will affect directly a value outside its domain. The first example of inside-out communication is the CIC assignment, where a local particles will contribute to an adjacent domain. The second example are the conservative updates due to hydro or radiative fluxes between cells at different levels and belonging to different processes. Because this update is asynchronous between levels, conservative values can be updated in a neighbor coarser cell and must be communicated to its home process. In this case the protocol is similar to the one described above except that there are no request stage and data is sent directly from the server to the client.

It should be noted that without load-balancing, the list of neighbors $\{P_j\}$ is static for each $P_i$. Hence the first stage of the communication protocol has to be performed only once. However the list of \textit{neighbor octs} is dynamic, because of mesh refinement. Therefore, for a given pair client/server $P_i/P_j$, the list of requests is changing on the fly and should typically be recomputed any time octs have been created or destroyed.

\subsection{Local vectorization}
\begin{figure}
\includegraphics[scale=0.25]{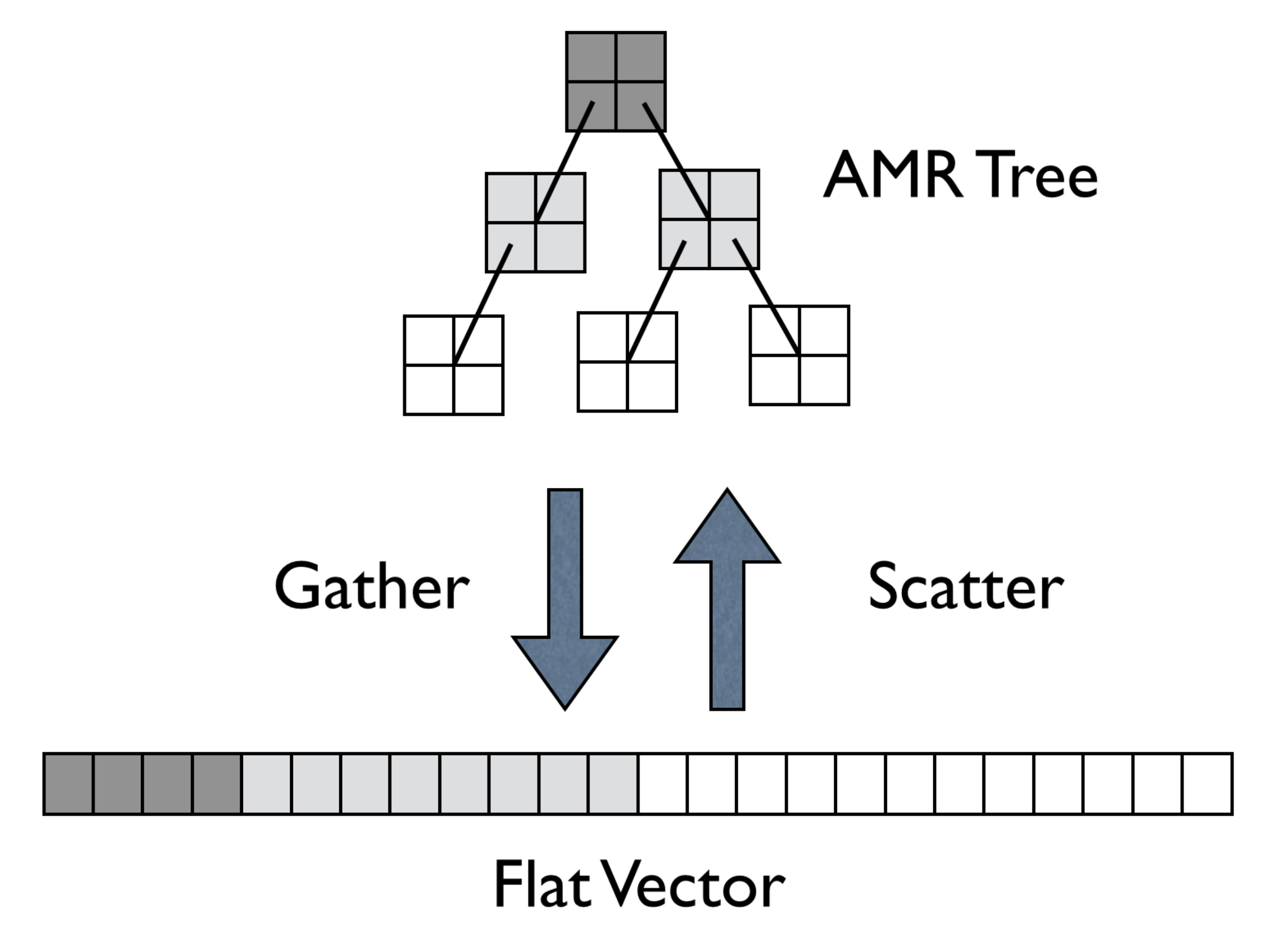}
\caption{Schematics of the conversion between oct-tree data management by the CPU for the AMR structure and array based calculations, for instance on GPU. A back and forth flow of data is performed through gather/scatter operations. Grey levels label different refinement levels both in the tree and the array. }
\label{f:gatscat}
\end{figure}

Local parallelism relies on a vector-based strategy, where arrays of data are processed through the same set of instructions and possibly on architecture with vectorization capability. The driver for this choice of design is the recent emergence of multi-core processors, graphics processing units or CPU-based co-processors, that all relies on this programming paradigm to be fully efficient. In the case of \texttt{EMMA}, this kind of parallelisation is focused on the physics solvers (i.e.  the relaxation of the Poisson equation, the conservative update of hydrodynamics and radiative transfer and the chemistry solver) which are fed with arrays of initial states and evolved into arrays of updated values.

Relying on vectors presents several pros. First it guarantees an optimal data layout in general by ensuring that it is accessed in a coalesced and aligned manner~: computation directly on tree stored values would induce random and unpredictable memory accesses, whereas an array-based organization ensures proximity of successive or concurrent calculations thus providing optimal performances. For GPUs in particular, enforcing this kind of memory access is a requirement to obtain a maximal throughput of the devices. Furthermore, arrays are generic and simple structures of data, that can be processed in a general manner~: each element of an array is computed like an other one and the implementation of this single-element flow of instruction can usually easily ported from one architecture to an another or even from one language to another. The difference usually arises on the details of the scanning operation on all the elements~: they can be parsed in sequence for a scalar calculator or by launching multiple threads of a single element computation on GPU or shared memory cores or by taking advantage of vector abilities of languages such as Fortran 90 or Python. Overall, vectorization provides an opportunity to choose easily a language or an architecture for the code computational modules, without any consideration on the design and layouts of the data structure. For instance, \texttt{EMMA} has been coded into both scalar CPU and CUDA GPU versions of the Poisson, hydro and radiative transfer solvers, both versions working in the same AMR framework. In fact, upcoming developments may lead to changes in the way AMR is handled and it would not impact the way physical engines are designed.

However, it becomes readily apparent that the AMR oct-tree being a non-vector based way to store data, the latter must be converted back and forth from a tree-based organization to an array-based one (see Fig. \ref{f:gatscat}). These \textit{gather} and \textit{scatter} operation are critical to the code performance as they constitute bottlenecks to the overall code performances. Nevertheless, if the amount of calculation is large enough, the cost of these operations can be hidden by computing or by overlapping transfers and calculations.
In practice in \texttt{EMMA}, when data is gathered from the tree to update a value in a given cell, all the necessary values from the neighbors are gathered too. For instance, Eq. \ref{e:jacobi} requests 7 values to update the potential of a given cell (6 from the cardinal neighbors and 1 for the density). Their related gather operation therefore organizes data in 7 arrays of $n_a$ values, required to update the potential in $n_a$ cells. Of course, since two adjacent cells share some neighbors, the data in these arrays can be redundant. \dom{Similarly, intercell fluxes (used during hydrodynamics and radiative transfer) are computed twice for two adjacent cells.} In principle such overheads could be avoided but at the cost of coding simplicity and at the current stage the data or flux evaluation has been kept redundant. Gather operations are also in charge of dealing with resolution jumps~: if a given cell requests data from a neighbor at an unavailable resolution, it is interpolated linearly from coarse data at the position of the fine virtual cell. Global boundary conditions are also dealt by these gathering operations As said previously, boundary conditions are periodic by nature~: if transmissive boundaries are required, the gather operation replaces the data from the periodic neighbor cell by the data of the current cell. If reflective boundaries are set up, the same operation is performed with an additional flux inversion.

\section{Coarse radiative transport approximation (CRTA)}
\label{s:coarse}
\begin{figure}
\includegraphics[width=\columnwidth]{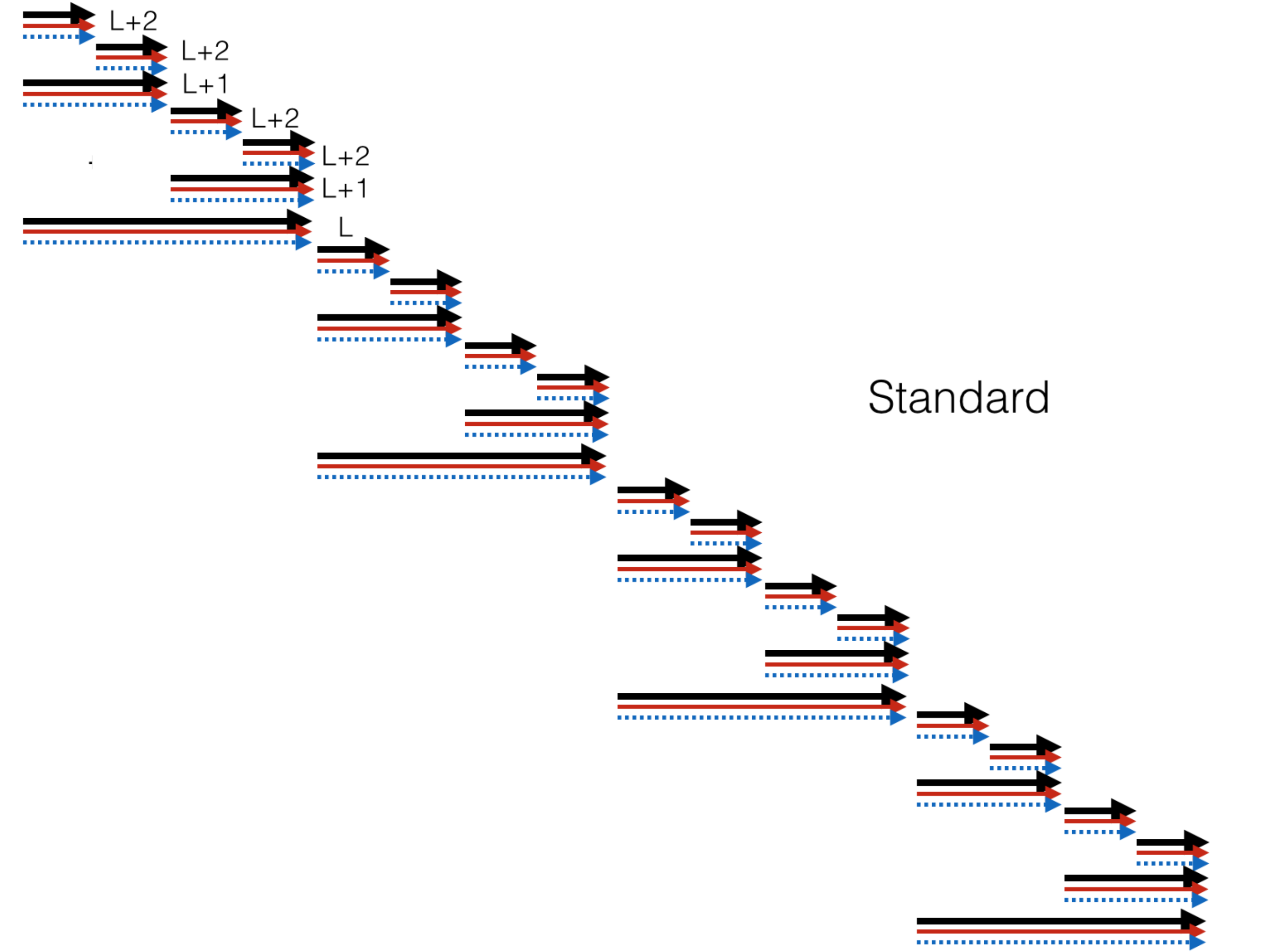}
\includegraphics[width=\columnwidth]{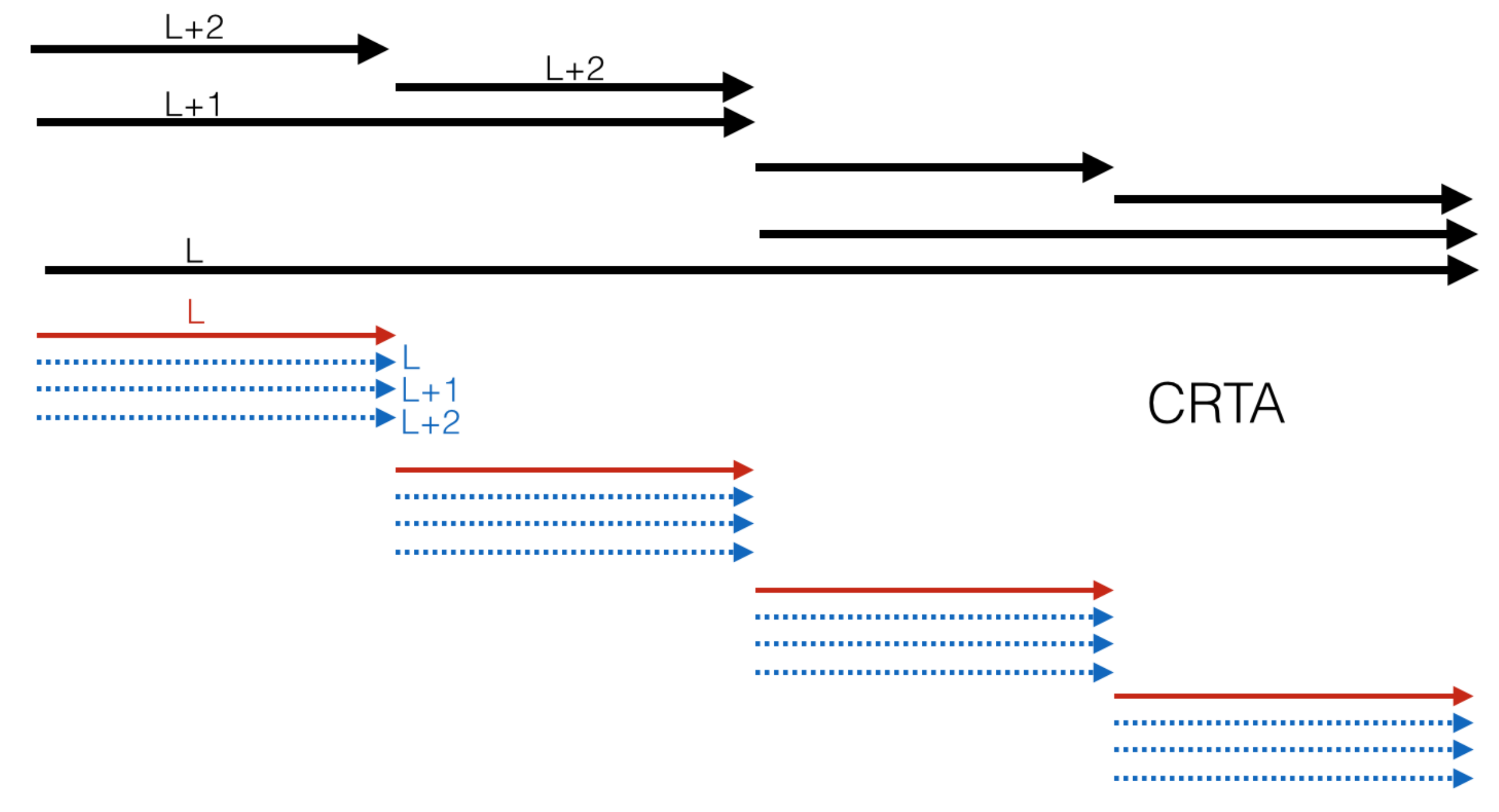}
\caption{Comparison of the time stepping and multi-physics coupling in the standard AMR approach (top, described in Sec.\ref{s:metho}) and in the coarse radiative transport approximation (CRTA) description (bottom, described in Sec. \ref{s:coarse}). Thick black/ thin red/dashed blue arrows stand respectively for dynamics (collisionless+hydro) operations, radiative transport and thermo-chemistry. The thermo-chemistry is shown as a dashed line to indicate that it is already sub-cycled with respect to the radiative transport. In both cases, the simulation is updated by a dynamical time step (from left to right), taken to be equal to 4 times the coarse radiative time step. Note how the red arrows have the same length in both scheme, corresponding to a coarse radiative transport time step. It is also assumed that 2 additional levels of refinement are enabled (L+1 and L+2). In the standard case, such an update takes 4 full updates of the AMR tree, because each update covers a radiative time step, resulting in a total of 83 engine calls. In the CRTA case, the dynamics are updated first over a dynamical time step (7 calls) and radiation + thermo-chemistry are treated in a second stage (16 calls) for a total of 23 calls. Note how the radiation transport is only performed at the coarse level.}
\label{f:corserad}
\end{figure}

Sec. \ref{s:metho} describes the standard methodology to couple the different physics within an AMR code with adaptive time stepping. Physical quantities and data structures are updated at the pace of the fastest evolving "dynamics" among the collisionless, the hydro- and the radiative ones. This implementation has been proven to be both accurate and practically sustainable (in terms of required computing resources) for hydro-dynamical codes in the past. However the inclusion of explicit radiative transport and out-of equilibrium chemistry severely impact the code's efficiency as it must track processes on time scales one or two orders of magnitude smaller than the pure hydrodynamical case. Hence an experiment covering a given physical duration must be sampled with a number of operations one or two orders of magnitude greater, including not only physical engines but also any kind of logistics functions or overhead. As such, any deviation to a perfectly optimized set of operations can see its magnitude multiplied by a factor 100 and reduces significantly the code efficiency.

In this section we suggest an additional level of approximation for the coupling between radiative processes and dynamics (hydro + collisionless). It can significantly reduce the resources necessary for a simulation with radiation, at the cost of a degraded (mostly spatial) resolution. It is summarized in Fig. \ref{f:corserad} and relies on two sets of additional approximations compared to the standard implementation of Sec. \ref{s:metho}:
\begin{enumerate}
\item Radiation transport and the associated thermo-chemistry is explicitly decoupled from the dynamics (collisionless+hydro-). Hence to advance the simulation, dynamical quantities are updated first on all the AMR levels on a timescale only constrained by the dynamics (usually set by the hydro- CFL condition). Then, matter is considered as "frozen" and radiation is propagated within this static distribution for the same duration. Since typical speeds encountered in dynamical processes are of the order of the local sound speed or free-fall velocities, which are much smaller than $c$, such decoupling should remain under control and provide results similar to the standard procedure.
Of course, radiation is subject to stringent CFL condition which implies that radiative quantities are updated through an intensive subcycling (typically 100-1000 cycles) of the dynamical time step with small radiative time scales.
 
\item Additionally, radiative transport is only performed at the \textit{coarse} level. However, thermochemistry is still computed on refined levels, but with a coarse-grained description of the radiation field that is simply injected from the coarse to the fine levels. Not only it reduces the number of transport operation but it also reduces significantly the number of thermochemistry steps : this engine is already subcycled even in the standard approach (see Sec. \ref{s:thermo}) and can therefore operate on a large radiative time step. Furthermore if an equilibrium situation is encountered in a given cell (a frequent situation in fully ionized regions for instance), this thermo-chemistry subcycling can be reduced to a few cycles.
\end{enumerate}
Fig. \ref{f:corserad} provides a simplified comparison of the standard and the coarse radiative transport approximation scheme (CRTA hereafter). We arbitrarily chose a situation where the coarse dynamical time step is 4 times larger than the radiative one at the coarse level. In the standard description the time step is set by the radiative CFL condition at all levels. Hence for a coarse+2 refined levels situation as the one described in Fig. \ref{f:corserad} the number of dynamics, radiative transport and thermo-chemistry engine calls are identical and equal to 28 (4 on the coarse level and 24 on the refined ones) for a total of 84 calls. In the CRTA case it reduces to 7 dynamics calls (including 6 calls on refined levels) + 4 radiative transport calls at the coarse level + 12 thermo-chemistry calls for a total 23 engine calls, i.e. a factor of 3 smaller in the number of operations. Bear in mind that a realistic case rather involves a ratio of 100 to 1000 between dynamics and radiative time steps and ~5-10 refinement levels: in such cases, the CRTA approximation essentially reduces the cost of hydrodynamics to zero and by neglecting transport on refined levels, it reduces the cost of a radiative time step by a few tens.
Additionally, AMR logistics, communications setups, analytics, etc.. are performed only at the dynamical time step and their costs are also essentially set to zero in the CRTA approach. Incidentally, this technique also increases the relative weight of physical engines over numerical overheads and among the engines it increases drastically the weight of radiative transport+chemistry over the others : it turns out that the latter engine is one of the most efficient in the use of vectorization (see Sec. \ref{s:para}) and therefore in the use of hardware accelerators such as GPUs. CRTA is expected to take a greater benefit of such devices to accelerate the code.

Of course this increased efficiency comes at a cost. The most evident one is the decreased spatial resolution for radiative transport, even though the thermo-chemistry is performed at the highest resolution available. It could be thought of as an intermediate approach between an homogeneous radiation field (as usually assumed in non-RT cosmological simulations) and a full AMR description. In the CRTA approach, spatial UV field fluctuations are existent but coarsened. However the impact could be limited. First, as shown in \citet{AUB10}, radiation fields (radiative density and flux) do not exhibit significant clumping factors compared to the ones of the matter distribution and are relatively smooth even in highly resolved simulations. In fact, the values of radiative densities span orders of magnitude between sources and dark voids and are therefore not very sensitive to local fluctuations. Furthermore, we use here a highly diffusive scheme (based on a LF intercell flux evaluation) which accentuates further the smooth aspect of radiative fields. One could therefore argue that having a fine spatial description of radiative density fields is not absolutely necessary. The other level of approximation is the imperfect coupling of radiation and matter, the latter being considered as still when light is being cast. Somehow, it relates to previous post-processing techniques but performed on the fly at every dynamical time step. Post-processing is known to provide satisfying results for large scale experiments on $\sim$ 50+ Mpc scales, hence we can be confident that this imperfect coupling can be controlled in such cases. On the other hand, it is clear that some coupling between matter and radiation in highly refined cells will be lost and it is difficult to evaluate properly this loss. As shown in Sec. \ref{s:valid} the CRTA returns very satisfying results for the tests shown here but in general using CRTA would require to perform an additional level of convergence study to ensure that this imperfect coupling is under control.

\section{Code Validation}
\label{s:valid}
\texttt{EMMA} has been submitted to a series of test to validate its implementation. For various combination of simulated physics, documented experiments are described here, as well as the code results.
\subsection{1D Hydrodynamics : Shock Tube}
\label{s:shock}
\begin{figure}
\includegraphics[width=\columnwidth]{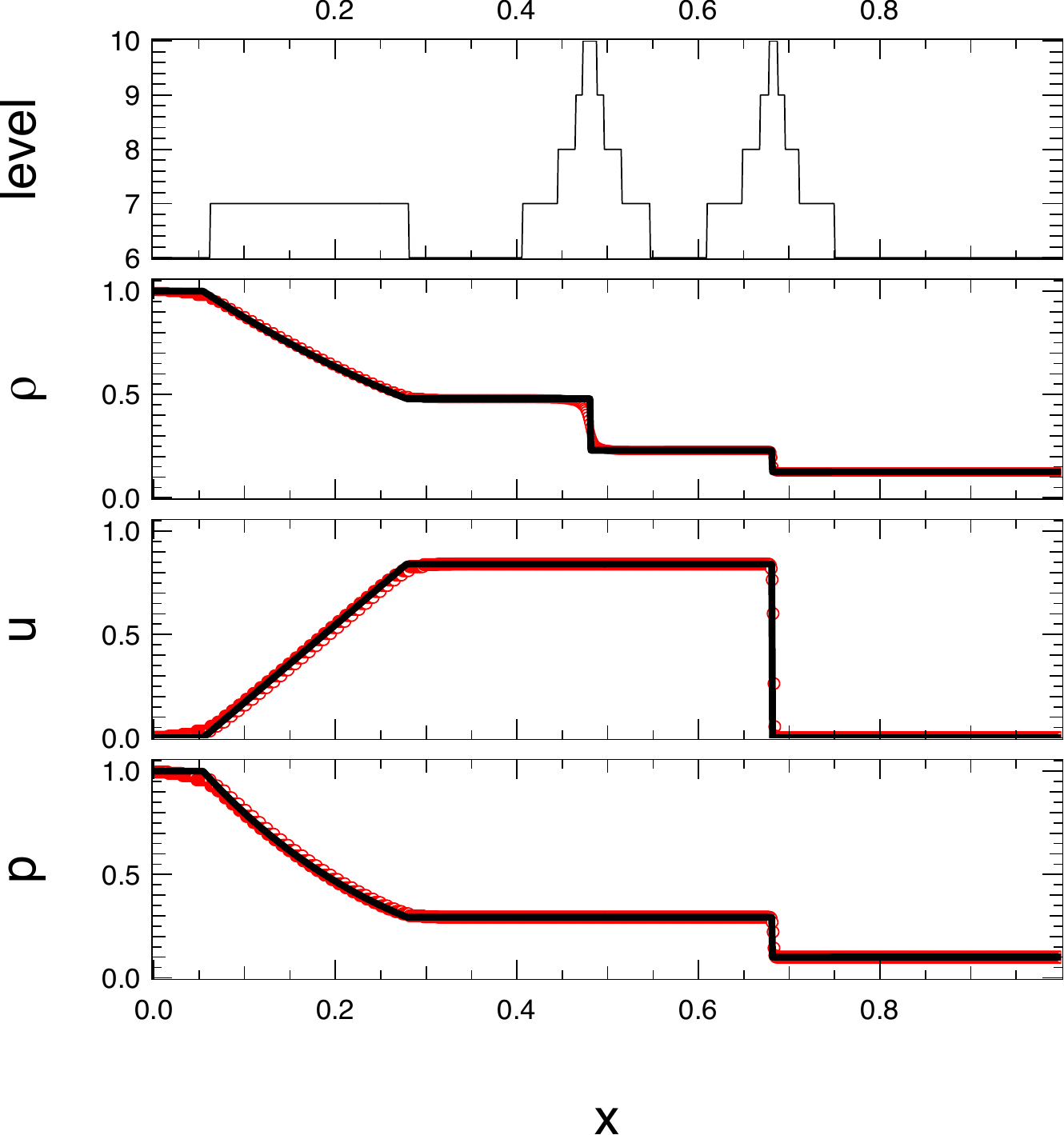}
\caption{Shock tube experiment. From top to bottom: refinement level, density, velocity and pressure as a function of position. Points stand for the simulation results and solid lines for analytic profiles.}
\label{f:sod}
\end{figure}
The shock tube is a 1D test where a Riemann problem is evolved by means of a simulation. It focuses on implementation of hydrodynamics and the ability of the MUSCL scheme and HLLC Riemann solvers to capture shock features. The initial conditions consist in a jump at $X_0=0.3125$ between two different states ($\rho_1=1, u_1=0, p_1=1$ and $\rho_2=0.125, u_2=0., p_2=0.1$, taken from \citet{TOR97}). The solution to this Riemann problem is known and can therefore be compared to the results delivered by \texttt{EMMA}.
 
The calculation has been performed using a $\ell=6$ coarse resolution with 4 additional levels of resolution, triggered by density gradients satisfying $\Delta \rho/\rho \ge 0.015$. Even though the problem is 1D, the calculation has been performed in 3D with the jump occurring along the x direction. Transmissive boundary conditions were retained along the $x$ direction and periodic ones along the two others.

Fig. \ref{f:sod} shows the density $\rho$, the velocity along the x-direction $u$, the pressure $p$ and the refinement level at $t=0.2$.  Also shown is the solution of the Riemann problem, as a solid line. Clearly, the match is satisfying with shocks being resolved on a few cells, thanks to both the shock-capturing scheme and the improved resolution allowed by the on-the fly refinement. It can also be noted that the contact wave is also present near x=0.5, even though some smearing can still be present at this resolution. Overall, this standard test demonstrates the ability of \texttt{EMMA} to solve classic hydrodynamical problems at high resolution.

\subsection{1D Gravity+ Hydrodynamics: Zeldovich Pancake}
\label{s:zeldo}
\begin{figure}
\includegraphics[width=\columnwidth]{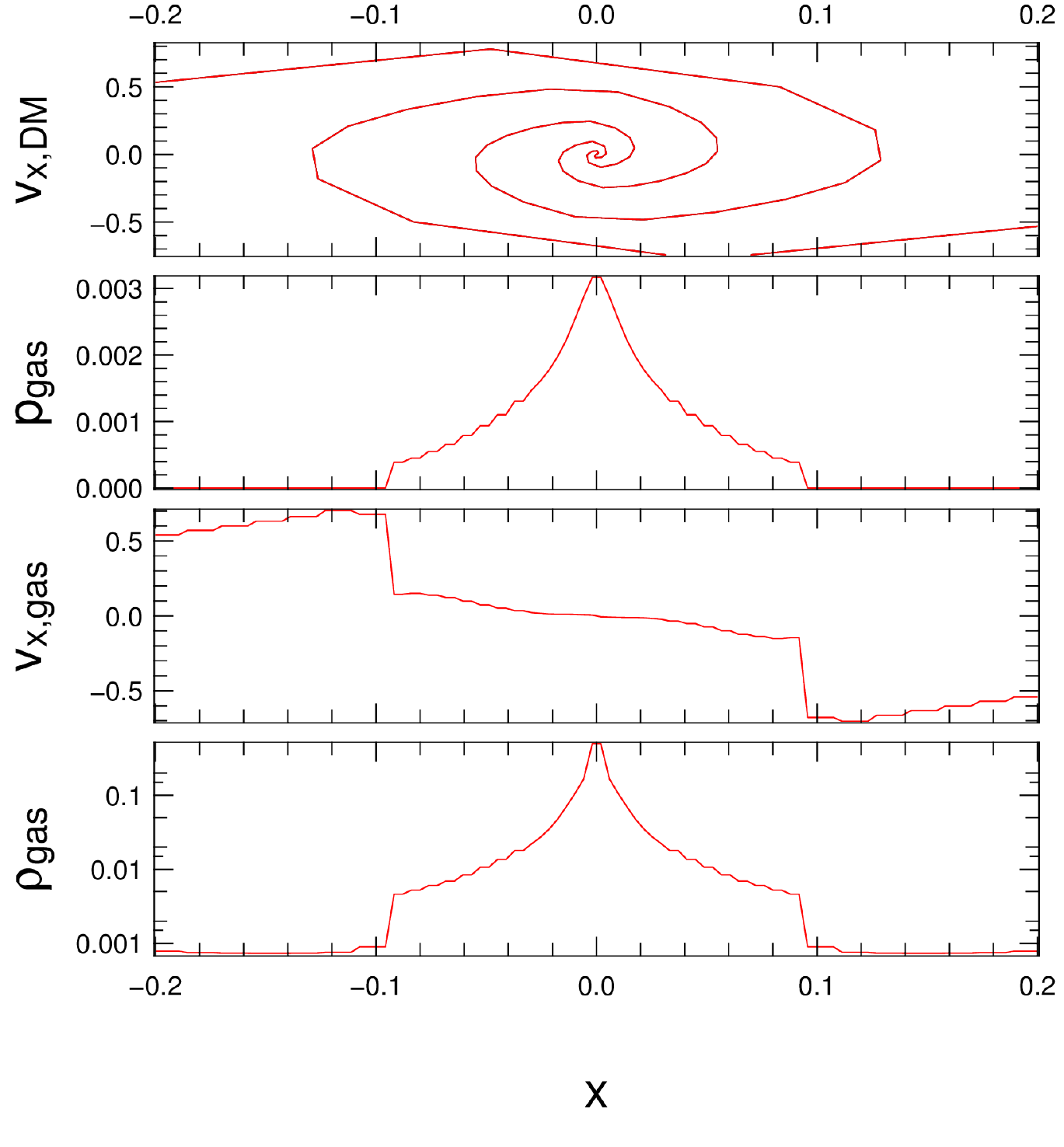}
\caption{Zeldovich Pancake experiments with baryons ($\Omega =1, \Omega_b=0.01$). From top to bottom: the dark matter  velocity $v_{x,\mathrm{DM}}$, the gas pressure $p_\mathrm{gas}$, velocity $v_{x,\mathrm{gas}}$, and density $\rho_\mathrm{gas}$,. All the quantities are shown as a function of the position, at z=0. The sine wave is initiated at  z=100 and collapsed at z=10. One can note the increasing resolution toward the center of the caustic, from $\ell=6$ to $\ell=8$. }
\label{f:zeldo}
\end{figure}

The Zeldovich Pancake test \citep{ZEL70}) tracks the evolution of a single planar mode in an $\Omega_m=1$ expanding Universe, where the linear stages of the evolution can be analytically predicted. The initial matter density is given by :
\begin{equation}
\rho(x)=1+\frac{1+z_c}{1+z_i}\cos(2\pi x)
\end{equation}
whereas the initial velocity is given by:
\begin{equation}
u(x)=\frac{1}{\pi}\frac{1+z_c}{(1+z_i)^{3/2}}\sin(2\pi x).
\end{equation}
Here the mode oscillates along the x direction. $z_i$ and $z_c$ stand for the initial and collapse redshift. 

As in Sec. \ref{s:shock}, this test case has been simulated with \texttt{EMMA} in 3D as a planar experiment. Both the baryons and dark matter were included with $\Omega_b=0.1$. \dom{The base resolution is $\ell=6$ (i.e. $64^3$ cells) and the dark matter field is sampled with $64^3$ particles. The initial temperature is chosen to be arbitrarily small at $T=10K$ and velocities orthogonal to the x-direction are taken to be zero.} The experiment has been conducted down to $z=0$ with $z_i=100$ and $z_c=10$. Two additional refinement levels were triggered on gas density gradients $\Delta \rho/\rho>0.1$. 
This setting provides a situation where the cosmological setting and the coupling between dark matter and baryons are tested. Linear stages were compared to the analytic solution and were found to match at better than the \% level (not shown here) until the redshift of collapse.

Fig. \ref{f:zeldo} shows the z=0 baryon density, velocity and pressure as well as the dark matter phase diagram. Clearly, being way later than the collapse redshift ($z_c=10$), it can be noted that several 'plane-crossing' occurred with a significant number of foldings for the dark matter (DM hereafter) phase space diagram. Baryons fell in the DM potential, creating shocks and an inner increase of temperature (via the pressure) within the collapsed region. In particular one can note how the infall velocity of the gas is strongly reduced as it enters the collapsed gas. Refinement levels were triggered as expected, providing a better resolution of the density peak and a smoother description of the phase space curve of DM in the innermost regions. Finally, direct comparisons with e.g. \citet{TEY02}, shows that the results of \texttt{EMMA} are consistent with other codes, even at this latest stages of the pancake collapse.

\subsection{3D Gravity + Hydrodynamics: Bertschinger's Self Similar Infall}

\begin{figure}
\includegraphics[width=\columnwidth]{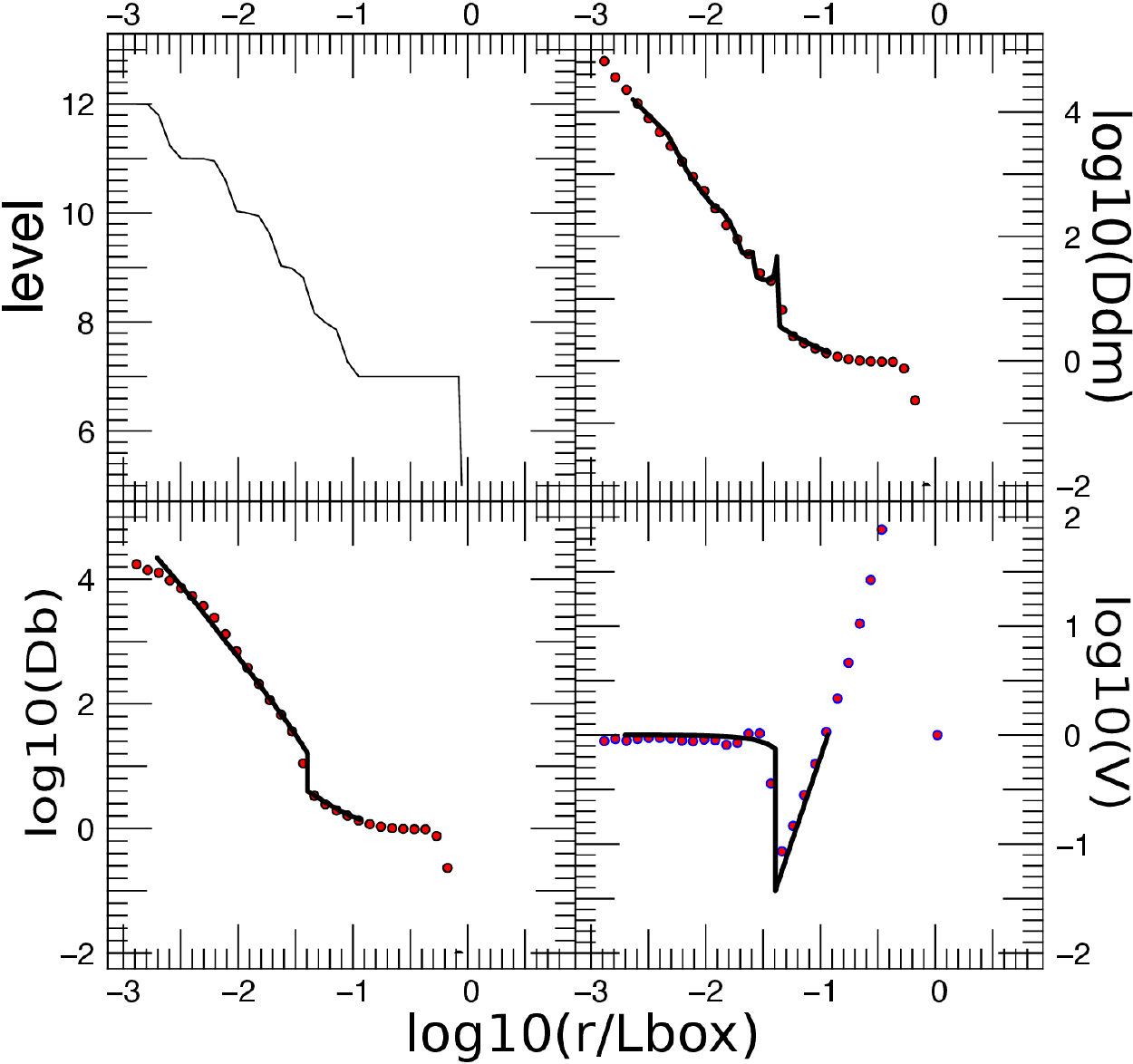}
\caption{Self Similar 3D Collapse of a Top-Hat density perturbation. Clockwise from top left: the radial profile of refinement level $\ell$, dark matter density $D_{dm}$ , baryons radial velocity $V$ and density $D_b$, at a=0.56. As in  \citet{BER85}, these quantities are expressed in units of turnaround quantities. Dots are for simulation results and lines stand for the analytic prediction of \citet{BER85}. Radii are in units of the box length.
}
\label{f:edbert}
\end{figure}

This experiment aims at reproducing the calculation made by \citet{BER85} where a top-hat overdensity within an expanding Einstein-De Sitter Universe ($\Omega_m=1$) collapses toward a scale-invariant density distribution in a self-similar fashion. Provided that radii are rescaled to the turnaround radius $\lambda$, this self-similarity can be fully predicted by analytic means. In the original paper, several configuration are explored and here we focus on the evolution of an overdensity that includes baryons in a dark-matter dominated potential. Compared to the test described in \ref{s:zeldo}, the situation here is three-dimensional with a spherical symmetry.

In practice, we generated a regular lattice of DM $128^3$ particles starting at $z=1000$ in a 1 Mpc box, with cosmological parameters $\Omega_m=1$, $\Omega_b=0.01$, $H_0=70$ km/s/Mpc. \dom{Two types of DM particles co-exist~: } particles within a radius $R_i=0.05$ Mpc to the center were assigned a greater mass (\dom{equals to $7.89\times10^4 M_\odot$}) than particles at larger distance \dom{(with a $6.37\times10^4 M_\odot$ mass)}, in order to produce a central overdensity $\delta_i=0.2$. Also an arbitrarily cold and motionless gas has been sampled on a coarse grid of $128^3$ (i.e. $\ell=7$), with the same central overdensity as dark matter~:\dom{ cells within a 0.05 Mpc radius were assigned a 797 $M_\odot$ mass and ones at greater radii were given a 604 $M_\odot$ mass}. Mesh refinement triggers when the mass within a cell is greater than 8 times the mass \dom{of a low-mass DM} particle~: this criteria is similar to the one used in cosmological simulations in order to provide a quasi-Lagrangian strategy.

Fig. \ref{f:edbert} shows the radial profiles of the DM and baryon densities as well as the baryon radial velocity, for the simulation and compared to the fits of the analytic solution provided by \citet{BER85}.  Also shown is the spherical average of the refinement level.  \dom{As in \cite{BER85}, the DM density $D_\mathrm{dm}$, the baryonic density $D_b$ and the baryonic radial velocity $V$ are expressed in units of `turnaround' values~:
\begin{eqnarray}
D_{\mathrm{dm}}&=&\frac{\rho_{\mathrm{dm}}}{\rho_{\mathrm{ta}}}\\
D_{\mathrm{b}}&=&\frac{\rho_{\mathrm{b}}}{\rho_{b,\mathrm{ta}}}\\
V&=&\frac{v_b}{V_{\mathrm{ta}}},
\end{eqnarray}
where $\rho_{\mathrm{ta}}=\left(6\pi G t^2\right)^{-1}$ and $\rho_{b,\mathrm{ta}}=\Omega_b\rho_{\mathrm{ta}}$.
\cite{BER85} gives the evolution of the turnaround radius~:
\begin{equation}
r_{\mathrm{ta}}(t)=\left(\frac{4\pi t}{3t_i}\right)^{8/9}\delta_i^{1/3}R_i,
\end{equation}
where $t_i$, $\delta_i$, and $R_i$ are respectively the initial time, the initial overdensity and the initial overdensity radius, from which the expression of the associated velocity can be obtained~:
\begin{equation}
V_{\mathrm{ta}}=\frac{r_\mathrm{ta}}{t}.
\end{equation}}
All the results are shown for $a=0.56$. Clearly we manage to reproduce the analytic solution in particular the predicted inner logarithmic slope of $-9/4$ for the DM density profile or the shock positions. A small amount of diffusion can be seen in the innermost regions in the baryon density and the velocity jump from the Hubble flow to the shocked region is not as sharp as the predicted one, but the overall features are well reproduced by our calculations and of the same quality described for \texttt{ART} or \texttt{RAMSES}.

\subsection{3D radiative hydrodynamics: Growth of an HII region}
\begin{figure}
\includegraphics[width=\columnwidth]{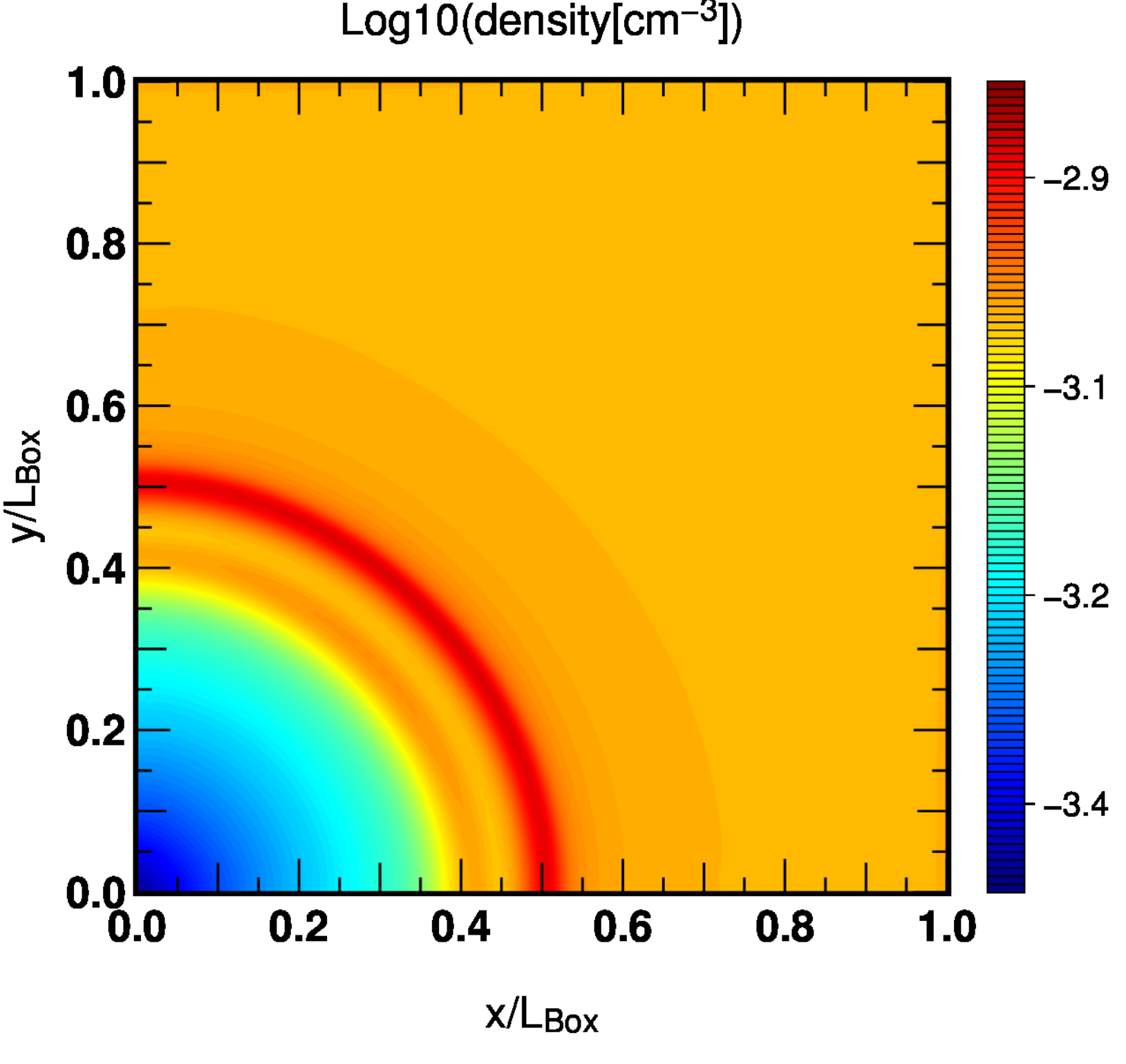}
\includegraphics[width=0.9\columnwidth]{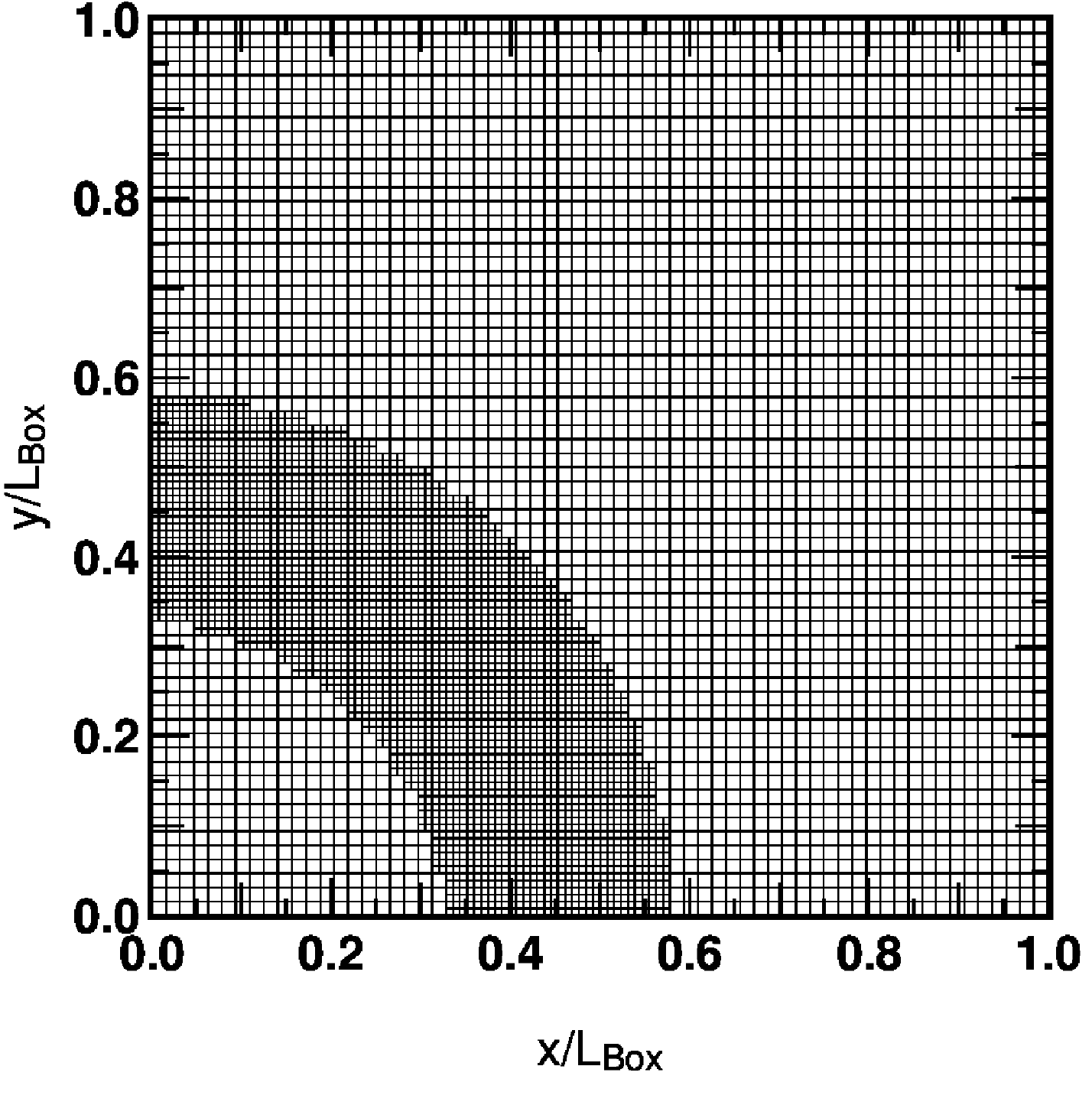}
\caption{Expansion of an HII region. Top: the log10 of density map (in cm$^{-3}$), Bottom: the AMR grid used for the computation. The coordinates are expressed in units of the box length which has a physical extent of 15 kpc. The source is located at the bottom left corner and has been ignited 200 Myrs ago.
}
\label{f:rh_200_map}
\end{figure}

\begin{figure}
\includegraphics[width=\columnwidth]{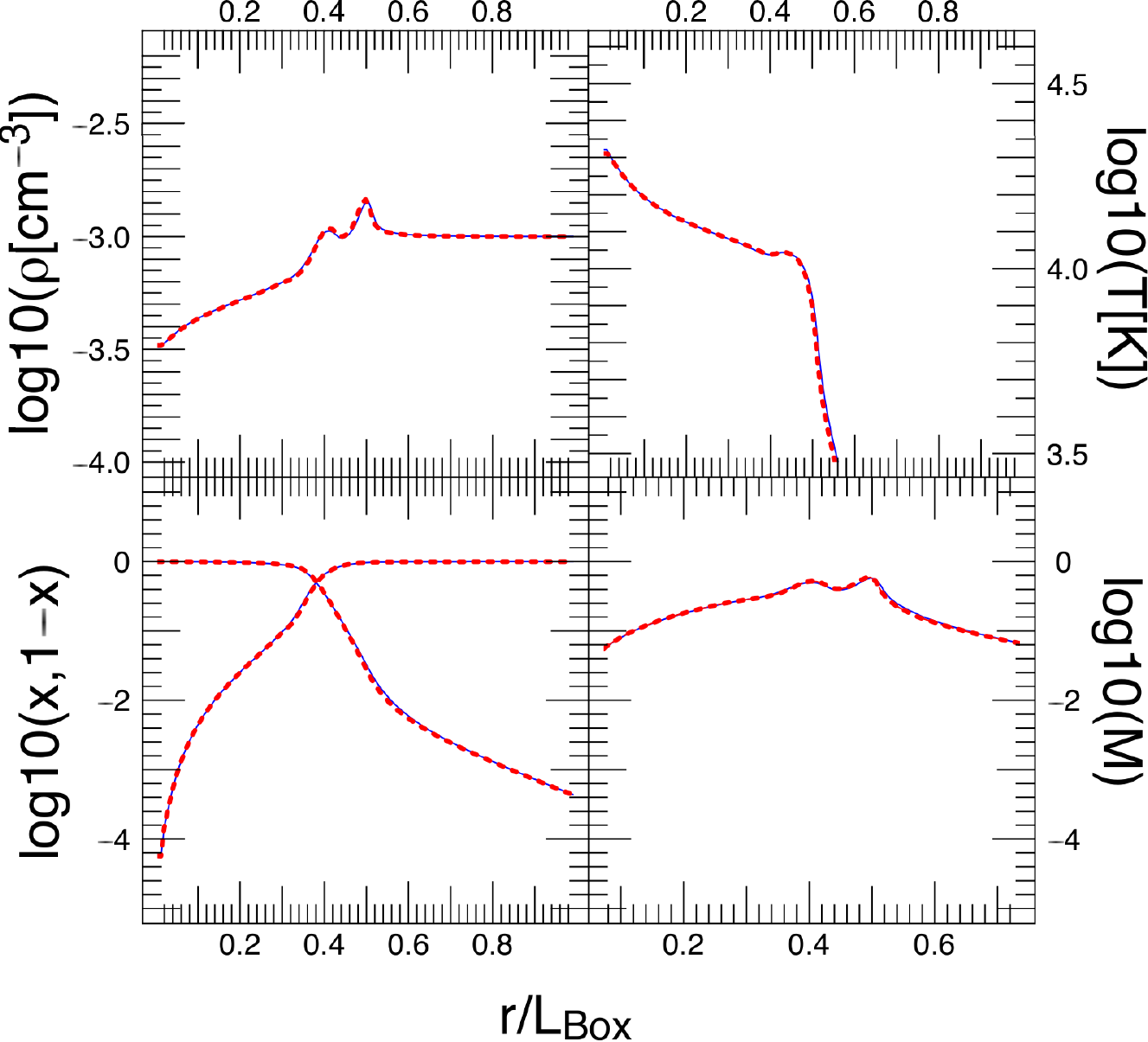}
\caption{Expansion of an HII region. Clockwise from top left: the gas density,the temperature, the Mach number and the ionized/neutral fraction. The coarse resolution is $\ell=6$ and refinement is triggered on the ionization front to $\ell=7$ in accordance with the Iliev et al. test 5. Red dashed lines stand for the radial average taken 200 Myrs after the source has been ignited. Blue lines stand for the same calculation, but performed with the CRTA approximation.} 
\label{f:rh_200}
\end{figure}

\begin{figure}
\includegraphics[width=\columnwidth]{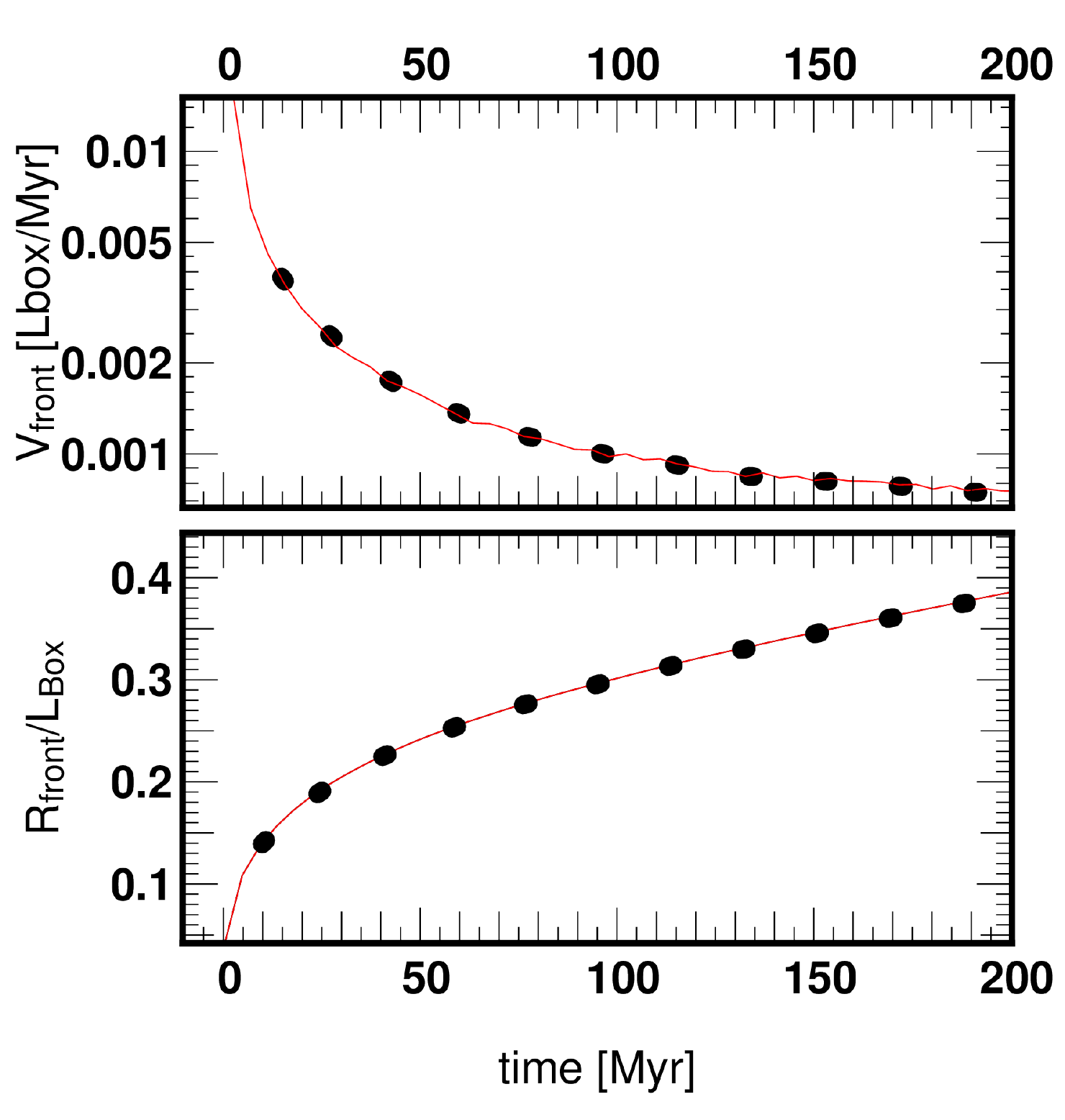}
\caption{Expansion of an HII region. Front propagation predicted by the standard AMR RT implementation (red line) and the CRTA approximation (dots). Top : the front velocity. Bottom: the front position.}
\label{f:rh_200_comp}
\end{figure}

\dom{This test consists of a corner source, powered by a 100 000 K black-body that send photons in a surrounding homogeneous hydrogen-only medium.
It belongs to the suite proposed by \citet{,ILI06, ILI09} and comes in two different versions. The first version deals with a static and uncoupled gas ('Test 2', \citet{ILI06}): we obtained a very good agreement with \texttt{EMMA} (not shown here), which does not come as a surprise since this test was also successfully passed by \texttt{ATON} or \texttt{RAMSES-RT}, that share a great number of details with \texttt{EMMA}. Here we focus on the second version, a coupled test known as 'Test 5'. The UV photons ionize and heat up the gas, leading to the creation of ionization fronts that propagate inside-out and putting the gas into motion. This test couples radiative transfer and hydrodynamics and has been performed by a whole serie of codes in \citet{ILI09}. }

A 100 000 K Black-body is located at the corner (x=y=z=0) of an 15 kpc box, emitting $5\times 10^{48}$ UV photons per second. We sample the frequencies with the 3 groups of photons given in Sec \ref{s:rt}. The surrounding gas has an homogeneous number density of $0.001$ hydrogen atoms per cm$^3$. The calculation is run on a $64^3$ coarse grid ($\ell=6$) and allowing for an additional level of refinement ($128^3$ i.e. $\ell=7$) to comply with the resolution requirements of \citet{ILI09}. The refinement is simply triggered for cells with ionized fraction $0.01<x<0.8$~: with the two-cells layer of neighbours being also refined, it provides a simple manner to track the ionization front. Boundary conditions are reflective for boundaries adjacent to the source and transmissive otherwise.

Fig. \ref{f:rh_200_map} shows the baryon density in the $z=0$ plane, as well as the AMR structure that tracks the ionisation front for t=200 Myr. The source being in the bottom left corner, the distant cells remain at the base resolution as the front has not yet progressed to these regions. The cells close to the source are also at the base resolution~:  \dom{this region has returned to low resolution as it does not contain an ionization fraction that satisfies the refinement criterion (chosen to track front-like features).}

Finally between these two ensembles of coarse cells, one can find the refined region at $128^3$ resolution, tracking the front. On the other panel, the number density field of hydrogen is also shown, presenting a typical double peak structure encompassing a void created by the energy injection by the corner source.

Fig. \ref{f:rh_200} shows the radial profile of the density at the same instant (i.e. 200 Myrs after the source ignition) as well as the temperature, ionisation/neutral fraction and the Mach number profiles. \texttt{EMMA} recovers the typical features already obtained by most of the codes of Iliev et al. 2009. The double peak is due to the presence of high energy photons in the spectrum of the $10^5$ black body: these photons with large mean free path can deposit energy behind the ionisation front, while lower energy photons deposit their energy at the base of the front. The input of energy at larger radii is also at the origin of the moderate temperature increase seen before its drop in the neutral region. The effect of hard photons can also be seen in the extent of the ionized fraction drop at the front, that would be much sharper in the case of a monochromatic incoming flux. Overall, the fact that \texttt{EMMA} reproduces these specific features seen by all other codes indicates that the coupling between radiation and matter and the handling of multi-frequency transfer is consistent with others implementations.

Regarding the CRTA approximation, Fig. \ref{f:rh_200} and \ref{f:rh_200_comp} also provides the different fields profiles as well as a comparison of the temporal evolution of the front position and velocity (the front being defined as having a 50\% ionised fraction) in both cases. Clearly the CRTA approximation provides the same results as the standard calculation~: \dom{in Fig. \ref{f:rh_200} the radial profiles taken at $t=200$ Myrs are almost indistinguishable and in Fig. \ref{f:rh_200_comp}, the front position and velocities of the CRTA calculation  follow the ones obtained from the standard procedure.}  The evolution is smooth enough both spatially (with features sampled on $\sim$ 15 fine cells or 7 coarse cells) and temporally (with terminal velocities as small as $0.1\%$ box lengths per Myr)  to ensure a good convergence of the CRTA toward the standard case.

\subsection{3D radiative hydrodynamics: Photo-Evaporation of a dense clump}
\begin{figure}
\includegraphics[width=\columnwidth]{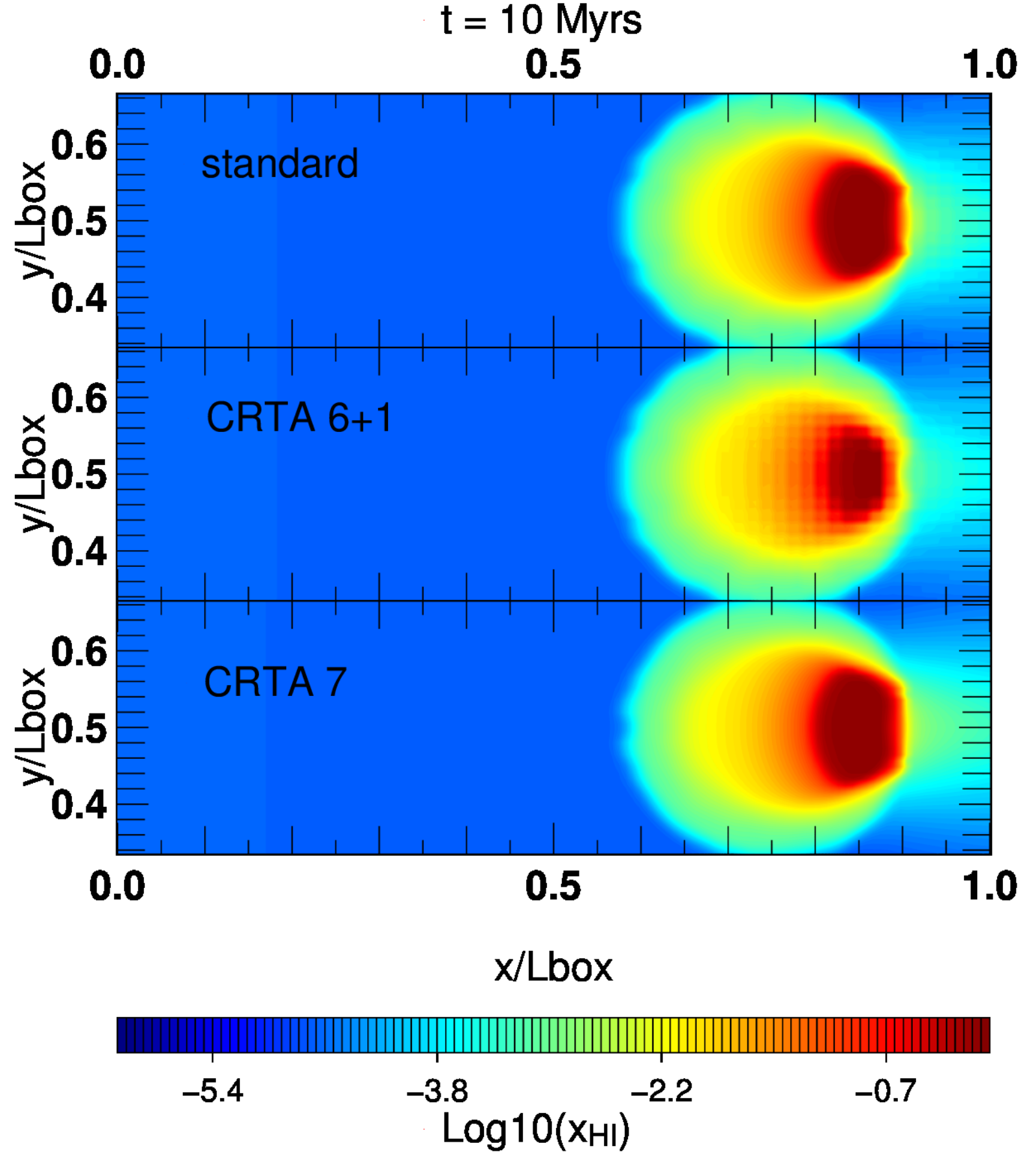}
\includegraphics[width=\columnwidth]{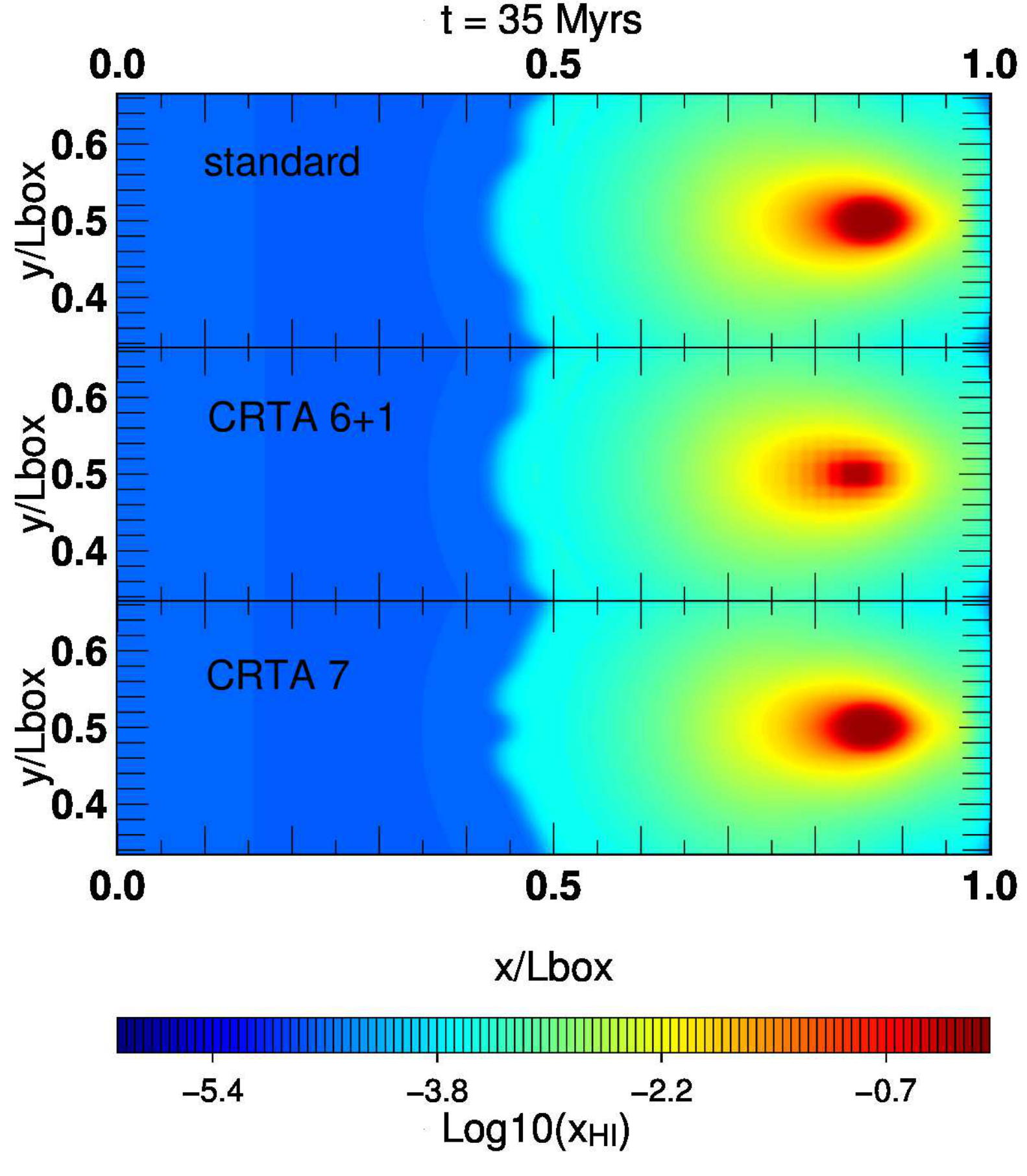}
\caption{Photo-Evaporation of a dense clump. Top: log10 of the neutral fraction along the plane of symmetry of the clump at t=10 Myrs, as predicted by the standard AMR RT implementation (top) and the CRTA approximation (middle), both assuming an $\ell=6$ base level + 1 level of refinement. The bottom row is the prediction of the CRTA approximation on a static $\ell=7$ grid. Bottom: the same quantities at t=35 Myrs. Blue stands as ionized and red as neutral. 
}
\label{f:photo_map}
\end{figure}

\begin{figure}
\includegraphics[width=\columnwidth]{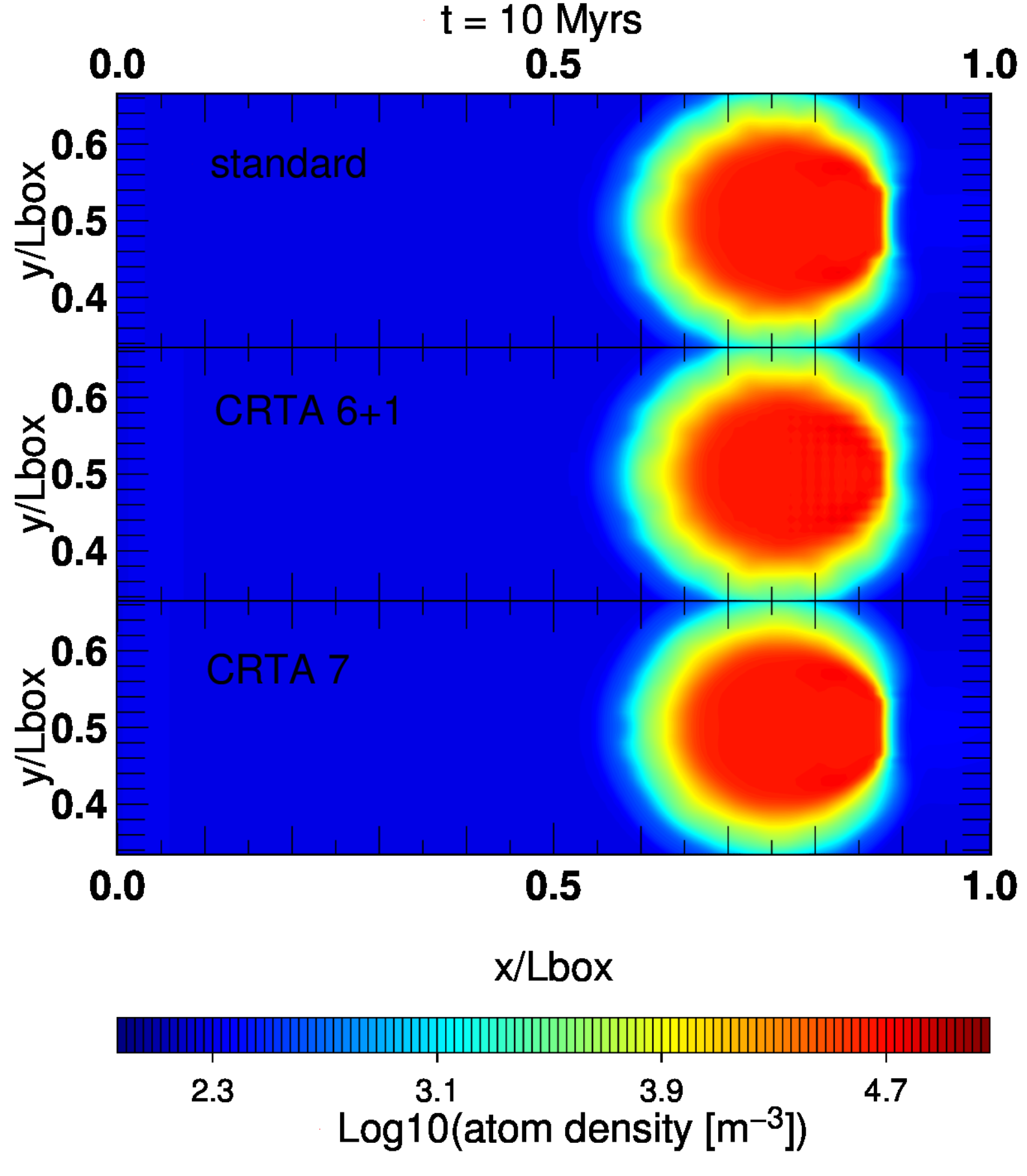}
\includegraphics[width=\columnwidth]{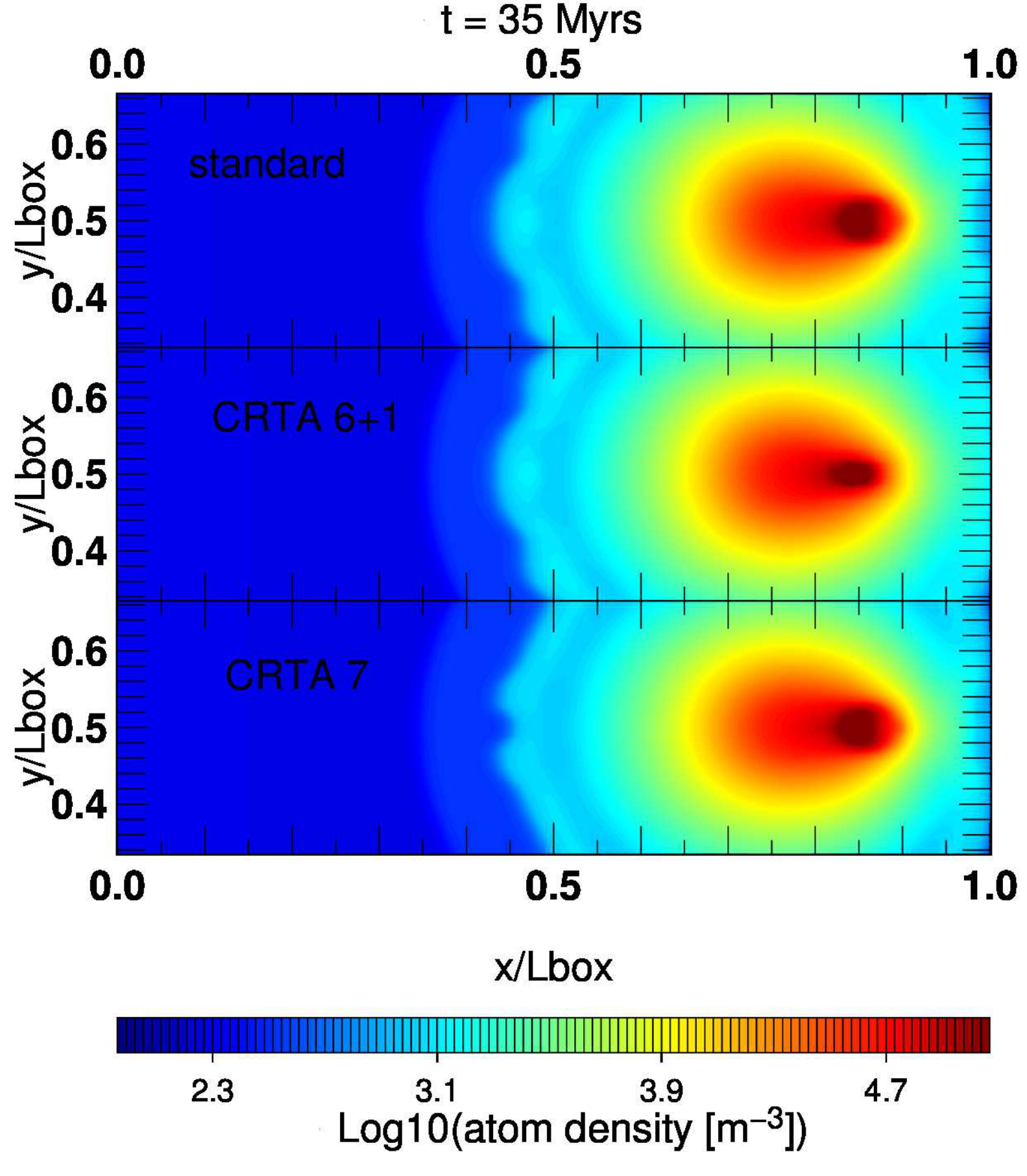}
\caption{Same as in Fig. \ref{f:photo_map} but for the gas density. Underdense regions are blue and dense regions are red.
}
\label{f:photo_map_d}
\end{figure}

\begin{figure}
\includegraphics[width=\columnwidth]{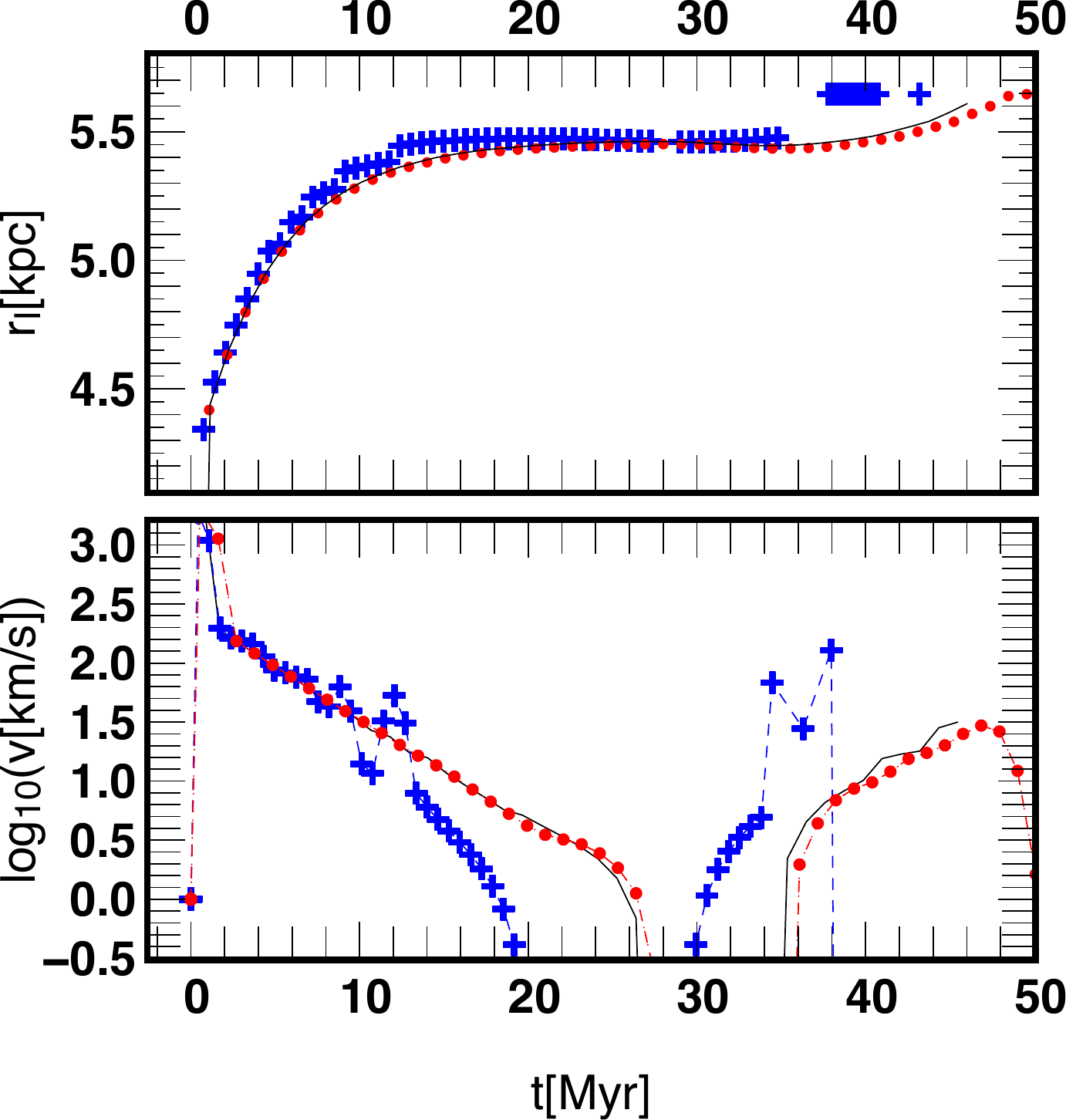}
\caption{Photo-evaporation of a dense clump. Position (top) and velocity (bottom) of the ionization front along the x axis as a function of time. The solid line stands for the standard AMR RT implementation results and blue crosses stand for the CRTA approximation results. Both were obtained using an $\ell=6$ base level + 1 level of refinement. Red dots stand for the CRTA approximation results with a $\ell=7$ static grid.}
\label{f:photo_front}
\end{figure}
This situation has also been suggested by \citet{ILI09} and consists of a dense cold clump irradiated by a planar UV front. As the ionization front encounters the cloud, the high density will slow down its progression, acting as a trap on the incoming photons. As a side effect, a shadow will also be cast in the trail region of the clump. Finally, the energy deposited in the clump will put the gas in motion, leading to a photo-evaporation process by the incoming photon flux.

The setup is given by \citet{ILI09}: a spherical clump of radius 0.8 kpc is centered on (5.4, 3.3, 3.3) kpc inside a 6.6 kpc box. Outside the clump, the gas has a 8000 K temperature with a density of 200 atoms/m$^3$. The clump itself has density of 40 000 atoms/m$^3$ and a temperature of 100 K. A UV flux with a 100 000K black body spectrum is incoming from the x=0 boundary at a rate of $10^{10}$ photons/m$^2$/s. In practice the simulation is performed on a $64^3$ grid ($\ell=6$) with one additional refinement level to comply with the \citet{ILI09} recommended resolution. Mesh refinement is triggered for cells with a density greater than the background density. The x=0 boundary is a source of flux of the required rate, whereas the x=6.6 kpc boundary is transmissive. Boundaries in the two other directions are periodic.

Fig. \ref{f:photo_map} and \ref{f:photo_map_d} show maps of the neutral fraction and gas density at t=10 and 35 Myrs. In each figure the top and middle row were obtained from $64^3$ simulations with an additional level of refinement, the top being obtained from the standard RT implementation on AMR and the middle panel being obtained from CRTA approximation simulations.

Globally a simple comparison to \citet{ILI09} results demonstrate their consistency. In particular, the evaporation is made obvious with the expansion of the cloud limits due to the energy injected by the UV front. However clear differences can also be noted~: first the shadow behind the clump, albeit existent, is much weaker than in other radiative transfer codes. This does not come as a surprise since \texttt{EMMA} implements the GLF flux to compute intercell exchanges and is known to be very diffusive. It therefore prevents the creation of clear cut shadows as a diffuse component of the flux eats the neutral gas in direction orthogonal to the incoming direction of UV photons. The same effect was already noted for \texttt{ATON} \citep{AUB08}. Second, at late time, the contours of the extended cloud are not as spherical as expected and present significant fluctuations around a mean radius. These fluctuations are artifacts of the cloud initial sampling on the coarse $\ell=6$ grid. The same artifact can be seen e.g. in the FLASH-HC results in \citet{ILI09} and also linked to the initial conditions. Comparing the standard RT and the CRTA approximation, it can be seen that the latter provides a faster ionization of the clump. Looking at the Fig. \ref{f:photo_map}, the shadowed neutral clumps are systematically smaller in the CRTA regime. It does not come as a surprise since the RT is performed at the base level only, which increases artificially the extent of the UV flux penetration into the clump and also increases the scheme diffusion. Fig. \ref{f:photo_front} provides a more quantitative insight on this aspect, describing the front progression inside the cold clump and its velocity. The front position is defined as corresponding with the position of the cell having $x=0.5$.  At early stages ($t<7.5$ Myr), the CRTA and standard description produce an identical front propagation. However it can be noted that the CRTA presents a step-like progression due to the coarser resolution of the radiative transfer which translates into a coarser resolution of the ionized front. Later on, the front is pushed back by the expanding cloud in both descriptions (as can be seen from the recessing velocities), but it happens earlier in the CRTA case. Finally the front cannot be tracked for $t>38$ Myrs, as no cells with a neutral fraction greater than 0.5 can be found anymore.  Let us  mention that comparisons of the standard calculations with the results presented in \citet{ILI09} confirms the capacity of \texttt{EMMA} to track correctly the front propagation within the clump. In particular, \texttt{EMMA} recovers as the other codes the phase where the front is pushed back by the expanding cloud, when $t\sim 35$ Myrs.
Globally a faster photo-evaporation of the cloud can be detected in the CRTA approximation compare to the standard AMR description, essentially due to the coarse description of the radiative fields.

Finally, we present in the lower panels of Fig. \ref{f:photo_map} and \ref{f:photo_map_d} the results of a CRTA calculation on a static $\ell=7$ grid. It allows us to probe the separate the effects of incomplete coupling between dynamics and radiative transfer and of the coarse description of radiation. In this experiment, the CRTA is equivalent to a radiative post-processing of the dynamics but performed on the fly, at the temporal scale of dynamical times and without any impact of a coarsened radiative transfer. Compared to the standard treatment, radiative transfer is subcycled with respect to dynamics, leading to a large number of radiative transfer+thermochemistry calls per single gravity and hydro calculation. Clearly the CRTA greatly reproduces the standard calculation in this case, both in the neutral fraction and density maps and in the propagation of the fronts. It confirms that the coarsened resolution is indeed the reason for an accelerated photo-evaporation of the clump and that the radiation subcycling does not induce significant deviation from the standard treatment.

\subsection{Cosmological runs}

\subsubsection{Preliminary Reionization Simulations}
\label{s:reionsimu}
 \dom{Finally, we present the results of  full simulations of cosmological reionization. The focus is put on hydrodynamical simulations with radiative transport  and on reionization-related quantities but additional tests on the dark matter haloes mass function or the energy conservation are presented in appendix \ref{s:addtest}.  We produced a set of 4 simulations with 4 different specific emissivities for the sources.  Each simulation consists in a 4 Mpc/h box sampled with $128^3$ base resolution cells and $128^3$ dark matter particles. These simulations will be referred as $X0.3$, $X1$, $X3$ and $X30$. The  $X1$ simulation is a fiducial run with sources emissivities that produce a reasonable ionization history. The three additional cases uses stars with boosted or depleted specific emissivities by the corresponding factor, $X0.3$ and $X30$ standing respectively for the dimmest and brightest source model.  Initial conditions were produced using Mpgrafic \citep{PRU08} with a Planck cosmology \citep{PLA13} ($\Omega_m=0.315, \Omega_\Lambda=0.685, \Omega_b=0.049, n_s=0.96, H_0=67$ km/s/Mpc) starting at $z=80$. Each DM particle weights $4\cdot 10^6 M_\odot$. AMR is triggered using a quasi-Lagrangian strategy and a cell is refined if it contains more than 8 DM particles. Radiative transfer is run with 3 groups of frequencies([13.6,24.6] eV, [24.6,54.4] eV and [54.4,1000] eV, dictated by the ionization thresholds of hydrogen and helium)}.

In addition to hydrodynamics and radiative transfer, we had to implement a simple star formation recipe in order to populate the simulated volume with the ionizing sources that drive the reionization process. This star formation model is briefly described here and will be the subject of a dedicated paper in the near future: its is widely inspired by \citet{KAT96,KAY02,RAS06, DUB08}. A cell is said to be prone to star formation if either its gas comoving density ($n_*$) or its gas density contrast ($\delta_*$) are greater than user-set thresholds. Once a cell is flagged to form stars, the number of stellar particles to be created is drawn from a Poisson law with the $\lambda$ parameter given by:
\begin{equation}
\lambda = \epsilon \frac{m_\mathrm{cell}}{m_\ell}\frac{\Delta t}{t_*}.
\end{equation}
$\lambda$ corresponds to the average number of stars created during a time step $\Delta t$ within a cell that contains a mass of gas given by $m_\mathrm{cell}$. The star formation process is controlled by a typical star formation time scale $t_*$ and an efficiency parameter $\epsilon$. The mass of a stellar particle is given by $m_\ell$, depends on the level of the cell and is equal to 
\begin{equation}
m_\ell=\bar \rho_b  \delta_*\Delta x^3.
\end{equation}
The following results were obtained with $\delta_*=150$, and $t_*/\epsilon=2 \mathrm{Gyrs}$. This values would be considered as 'standard' even though we won't discuss them here and we will explore thoroughly the results dependence on these values in a forthcoming paper.As shown hereafter they nevertheless lead to a star formation and a reionization process in reasonable agreement with constraints. Each stellar particle emits photons for 20 Myrs, with a constant emissivity and assuming a 50 000 K black body spectrum (see \cite{BAE09}). \dom{In practice the source emissivity has been tuned by trial and error to produce a reasonable reionization history, complete at $z\sim 6$ and for the fiducial $X1$ model, it results in an emissivity of $1.5\times 10^{16}$ ionizing photons/sec/stellar kg. Taking the calculation of \cite{BAE09} as a reference, which assumes a Salpeter IMF and $1-100 M_\odot$ mass range, it corresponds to a 15\% escape fraction. Again, X0.3, X3 and X30 use emissivities multiplied by the corresponding factor, the X30 model being clearly over-powered and merely used to probe the qualitative behaviour of the code in the regime of strong radiation. }

At the current stage we restrict ourselves to this simple model that obviously lacks important ingredients. For instance SN/AGN feedback has not been implemented yet, chemistry is limited to the simple hydrogen and no modification of equation of state is assumed at very high densities. As a consequence the star formation rate is essentially not regulated in this cosmological toy model. Hence the following results could not be considered as definitive regarding what the code could do but should rather be seen as tests on experimental configurations close to production runs.

\begin{figure}
\includegraphics[width=\columnwidth]{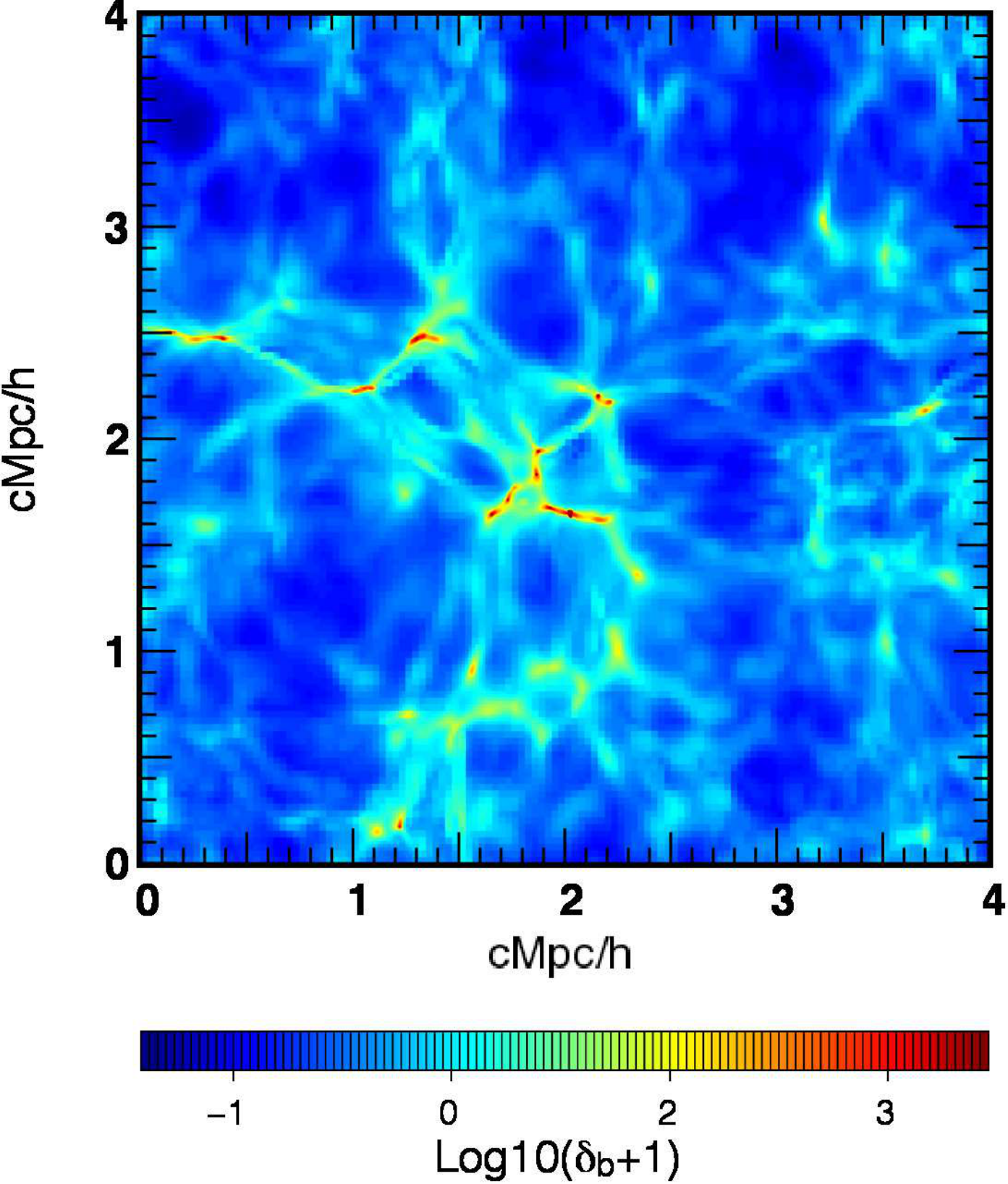}
\includegraphics[width=\columnwidth]{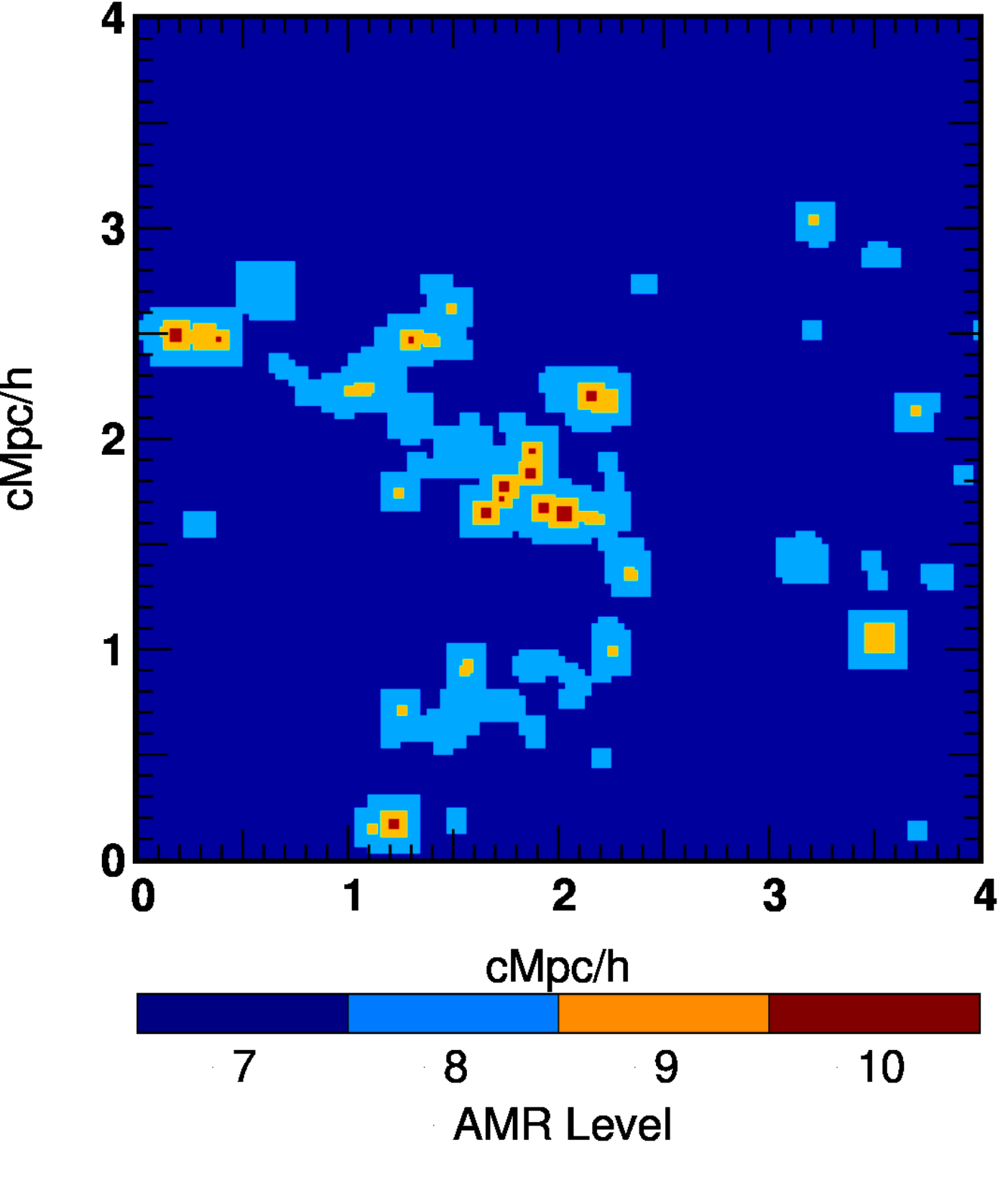}
\caption{Structuration of matter in a cosmological radiative transfer run of a comoving 4 Mpc/h-$128^3$ box, taken at $z=6.8$. Top: the baryon overdensity map. Bottom: the AMR levels. The shown region has a thickness of 320 comoving kpc/h.
}
\label{f:mapcosmo_d}
\end{figure}

\begin{figure}
\includegraphics[width=\columnwidth]{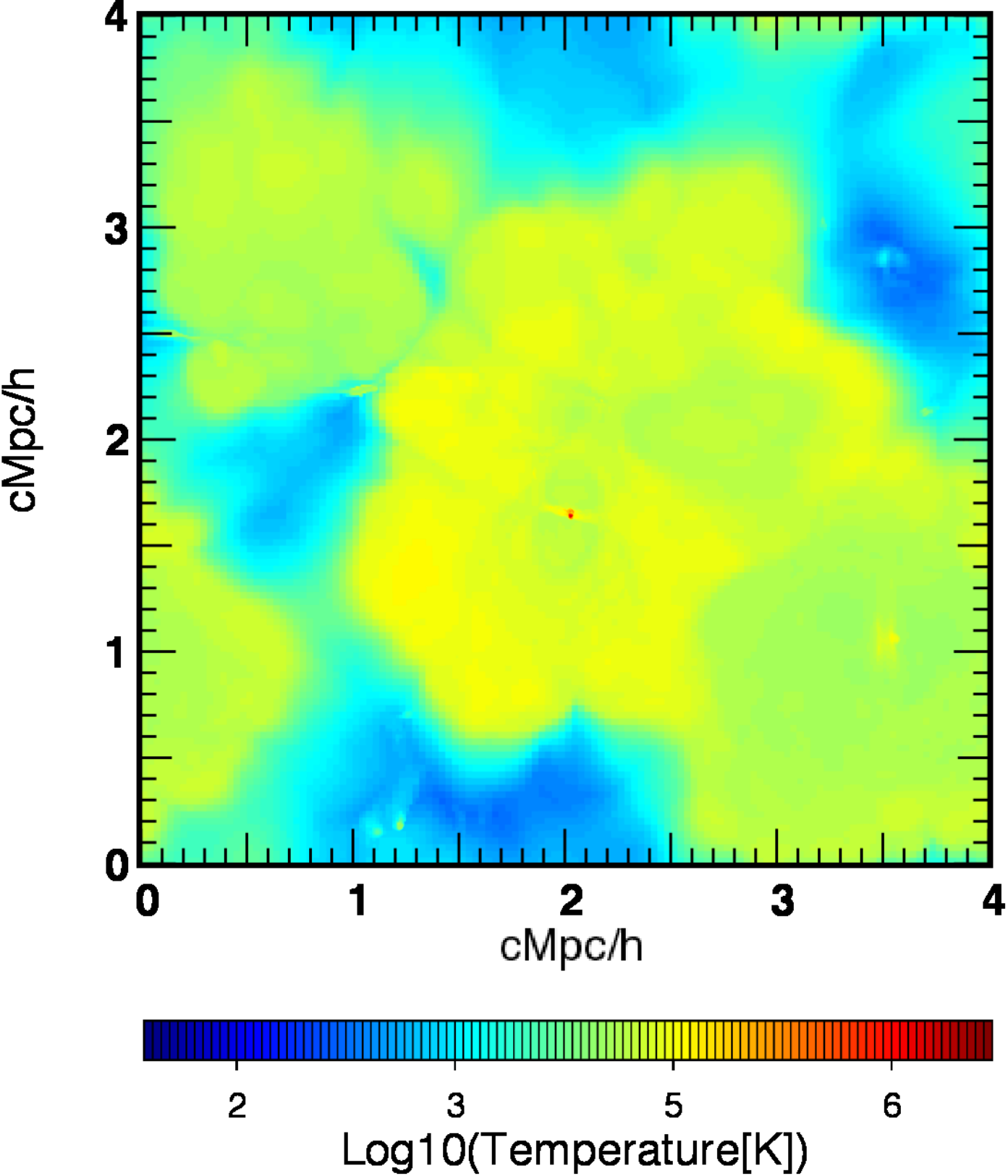}
\includegraphics[width=\columnwidth]{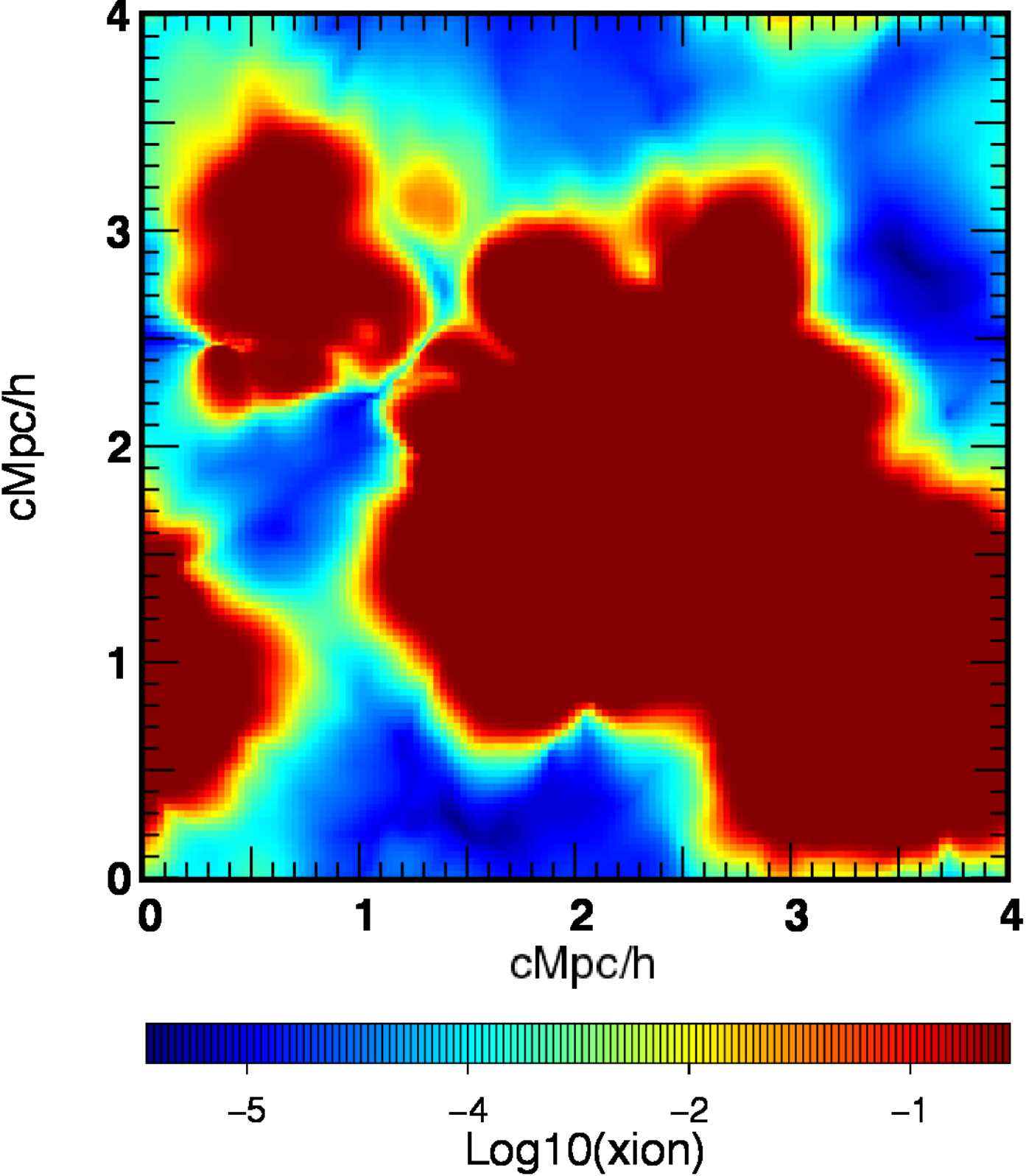}
\caption{Same experiment and region as in Fig. \ref{f:mapcosmo_d}. Top: the temperature map. Bottom: the hydrogen ionized fraction map.
}
\label{f:mapcosmo}
\end{figure}

Fig. \ref{f:mapcosmo_d} and \ref{f:mapcosmo} present the distribution of matter, AMR levels, temperature and hydrogen ionized fraction in a 320 kpc/h thick slab of the $X1$ run at $z=6.8$. Clearly, the matter on these scales is already highly structured at $z=6.8$, with regions having a density contrast greater than 1000. These regions are effectively tracked by the AMR grid and the overall distribution of high resolution grids follow the main features of the filamentary structure in this simulated volume. Sources are created in this overdensities and their radiation leads to large HII regions. One can note how the fronts are locally prevented to progress into the IGM by filaments and dense clumps, leading to complex features in their geometry.  It can also be seen that the ionization fronts present a certain extent induced by the larger mean free path of high energy photons. It also leads to a preheating of the gas, behind the ionization fronts, to temperatures close to a few thousands K. Within the ionized regions, a quasi homogeneous temperature close to 10000 K is set by the UV radiation with local fluctuations correlated with the density field. Some shock-heated gas (located at $\sim [2,1.7]$ Mpc/h) with temperatures greater than 100 000 Kelvins can also be seen.

\begin{figure}
\includegraphics[width=0.8\columnwidth]{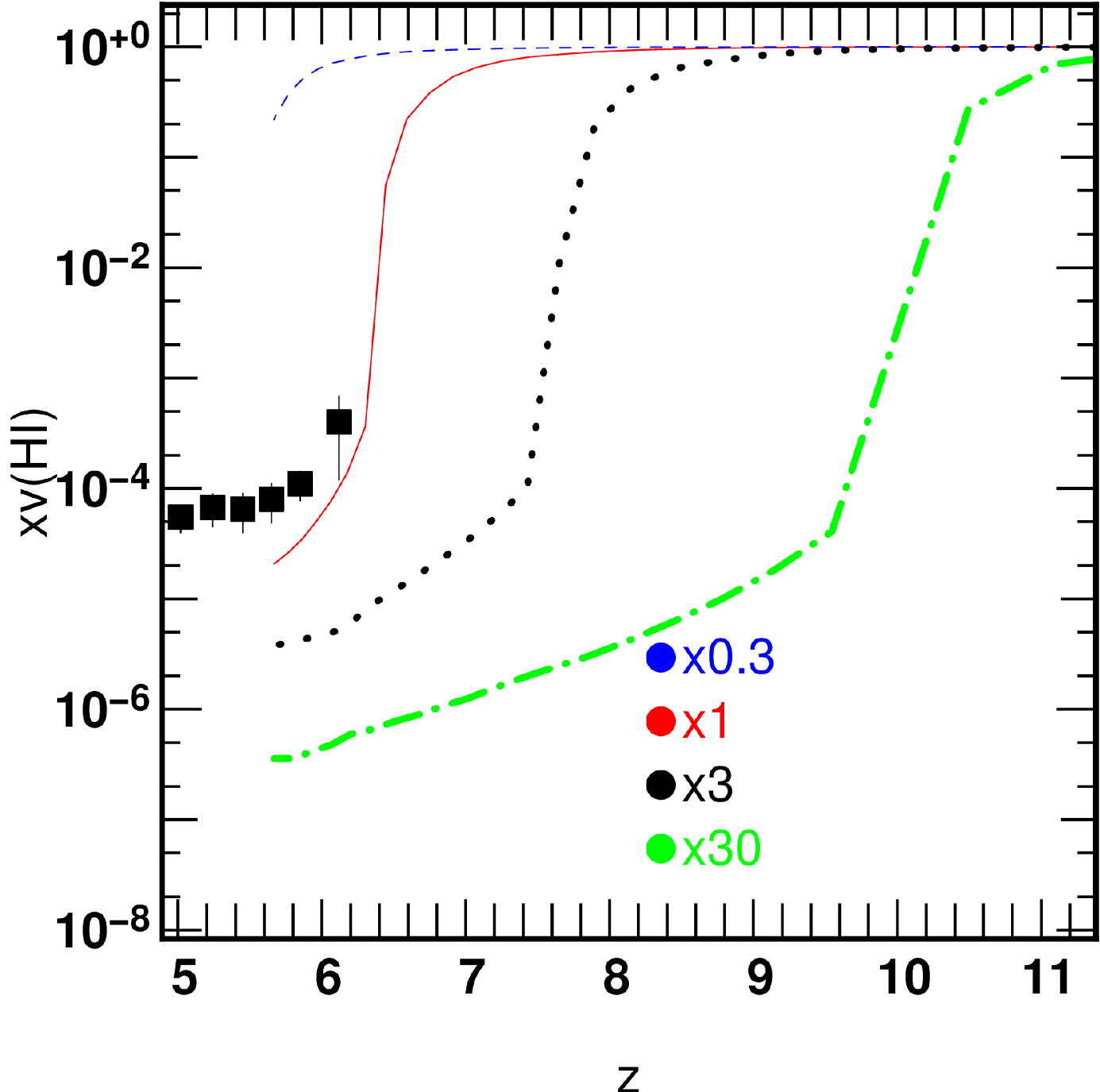}
\includegraphics[width=0.8\columnwidth]{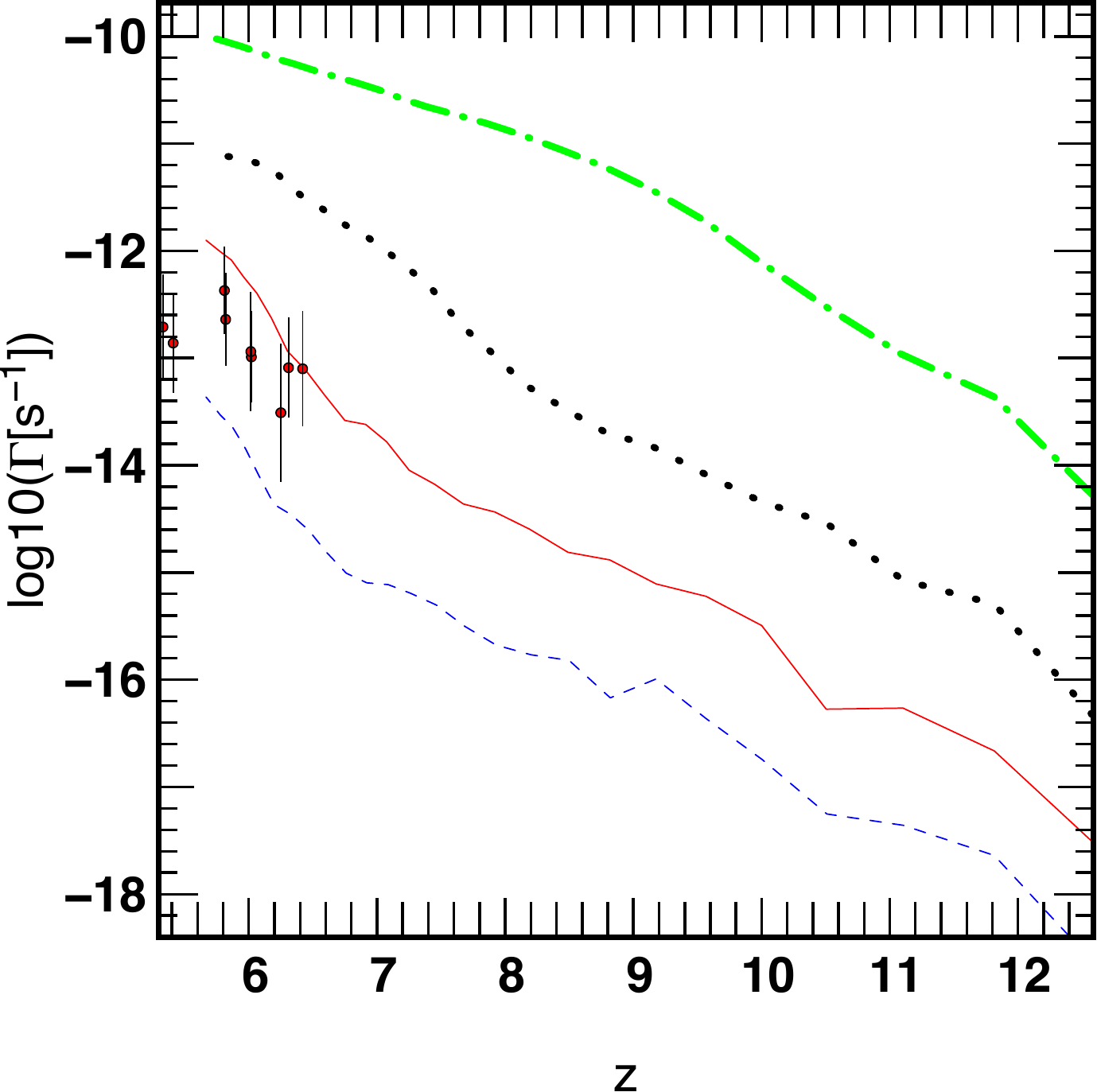}
\includegraphics[width=0.8\columnwidth]{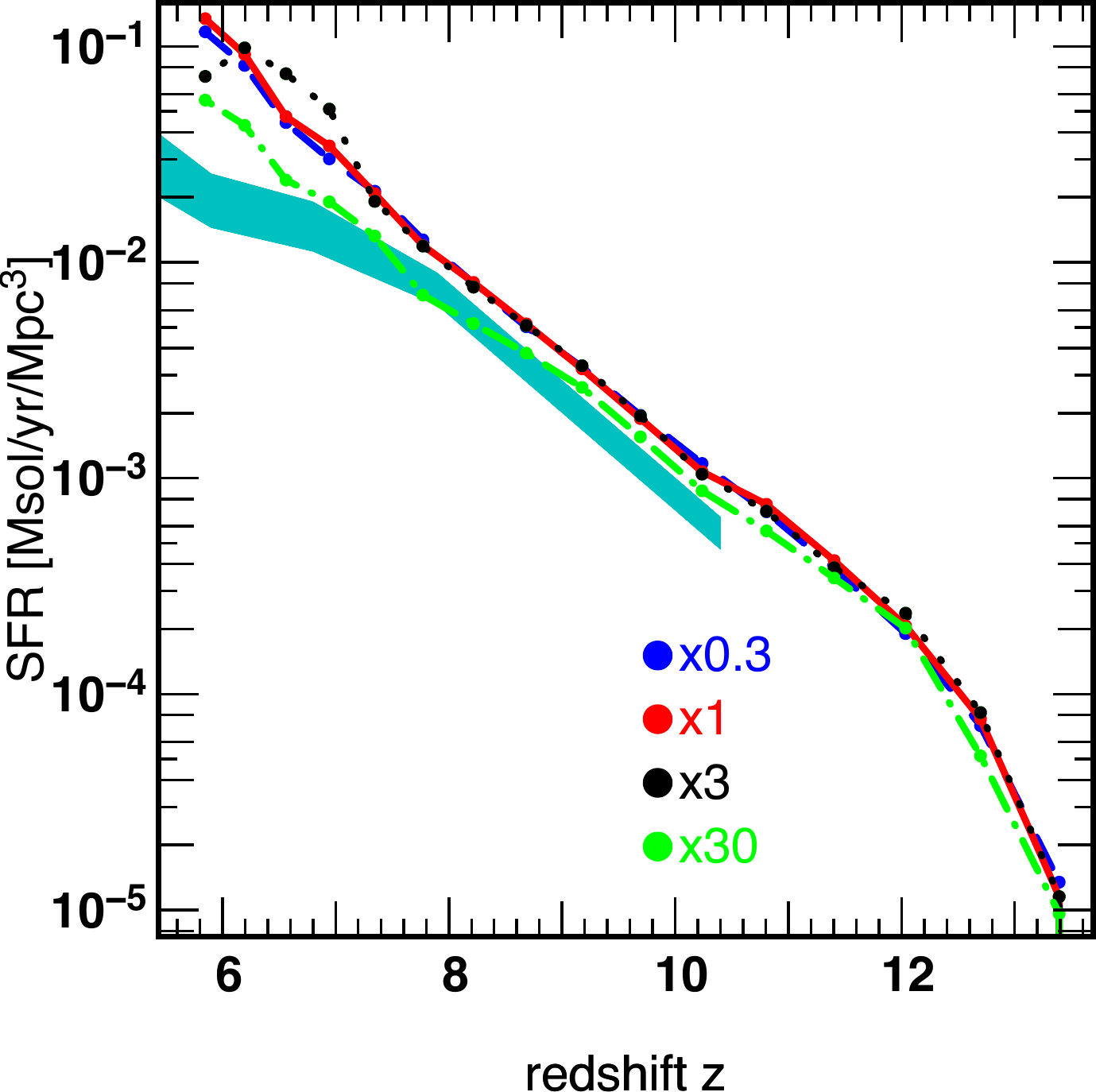}
\caption{Global evolutions of the average volume weighted neutral fraction (top), photoionization rate (middle) and star formation history (bottom) in 4 Mpc-$128^3$ reionization simulations. The fiducial model is shown in solid-red (labeled as having an $X1$ emmissivity) while simulations with different emissivities are shown in dashed-blue (with an emissivity equals to 30$\%$ the fiducial one, $X0.3$), dotted-black ($X3$) and dash-dotted-green ($X30$). Observational constraints from \citet{FAN06} (Top panel, squares), \citet{CALV11} (middle panel, points) and \citet{BOU14} (bottom panel, blue shaded area) are also given.
}
\label{f:xion_cosmo}
\end{figure}

The fiducial model $X1$ presents a reasonable reionization history and SFR, in broad agreement with observational constraints (see Fig. \ref{f:xion_cosmo}). Compared to \citet{FAN06}, the ionization happens slightly earlier than observed and correspondingly the photoionization rate at $z\sim 6$ is overestimated when compared to \citet{CALV11}. The cosmic star formation history is also in excess compared to the observationnally deduced rates given by \citet{BOU14}. This fiducial model could have benefited from a slightly improved calibration to reproduce the observed data points, however we consider that the current level of agreement is good enough at this stage : let us recall for instance that these simulations lack SN feedback and the small simulated volume could also be inadequate to make quantitative prediction on cosmic averaged quantities. At this stage we merely aim at looking for qualitative and not quantitative clues of the impact of radiation within a cosmological settings. 

These clues can be obtained by comparing this fiducial simulation with the 3 other models $X0.3$, $X3$ and $X30$. As expected these models result in different reionization histories that are in place at earlier (resp. later) times for larger (resp. lower) emissivities (see Fig. \ref{f:xion_cosmo}). The SFR however remains essentially unaffected by the change in emissivity, except at later times ($z>10$) for the models with the brightest sources ($X30$), leading to a depleted star formation rate.

\begin{figure}
\includegraphics[width=\columnwidth]{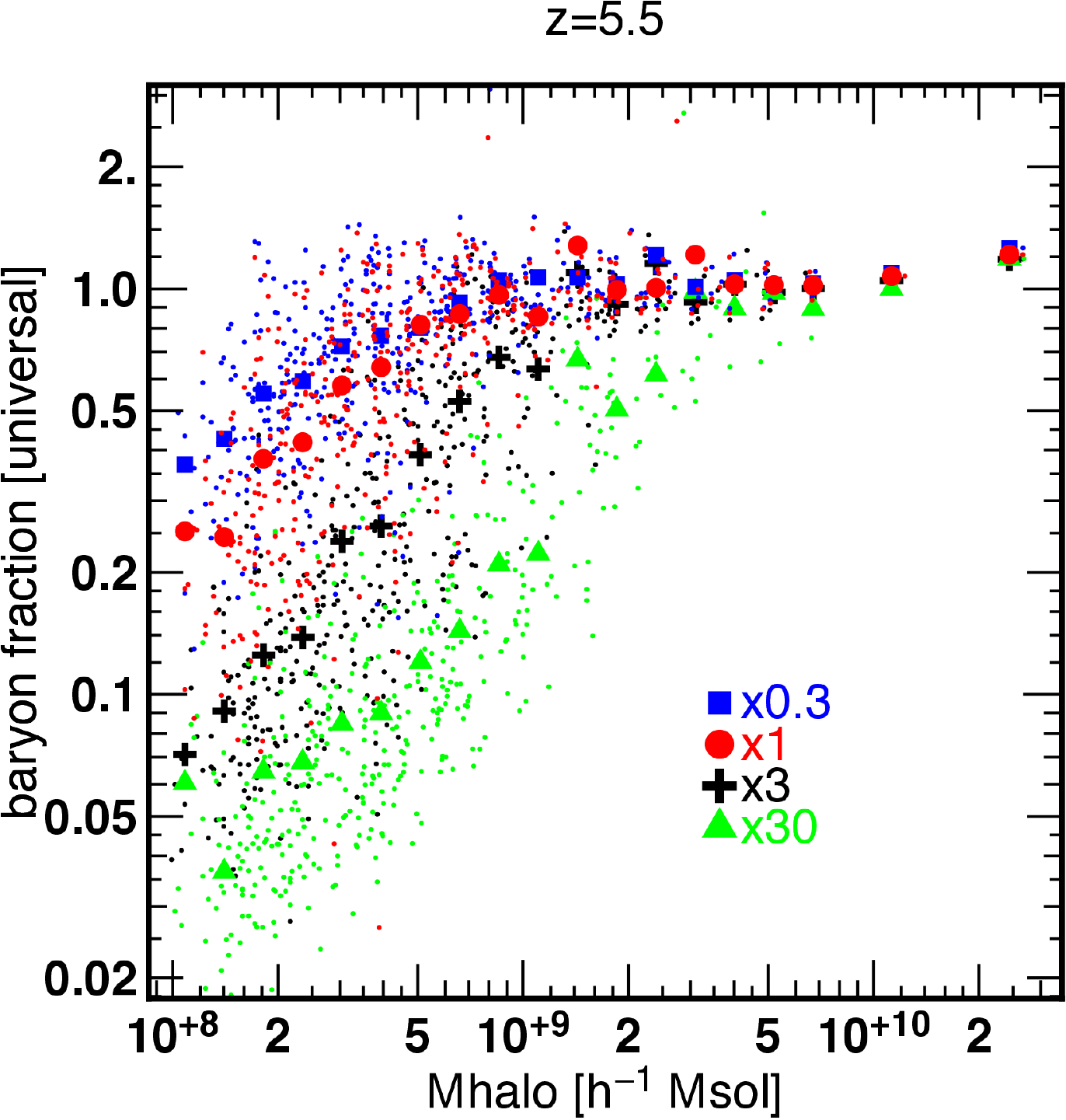}
\includegraphics[width=\columnwidth]{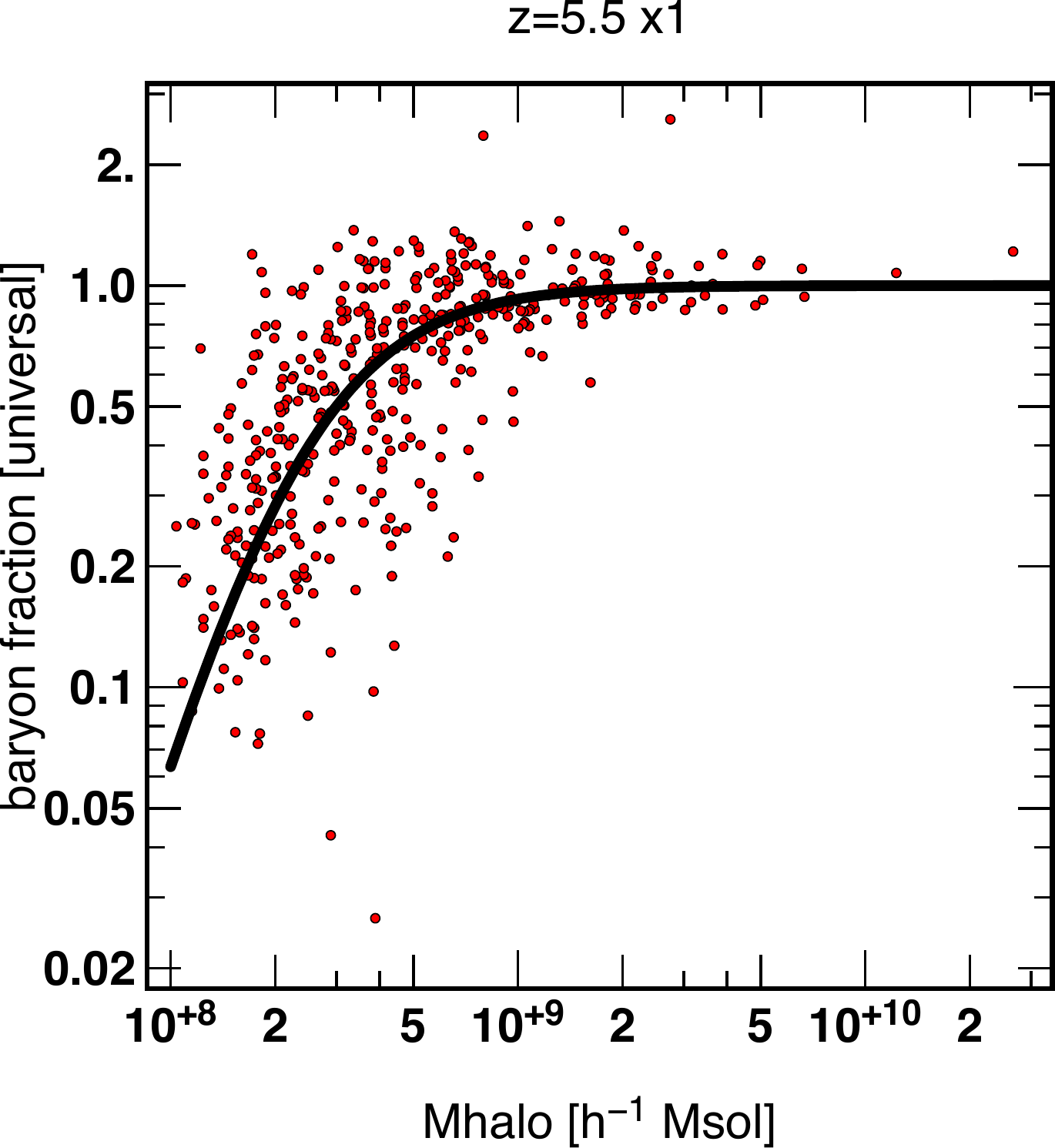}
\caption{Top: Baryon fraction in DM halos as a function of their mass at z=5.5 computed in the 4 models of emmisivities. Small dots stand for the values for each individual halo whereas large symbols stand for the average baryon fraction within a bin of halo mass. Bottom: the same quantity but for the fiducial model only (dots) compared to the \citet{OKA08} fit.   
\label{f:bfhalo_cosmo}
}
\end{figure}

Fig. \ref{f:bfhalo_cosmo} presents the baryon fraction and the instantaneous SFR measured in the dark matter halos found at $z=5.5$ in the different models. At this redshift, the reionization is well advanced in most models except in $X0.3$ where an $75\%$ ionization level is only achieved. Halos have been detected using the HOP halo finder (\citet{EIS98}) and baryons are counted within $R_{200}$ i.e. the radius of the spherical region around each halo with an average density 200 times greater than the cosmic average matter density. $\sim 450$ halos with a mass greater than $10^8 h^{-1} M_\odot$ (corresponding to 45 particles) are found. Clearly a significant scatter can be found in the distribution of the baryon fraction but general trends can nevertheless be observed in the data~: halos with a mass greater than $10^9 M_\odot$ basically present a universal fraction whereas lighter objects are more dark matter dominated as expected. A comparison of the fiducial model distribution to the fit provided by \citet{OKA08} shows a reasonable agreement with a correct transition mass at $M\sim 3\cdot10^8 M_\odot$, even though a significant scatter is obtained. If all the models are considered, a clear trend can be noted : the dimmest models ($X0.3$, $X1$) share the same global behavior (or qualitative functional form) even though the fiducial model presents baryon poorer low mass halos.
Meanwhile, the brightest models can produce baryon fractions 10 times smaller than the fiducial case with a different functional relation between the baryon fraction and the halo mass. The impact of radiation on this quantity seems therefore well established in this series of models, where brighter sources have a strong impact on the gas within shallow potentials. 

\begin{figure}
\includegraphics[width=\columnwidth]{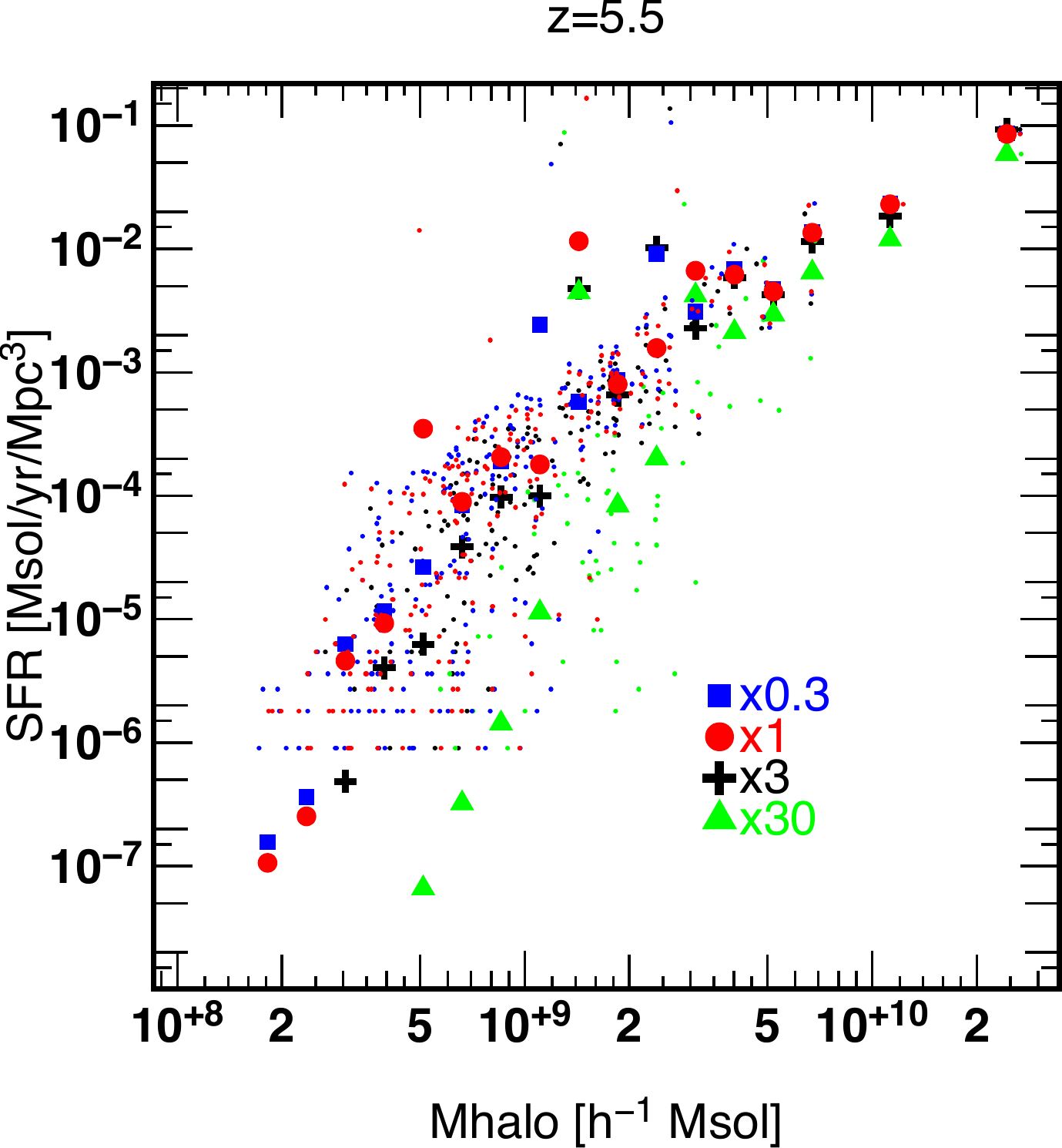}
\caption{The instantaneous star formation rate per DM halo mass for the 4 emissivity models.  Small dots stand for the values for each individual halo whereas large symbols stand for the average baryon fraction within a bin of halo mass.
\label{f:sfr_cosmo}
}
\end{figure}

Interestingly, this impact does not directly translate into a modified SFR  inside the halos (see Fig. \ref{f:sfr_cosmo}). Again the scatter is quite important and finite mass effects can be seen in halos with small formation rates and the interpretation can therefore be difficult. Still, it appears that the 3 dimmest models ($X0.3$, the fiducial $X1$ and $X3$) are not significantly different and present the same mass dependence of the star formation rate within their halos. Since we found that the global baryon quantity is indeed affected, it seems to imply that the \textit{star forming} baryons are unaffected by the source emissivity and the presence of radiation. Only the most extreme case of source emission shows a significant dip in the SFR of low mass halos : in our simplistic model of star formation, a certain level of gas depletion must be achieved to impact the production of stellar particles.

\begin{figure}
\includegraphics[width=\columnwidth]{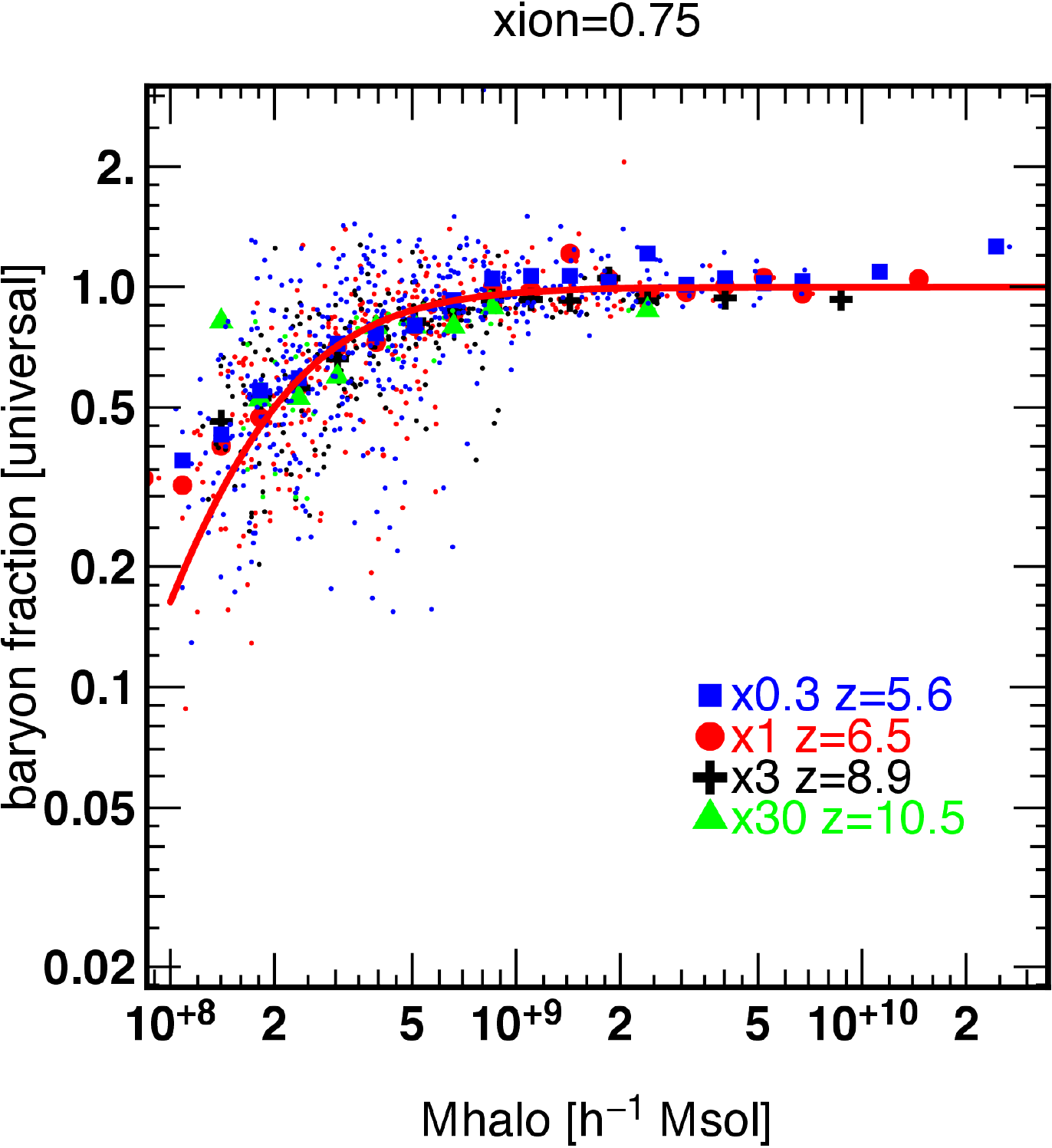}
\caption{The baryon fraction as a function of the halo mass for the 4 different models measured at the same ionization fraction $x=0.75$. The corresponding redshifts are given in the labels. The red line stand for the  \citet{OKA08} fit taken at the redshift of the fiducial model ($X1$).
\label{f:bfhalocstx_cosmo}
}
\end{figure}

As a final note, we present in Fig. \ref{f:bfhalocstx_cosmo}  the halo baryon fraction in the 4 models at the same cosmic average ionization fraction, $x=0.75$. Of course this level of ionization is achieved at high redshift ($\sim 10$) for the brightest model and corresponds to the last snapshot at $z=5.5$ for the dimmest one. It can be easily seen that the baryon fraction mass distribution is essentially identical in the 4 models, taken at 4 different redshifts but at the same ionization level. It could hint that an essential ingredient of the baryon depletion is not only the source intensity but also the exposition duration to the UV background. In the previous analysis at $z=5.5$ not only the brightest model contains the brightest sources but it also provided the longest duration over which halos are in an optically thin Universe since such models provide an early reionization. Conversely, dimmer models produce a late reionization and therefore a shorter exposition duration to the UV flux in a transparent Universe. It could impact the baryon fraction in low mass halo measured at a given time. At a given average ionization fraction, we somehow get rid of the scatter in flux exposition and look at halos from different simulations at a similar stage of their 'Universe' ionization history, with a similar structure for the UV field. And indeed, in our model, it significantly reduces the differences observed previously.

Let us recall that several important ingredients are missing in our models like the inclusion of supernovae feedback, which may enhance the SFR suppression in low-mass halos or the presence of $H_2$, which fraction can greatly differ from the baryon fraction. Hence the results presented in this section indicates in a qualitative manner that \texttt{EMMA} is able to handle cosmological reionization simulations. Further investigations and implementations are necessary to quantitatively assess the subjects discussed here.





\section{Performances}
\label{s:perf}
\subsection{Preamble}
\dom{As a closing chapter to this description of \texttt{EMMA}, we now discuss the performances of the code.  As shown hereafter, the comparison of performances on different architectures is a complex matter as it depends on how architecture-dependent optimizations are implemented. 
Such a comparison also depends on the context it has been performed~: as such it will evolve in time (as hardware improves for instance) or in 'space' (from one computer to another one at a given time). We nevertheless think that the following sections will shed some light on how the code behave and how these behavior can significantly vary depending on the compilers or the architecture. More generally it is also an opportunity to demonstrate that codes performances  should be carefully considered, and not only for \texttt{EMMA}.}

\dom{In the following sections, the calculations involved a 4 Mpc/h - $128^3$ cosmological simulation with full physics and the same parameters and simple star formation recipe as the ones described in section \ref{s:reionsimu}. They used the CRTA approximation with a $c/10$ speed-of-light, which should not affect the discussion on raw performance and scaling issues. We compared 3 types of \texttt{EMMA} binaries on the Curie-CCRT supercomputer hybrid nodes, compiled using single precision arithmetic:
\begin{itemize}
\item a GPU binary produced by the NVCC compiler from the CUDA 5.5 SDK with O2 optimization level, called \textit{gpu-O2}. This version runs the vectorized physical engines on M2090 Nvidia GPU devices while still relying on a single core to perform the other tasks, such as AMR logistics, vectorization, particles operations, etc... In the case of a multi-GPU run, each MPI process is attached to a single CPU core associated to a single distinct GPU.
\item a CPU binary produced by GCC 4.4.7, called \textit{gcc-O2} hereafter. This version fully runs on 2.7 GHz Sandybridge Westemere processors and uses the standard O2 optimization level.
\item a CPU binary produced by ICC 14.0.3. with the same O2 optimization level, called \textit{icc-O2} hereafter. Such \texttt{EMMA} binaries are usually faster than the ones provided by GCC by a factor close to 4~: this difference is essentially the result of optimizations on floating point operations that are enabled by default. Such optimizations can be disabled by setting an additional \texttt{fp-model=strict} flag and produce \texttt{EMMA} binaries with reduced performances at the level of the ones produced by gcc (not shown here).
\end{itemize}
ICC is available in most supercomputing and institutional facilities. GCC on the other hand is widely distributed and could be the only option on small configurations (e.g. on laptops, desktop machines or local shared memory calculators). Since they produce binaries with different performances, the resulting GPU acceleration will also depend on the CPU-version taken as a reference. }

\dom{Comparisons of GPUs and CPUs are done by considering one graphics device against one CPU \textit{core}. Obviously, it biases performances in favor of GPUs which are essentially parallel devices. Nevertheless, we argue that it is the simplest way to do the comparison, since a given GPU can be associated with a variety of different CPU nodes with different core numbers. However some care must be taken when considering acceleration rates.  If an acceleration factor of x80 is found, it should be seen as considerable since 80 CPU cores per GPU is already a significant configuration and codes usually don't follow strong scaling laws at such levels of acceleration (i.e. an x80 acceleration cannot be obtained using 80 cores). On the other hand if an x4 acceleration factor is found, it should be considered as low since 4 cores are easily obtained and x4 strong scaling factors can usually be achieved. In the case of Curie hybrid nodes, 1 GPU is associated with 4 cores but other configurations exist (e.g. Titan-ORNL associates 16 cores with 1 GPU).}

\subsection{Computing time consumption}
\begin{figure}
\includegraphics[width=\columnwidth]{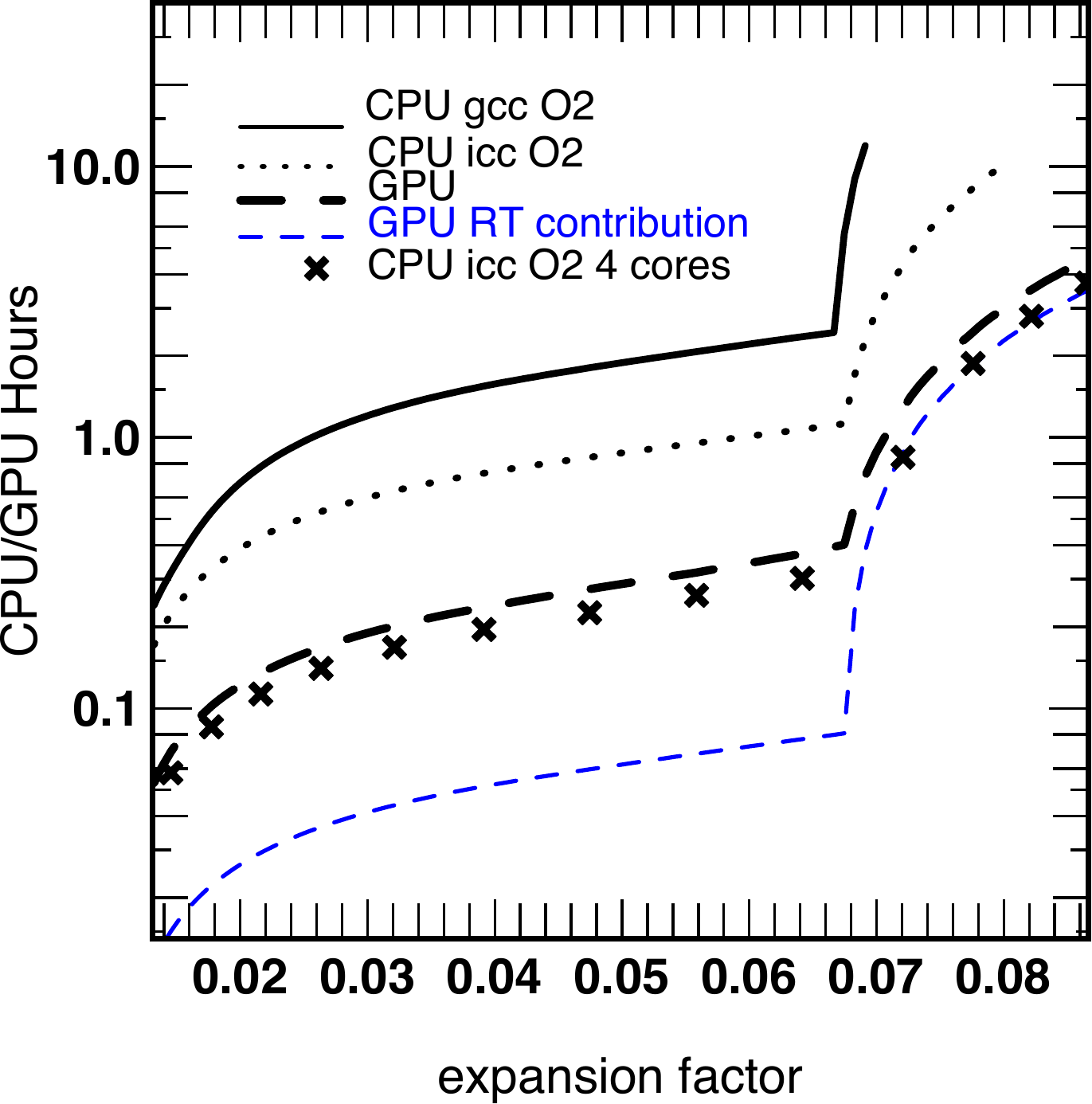}
\caption{Comparison of the cumulative time spent to reach a given expansion factor for a 4 Mpc/h-$128^3$ cosmological simulation of the reionization. Times are given for a single computing device (i.e. 1 GPU or 1 CPU core). The thick black dashed line stands for the GPU run performed on a M2090 Nvidia GPU whereas the thin dashed blue line stands for the contribution of radiative transfer to this cost. The black solid (resp. dotted) line stands for a single CPU core (2.7 GHz Sandybridge Westemere) using the \textit{gcc-O2} (resp. \textit{icc-O2}) binary. The symbols stand for a 4-core CPU calculation using \textit{icc-O2} on a Curie node.
\label{f:gpuvscpu}
}
\end{figure}

 Fig \ref{f:gpuvscpu} presents the time spent by a calculation to reach a given expansion factor in a cosmological simulation. Whichever code version is considered, two major phases can be distinguished : for an expansion factor $a=(1+z)^{-1}<0.065$ the code achieves a stable regime with small and  regular time steps (given by the slope of the Fig. \ref{f:gpuvscpu} curves).  At this stage, no source has been created yet and no light has to be propagated : the radiative engine (which also includes thermo-chemistry modules such as cooling processes) is not limited by the CFL condition and is called once per dynamical time step. Furthermore, non linearities are small and AMR has not been deployed yet, hence the work per coarse cell is naturally close to balance. At $a\sim0.065$ the first source appears and radiative transport must be computed while satisfying the stringent CFL condition. The number of RT calls per dynamical time step increases to typical levels of 150 calls per step. In Fig \ref{f:gpuvscpu} the time spent increases by orders of magnitude with a much greater slope, i.e. a much greater time spent per time step. This contribution of RT to the computing time is further emphasized by the dashed blue line in Fig. \ref{f:gpuvscpu} which stands for the RT-only contribution in the \textit{gpu-O2} calculation (similar curves are obtained for the CPUs calculation albeit not shown here): clearly the dramatic increase in the computing time is driven by this specific module.

In the same plot, solid lines stand for the computing time required for \texttt{EMMA} to reach a given expansion factor using a M2090 Nvidia GPU Device  with \textit{gpu-O2}(black dashed line) and using a single CPU core with \textit{gcc-O2} (black solid line) and \textit{icc-O2}  (black dotted line) binaries. 
Comparing these different versions of \texttt{EMMA}, it can be seen in Fig. \ref{f:gpuvscpu} that for the \textit{gpu-O2} version  $a=0.07$ is achieved in 50 minutes, whereas 16 hours are required for the \textit{gcc-O2} version, providing an x16.9 acceleration factor. In the pre-source regime (for $a<0.065$) this acceleration factor drops to x6.4: in this regime the contribution of the radiative transfer engine is much smaller and so is the level of potential acceleration. In the very first stages of the calculation this acceleration rate even drops further (to factor close to x4) as the cooling induced by dynamical effects is small and hence the need for associated calculations that could have benefited from hardware acceleration. \dom{If the GPU version is compared to the \textit{icc-O2} run,  the maximal acceleration rate of the \textit{gpu-O2} code drops to x3.9. Clearly the removal of strict value-safe floating point operations (which allows greater optimization from the CPU compiler) results in a more competitive CPU code performance-wise. 
}
\dom{Moreover, it should be noted that the current comparisons deal with a GPU device against a  single CPU core as we argued that it provides the simplest mean of comparison. However cores are usually part of multi-core nodes, connected to one or two GPU devices. Hence an acceleration rate of a few can be seen as not sufficient if it does not exceed the core per GPU ratio. For instance we show in Fig. \ref{f:gpuvscpu} the time required for the 4 cores of hybrid Curie \textit{node} to run the same test (symbols). As can be seen here, the strong scaling behaviour of \texttt{EMMA} is sufficient to further improve the CPU-consumption by a factor of almost 4 and the parallel \textit{icc-O2} binary slightly outperforms the \textit{gpu-O2} performances. As described in the next subsection, these diverse performances result from the impact of CPU optimization and GPU acceleration that are not uniformly distributed among the different modules.}

\subsection{Detailed computing cost breakdown}
\begin{figure}
\includegraphics[width=\columnwidth]{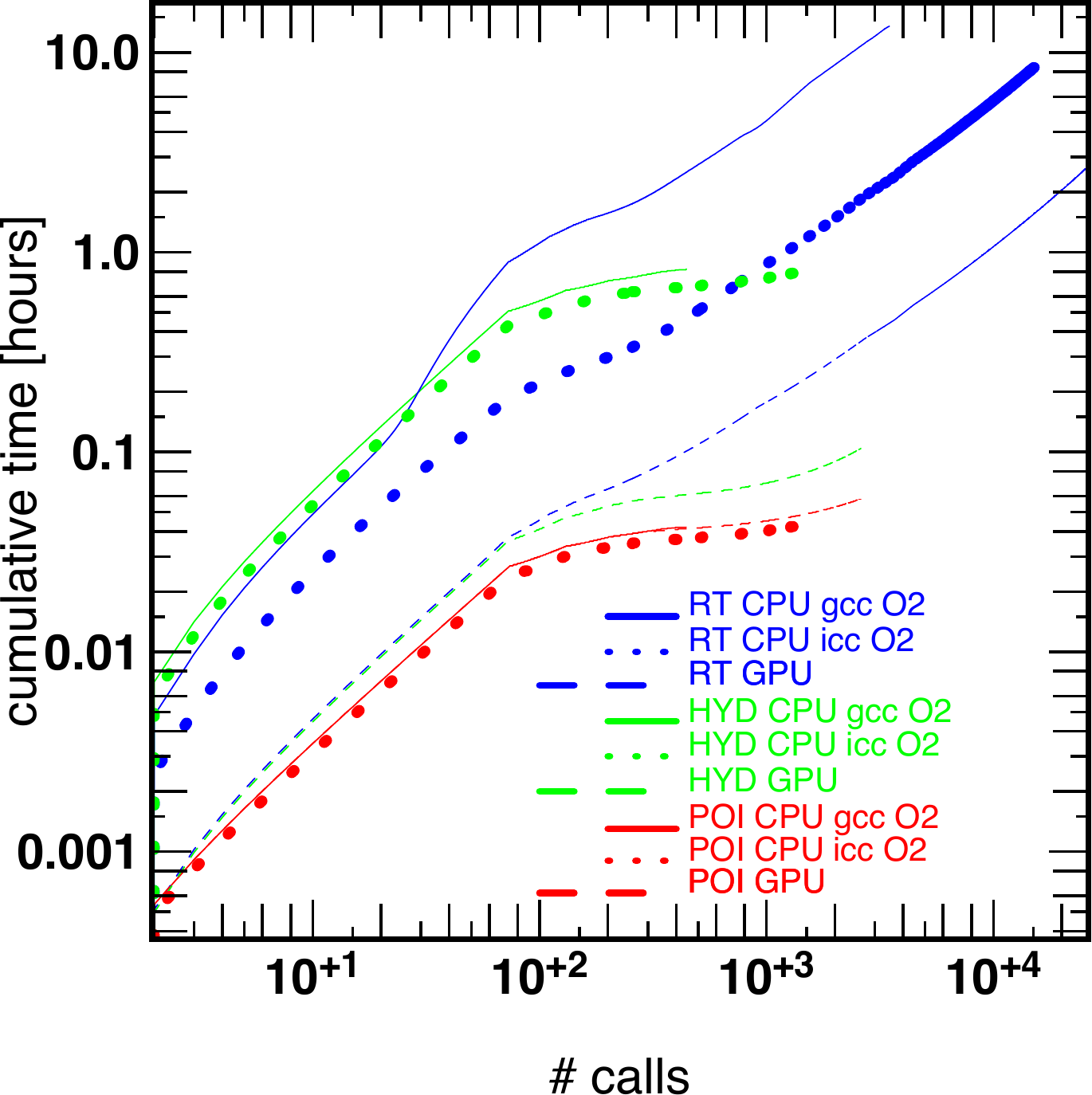}
\caption{Cumulative time spent in the Poisson (red), hydro (green) and radiative transfer (blue) physical engines as a function of number calls. Solid lines stand for single core CPU calculations on a 4Mpc/$128^3$ simulation using \textit{gcc-O2}, dotted lines stand for single CPU core calculations using \textit{icc-O2} and dashed line for calculations driven by a single M2090 GPU using \textit{gpu-O2}.  
\label{f:calls}
}
\end{figure}

For the same experiments, Fig. \ref{f:calls} presents the cumulative time spent in the 3 principal modules (Poisson, Hydro and Radiative solver) of \texttt{EMMA} as a function of the number of successive calls made to these modules. Solid lines stand for the single core CPU calculation obtained with \textit{gcc-O2}, \dom{dotted lines stand for single core CPU calculations produced by \textit{icc-O2}} and dashed lines stand for the GPU-driven experiments. . 

Focusing first on the \textit{gcc-02} results, it appears clearly that hydro and RT calculations dominate the overall time budget of \texttt{EMMA}. Unsurprisingly, the Poisson solver only contributes marginally to the overall cost~: first, the amount of calculation involved in this stage is small compared to the complex hydrodynamical solvers or thermo-chemistry calculations. Second, it relies on an iterative solver, where the solution does not evolve quickly from a time step to another or only in a few hyper-refined cells, ensuring a rapid convergence and hence a low computational cost. It can also be noted that the hydrodynamics are the dominant stage at early times, being overtaken by RT only as thermo-chemical computations start to contribute and obviously at later time when the CRTA approximation execute $\sim 150$ RT calls per hydro call. This effect due to CRTA is also evident in the number of RT calls which is much greater than the identical number of hydro and Poisson calls.

Looking at the performance of the GPU -driven binaries \textit{gpu-O2}, the time spent in the hydrodynamics and the RT is reduced to the levels of the Poisson solver~: \dom{compared to \textit{gcc-O2} the RT module is accelerated by a factor x32 and the hydrodynamics by a factor x14 }. Interestingly we could not achieve any acceleration with the Poisson solver on GPU architecture. The reason is the poor computation/transfer ratio for the Poisson solver: our measurements show that gathering the data from the AMR to vector-like structure on CPU takes $\sim 75\%$ of the time required by the Poisson solver: the room for acceleration is therefore extremely small whereas in hydro and RT this gathering stage only represents $\sim 5-10\%$ of the computation. In general, the acceleration can be efficient on computation-dominated modules, and in our implementation the Poisson iterative solver does not belong to this family of functions and represents therefore an intrinsic limit to \texttt{EMMA} performances on GPUs. 

Finally, dotted lines show the cumulative time per calls of a given module but using \textit{icc-O2}. No differences can be noted for the hydro and Poisson solver compared to the timings obtained from \textit{gcc-O2}, but the time spent into the radiative transfer module is greatly reduced. It is easily explained by the important contribution of non trivial mathematical operations in cooling rates, cross-sections, ionization rates, etc... present in the thermo-chemistry operations handled by the RT module. Such operations have a clear benefit from the optimizations made by the compiler.  This is not the case for hydrodynamics~: even though a MUSCL scheme involves a great number of operations, they essentially rely on simple arithmetic operations, which are less prone to optimizations. \dom{For hydrodynamics, \textit{gpu-O2} still provides a x12  acceleration compared to \textit{icc-O2} but RT acceleration rate drops to x5.5 : since it is the dominant process, it strongly affects the overall GPU acceleration.}
\begin{table}
\begin{tabular}{|c|c|c||c|c|}
\hline 
& HYD VT &HYD Cal&RT VT&RT Cal\\ 
\hline 
\textit{icc-O2} & 1.2 & 26.4 & 0.9 & 13.1 \\ 
\hline 
\textit{gpu-O2} & 1.76 & 0.69 & 1.66 & 0.85 \\ 
\hline 
\end{tabular} 
\caption{Typical time spent (in seconds) in the vectorization+transfer operations (VT) and in calculations (Cal) for the hydrodynamics (HYD) and radiative transfer modules (RT). Times are given for time step $\#$10 of the benchmark simulation described in Sec. \ref{s:perf}, corresponding to a regime without sources and without AMR.}
\label{t:time}
\end{table}

\dom{The current limiting factor of GPU performance is the cost of vectorization and data transfer to and from the device. In fact, the near identical floor performance obtained by the three GPU modules is due to the irreducible cost of these operations. In Tab. \ref{t:time}, we list the time consumption for the vectorization-transfert stages as well as for the actual computations, measured in a typical early-stage step and for \textit{icc-O2} and \textit{gpu-O2} binaries. Poisson Solver results are not discussed as they are already dominated by vectorization on CPU.  We find that for the hydrodynamics and RT the cost of these operations is close to 70\% on GPUs (see Tab. \ref{t:time}) whereas they contribute to less than $10\%$ of the CPU calculation~: the acceleration potential of \textit{calculations} is thus almost fully exhausted. It can also be noted from Tab. \ref{t:time} that these vectorization/transfer steps are actually more expensive on GPU, because of the additional transfer of data from the CPU host to the GPU~: the cost of transfer is broadly equivalent to the vectorization. It doubles the time spent in non-calculations operations which end up dominating the cost of the hydrodynamical and radiative transfer modules.}

\subsection{Parallel scaling}

\begin{figure}
\includegraphics[width=\columnwidth]{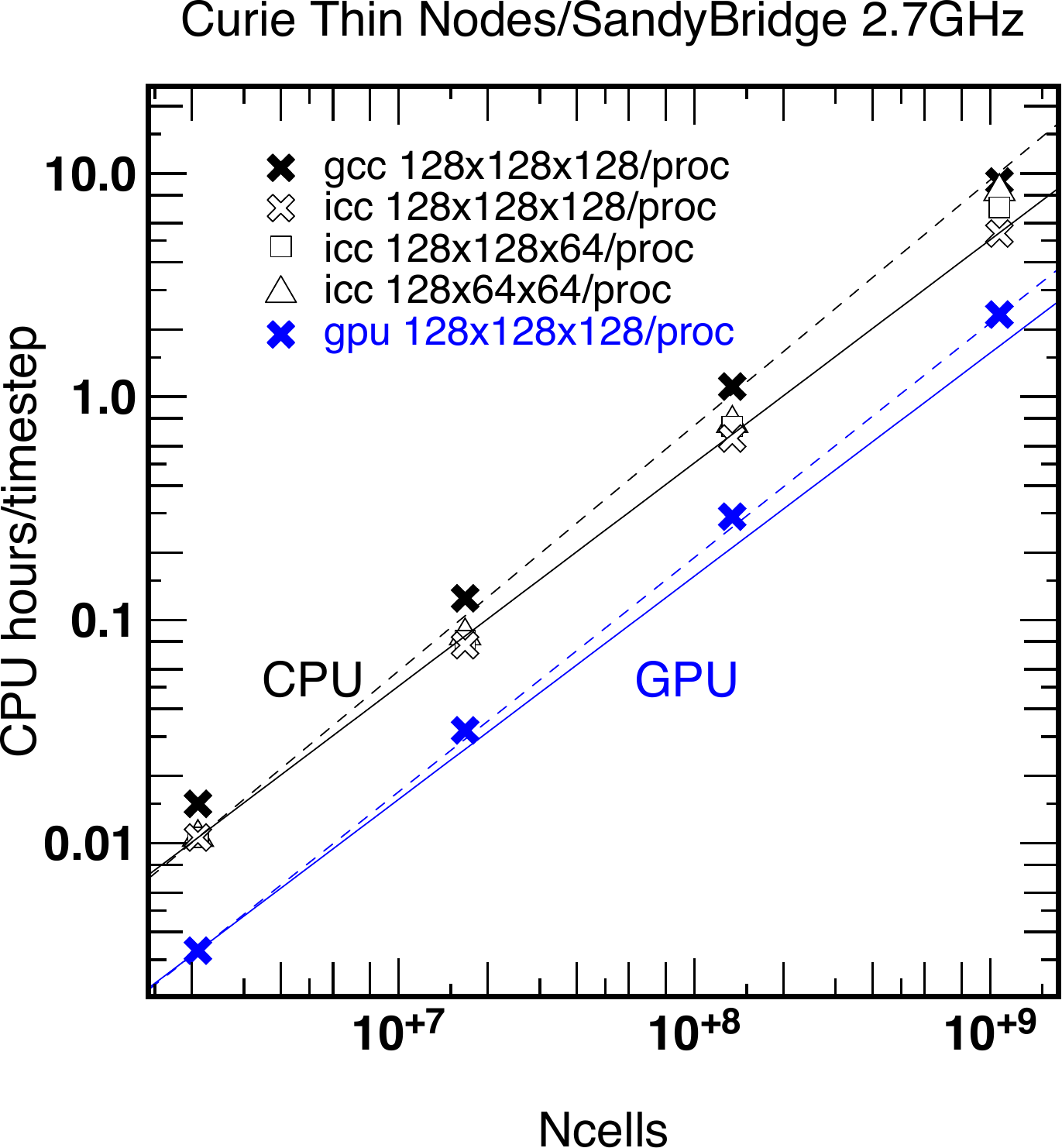}
\caption{Empty symbols: scaling curves for different loads per process configurations using the \texttt{icc-O2} binary, given as the CPU hours per time step as a function of the total number of coarse cells. The black solid line represents the perfect CPU scaling (i.e. $ t \sim N_c^\alpha$ with $\alpha=1$) while the black dashed line stands for the worst scaling law obtained here with $\alpha=1.1$. Filled black symbols stand for the same measurements but made with the \texttt{gcc-O2} binary and a constant load of $128^3$ per process. Filled blue symbols stand for the same measurements on GPUs (using \textit{gpu-O2} binaries) and a constant load of $128^3$ per process. The blue solid line stands for the perfect scaling expected for GPU whereas the dashed blue one stands for the one actually measured with $\alpha=1.05$.
\label{f:curiescal}
}
\end{figure}

Fig. \ref{f:curiescal} shows the scaling properties for \texttt{EMMA}, where a fix load per process is chosen and the number of process is increased, thus increasing the volume and total number of coarse cells or particles handled by the code. For the \textit{gpu-O2} and \textit{gcc-O2} scaling measurements, we stick to a 4Mpc - 128x128x128 coarse cells per process load, similar to the one used above. For the \textit{icc-O2} Intel CPU scaling we varied the load per core and used non cubical sub-domains. Configurations from 1 to 256 GPUs and from 1 to 2048 cores have been used. Times were measured in the initial stages of a cosmological run, during the first 20 steps. At these early stages, non linearities are weak and AMR is not triggered:  load imbalance is minimal and therefore allows a better estimation of parallelism-induced deviations. It also corresponds to the epoch where the acceleration is the weakest but it should not affect our conclusions regarding the scaling abilities of the code. Clearly the (weak) scaling trends are satisfying with a CPU/GPU computing time that scales as $t\sim N^{[1-1.1]}$ where N is the number of cells. In fact on CPU the scaling is almost perfect in configurations with a 128x128x128 load per process and drifts away for smaller problems per process : it is a standard strong scaling issue, where a smaller local load increases the weight of parallelism overheads. Furthermore, as the sub-domains become non-cubical the number of neighbors and therefore the amount of communication varies from one process to another if a Peano-Hilbert segmentation is used, leading to a small unbalance of communications between processes. Nevertheless, the scaling of \texttt{EMMA} running on multiple CPUs and GPUs seems satisfying, aside from load-balance issues that will inevitably arise later on as structures will emerge. We plan to implement balancing procedures in forthcoming developments.

\subsection{Discussion}
Overall, the performance achieved by GPU driven calculations is promising and acceleration rates greater up to x16.9 can be obtained, but such rates should be discussed as performance comparison is a complex matter. 
First the large acceleration rates are obtained in the regime where the CRTA is used and effective (i.e. after the first star has appeared): the CRTA is somewhat designed to favor hardware accelerators as it increases the weight of pure and heavy calculation on the overall budget of \texttt{EMMA}. Furthermore it remains an approximation with a coarse description of radiative transport, hence it stands as a lower order approximation compared to the standard AMR coupling of radiation to matter. Note that even in the CRTA regime this standard coupling is naturally enforced in the pre-source stages, and we demonstrated that only moderate acceleration is achieved in this regime (x6.4). Second, we also demonstrated that properly optimized CPU binaries can be only a few times slower than GPUs. The overall code acceleration rate drops then to x3.9 even in the CRTA dominated regime, as can be seen by examining  Fig. \ref{f:gpuvscpu}. Finally the performance gap can further be reduced by increasing the number of CPU cores used or by using full nodes capacities. 

\dom{ It should be noted that these acceleration are global ones, i.e. they rely on global timings of \texttt{EMMA} where a significant number of modules remain to be ported on GPUs. This is for instance the case for particles-related operations which are currently only handled by the CPU. Even if they are sub-dominant, optimizing these tasks or porting them for GPU architecture could provide a moderate additional acceleration. }

\dom{However, the most obvious way to increase the GPU acceleration is to reduce the cost of vectorization and data transfer to the device. The data transfer bottleneck is expected to evolve naturally as new standards are being developed to increase the CPU to GPU bandwidth \footnote{http://www.nvidia.com/object/nvlink.html} or by using architectures such as AMD's Accelerated Processor Units (APUs) where CPU and GPU share the same memory, thus nullifying the cost of transfer. It should be noted that thanks to the vectorization strategy of \texttt{EMMA}, future ports on new architectures should be of limited complexity. Regarding the cost of the vectorization (driven by gather/scatter operations), let us note that this operation is currently purely sequential and dealt with by the CPU. Gather/scatter operations could in principle be parallelized to reduce its imprint on the overall costs and to lower the intrinsic floor that limits the GPU performance.  Such parallelization is limited by concurrent memory accesses but current CPU architectures designed with Non Uniform Memory Access (NUMA) should in principle alleviate this issue.  Another option would be to deport the vectorization on the parallel computing device ( a GPU in our case)~: performance gains are expected to be limited, since gather/scatter operations rely on non `gpu-friendly' tree-walk operations, but even a weak acceleration of the vectorization process would provide a welcome boost to the calculation acceleration.}


The tests presented here were made on CPU and GPU architectures of the same generation (M2090+2.7 GHz Westemere on Curie), but newer hardware is and will be available and similar tests will be necessary to reassess the results shown here.  \dom{However preliminary tests on more recent devices show no significant improvement in performance (see appendix \ref{s:K20}) . It does not come as a complete surprise since our GPU calculations are currently limited by vectorization+transfer and the more recent hardware does not provide significant progresses on these aspects.}

On a broader perspective, the results presented here seems to make a strong case for a full usage of hybrid installations, where \texttt{EMMA} would distribute its different tasks/modules simultaneously on the different types of hardware (multicore CPU, GPU) available on a node. 
\dom{For instance gather/scatter operation on a CPU represents typically 5-10\% of the time spent in a physics engine, hence there is room for potential acceleration on a multi-core node through OpenMP directives, probably of a factor of a few, making it competitive with current GPU accelerations. For instance, multiple tasks could be done in parallel such as e.g. thermo-chemistry on multi-core CPU and radiative transport on GPU and current performances could be increased by an overall factor of 2. Of course, these estimates need to be confirmed by experiences.}

\section{Conclusions}

\texttt{EMMA} is a cosmological simulation code which handles simultaneously gravity, hydrodynamics and radiative transfer on an adaptive grid that can be refined on the fly (AMR). Written in C, this code is parallel (via the MPI protocol) and can deploy its physics modules on graphics processing units using CUDA. Designed for the study of the reionization epoch, \texttt{EMMA} is nevertheless a versatile code for structure formation. 

The code passed a variety of test cases and can confidently produce accurate and relevant simulations. A first comparison of cosmological reionization simulations with different source parameters presents the expected qualitative behaviour of the physics at play.

 \texttt{EMMA} has been tested in a wide variety of parallel configurations and it demonstrates satisfying scaling properties. It is able to use graphics processing units (GPUs) to accelerate hydrodynamics and radiative transfer calculations. Depending on the optimizations and the compilers used to generate the CPU reference, global GPU acceleration factors between x3.9 and x16.9 can be obtained.  Vectorization and transfer operations currently prevent better GPU performances and we expect that future optimizations and hardware evolution will lead to greater accelerations. Overall we demonstrate that \texttt{EMMA} is able to cope efficiently with a variety of hardware.


Aside from optimization to improve the code performance and GPU-driven acceleration factors, additional features will be included into \texttt{EMMA} in a near future. Among them, star formation and supernovae feedback is a major priority as it is an essential ingredient for galaxy formation theories and models. Their implementation is on the way and will be the described in a forth coming paper. Molecular chemistry is envisioned too as it is physically relevant to understand the formation of the first and smallest objects during the reionization epoch (like mini-halos with $M<10^7 M_\odot$) and also numerically interesting as such calculations can be easily accelerated. Full documentation of the code is also on the way in order to publicly release \texttt{EMMA} on a finite, maybe short, term.

\section*{Acknowledgments}
We are grateful to B. Semelin, N.Gillet, J. Blaizot, J. Rosdahl, C. Pichon, R. Teyssier, T. Stranex, P. Shapiro for discussions over the years that provided the basis for the \texttt{EMMA} code. We are also grateful to the anonymous referee who helped to improve the quality of the article. This work has been supported by the ANR grants ANR-12-JS05-0001 (EMMA), ANR-14-CE33-0016-03 (ORAGE) and ANR-09-BLAN-0030 (LIDAU).This research used resources of the Oak Ridge Leadership Computing Facility (INCITE 2013 Award AST031, INCITE Prep project AST105), which is a DOE Office of Science User Facility supported under Contract DE-AC05-00OR22725, from the Meso-Centre de l'Universite de Strasbourg and from the Centre de Calcul Recherche et Technologie-CCRT (Curie-CPU \& Curie-GPU, DARI Grant  2015047393). The authors thank D. Munro for freely distributing his Yorick\footnote{http://dhmunro.github.io/yorick-doc/} programming language and its yorick-gl extension.

\appendix
\section{Additional tests on cosmological simulations}
\label{s:addtest}
\subsection{Mass function in a pure DM simulations}
\label{s:fmass}
\begin{figure}
\includegraphics[width=\columnwidth]{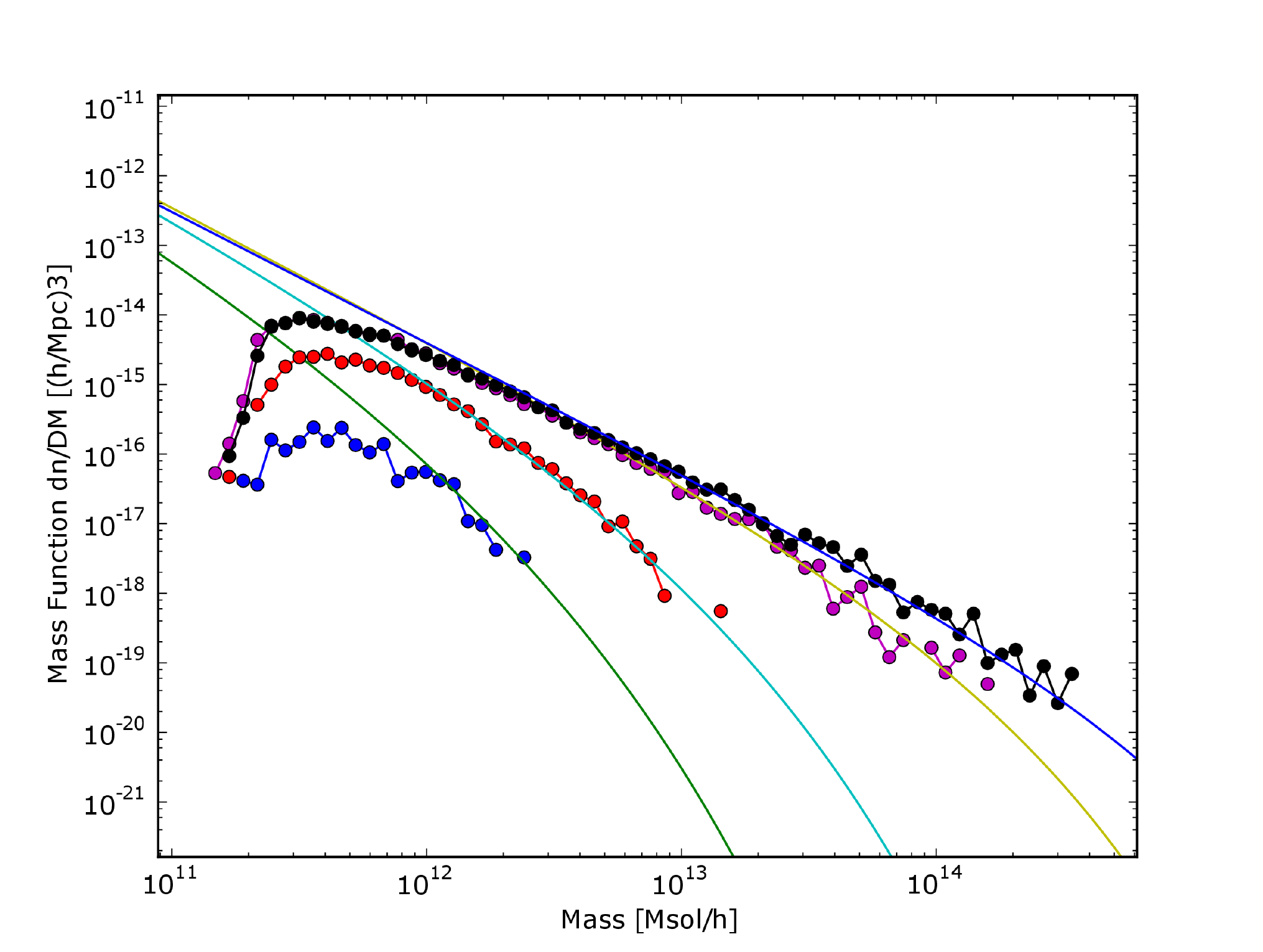}
\caption{The halo mass function of a pure dark matter $100 h^{-1}\mathrm{Mpc}/256^3$ cosmological run. Dots stand for the mass functions measured in the simulation at z=5.4, 3.3, 1.0 and 0.0 (from bottom to top). Lines stand for the \citet{SHE01} expression for the halo mass function.}
\label{f:fmass}
\end{figure}
First, we try to recover the halo mass function in pure dark matter cosmological simulations. Initial conditions were produced with the MPGrafic package \citep{PRU08} for a $100h^{-1}$ comoving Mpc box sampled with $256^3$ particles. The gravitational potential is computed on a $256^3$ coarse grid ($\ell=8$) and refinement up to $\ell=12$ is triggered in a quasi-Lagrangian manner when a cell contains more than 8 particles. The spatial resolution is equivalent to a $4096^3$ grid.
Cosmological parameters were taken from \citet{PLA13} (setting $\Omega_m=\Omega_c$)  and used as inputs to the \citet{EIS98} transfer function. The halos were detected using the HOP halo finder \citep{EIS97} and their mass function is compared to the formulation of \citet{SHE01}. Only halos with a number of particles greater than 10 particles were kept, corresponding to a minimal mass of $5.2 \times 10^{10} M_\odot$.

Fig. \ref{f:fmass} presents the halo mass function obtained at different redshifts, directly compared to \citet{SHE01}. A good agreement is obtained at all redshift, with massive halos kicking in only at later time as expected. On the low-mass end of the mass function, \texttt{EMMA} is complete for halos with at least $\sim 500$ particles and a factor of two below full completeness for halos with $\sim 100$ particles. These numbers are standard for such AMR codes and overall we can conclude that \texttt{EMMA} tracks correctly the assembly history of dark matter halos.

\subsection{Energy 'Conservation' in DM+gas adiabatic simulations}
\begin{figure}
\includegraphics[width=\columnwidth]{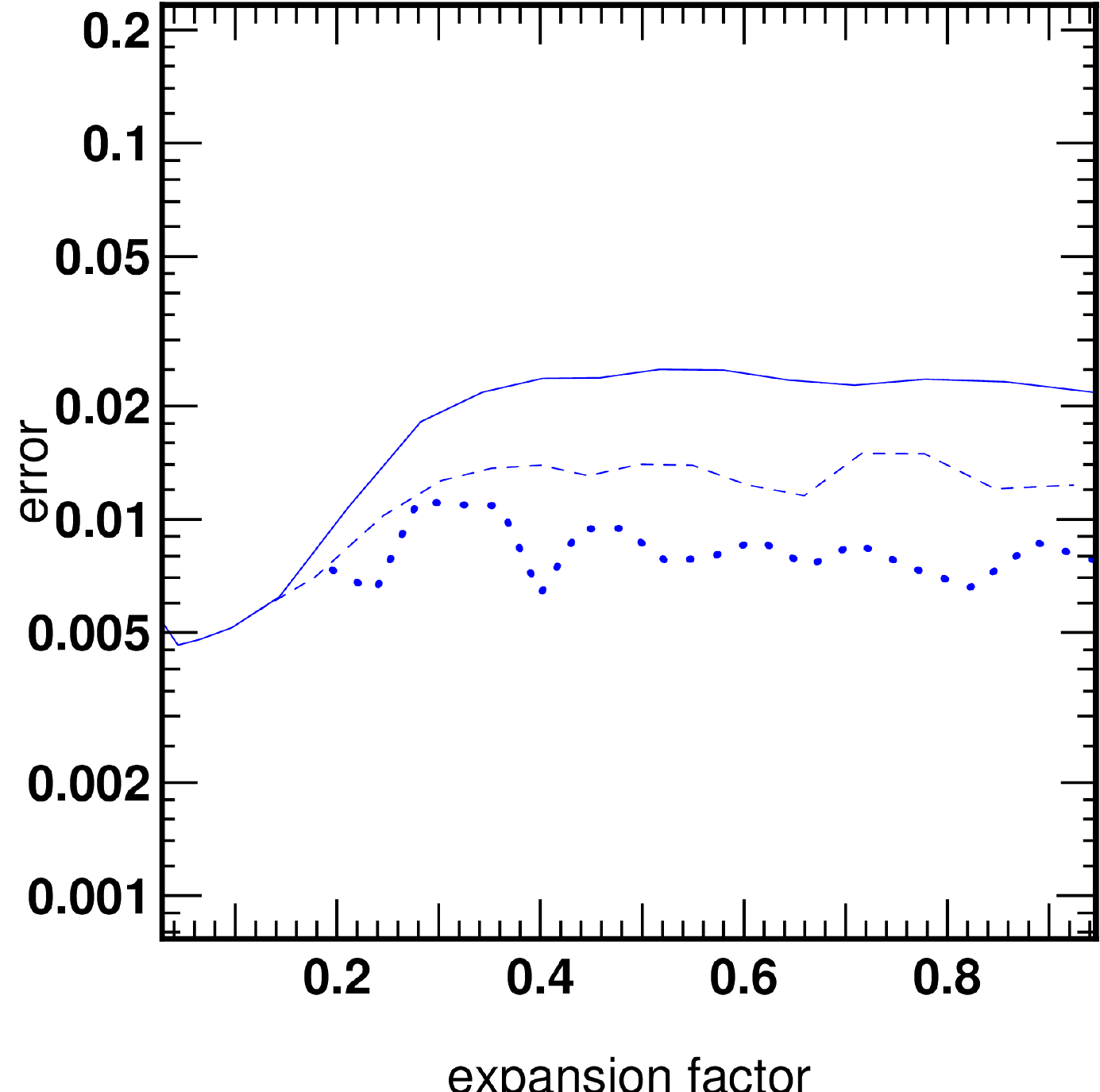}
\caption{Error on cosmic energy variation as defined in Eq.  \ref{e:egycons} for three 100 Mpc/h adiabatic DM+gas cosmological simulations with $128^3$ coarse resolutions ($\ell=7$). From top to bottom a simulation without AMR (solid), with $\ell_\mathrm{max}=8$ (dashed) and $\ell_\mathrm{max}=10$ (dotted). 
}
\label{f:egy}
\end{figure}

In this section, we probe the energy conservation of an adiabatic cosmological run (DM+gas). In supercomoving variables, the cosmic energy varies between expansion factors $a_1$ and $a_2$ according to~
\begin{equation}
(\tilde T +\tilde U +\tilde E)|_{a_1}^{a_2}=\int_{a_1}^{a_2} \frac{\tilde U}{a}da
\label{e:egycons}
\end{equation} 
for a $\gamma=5/3$ gas. $\tilde T$, $\tilde U$ and $\tilde E$ stand for the total supercomoving kinetic energy, potential energy and internal gas energy.
Three 100 $h^{-1}$Mpc $128^3$ ($\ell=7$) simulations with $\ell_\mathrm{max}=7,8,10$ where run, using the same cosmological parameters and refinement strategy as in Sec. \ref{s:fmass}. In the same spirit as \citet{KRA97}, we check Eq. \ref{e:egycons} against the change in potential energy, i.e.:
\begin{equation}
\mathrm{error}=\frac{(\tilde T +\tilde U +\tilde E)|_{a_1}^{a_2}-\int_{a_1}^{a_2} \frac{\tilde U}{a}da}{\tilde U|_{a_1}^{a_2}}
\end{equation}

In Fig. \ref{f:egy}, the error is shown for 3 maximum level of refinement~: $\ell_\mathrm{max}=7$ (i.e. no refinement), $\ell_\mathrm{max}=8$ and $\ell_\mathrm{max}=10$. The error is found to be under control at $2\%$, $1.2\%$ and $0.7\%$ respectively. One can note how the 3 tracks diverge as refinement levels are installed (e.g. $a=0.12$ for $\ell=8$ and $a=0.2$ for $\ell=9$)  providing greater resolution and smaller energy drifts. Overall, these levels of error is consistent with e.g. \citet{KRA97,TEY02}.

\section{Preliminary comparison of \texttt{EMMA} performances on different GPU devices}
\label{s:K20}
\dom{We briefly describe the timings obtained by \texttt{EMMA} on K20c devices, more recent than the M2090 GPUs available on Curie.  K20c devices differ by the number and type of cores (2496 Kepler cores versus 512 Fermi cores for the M2090), core clock (706 MHz Vs 1.3GHz for the M2090), memory frequency (2.6 GHz versus 1.9 GHz for the M2090) and bandwidth (208 Gb/s versus 177 Gb/s for the M2090). In terms of single precision floating point operations, the theoretical peak performance of K20c is a factor of 2 greater than the M2090.}

\begin{figure}
\includegraphics[width=\columnwidth]{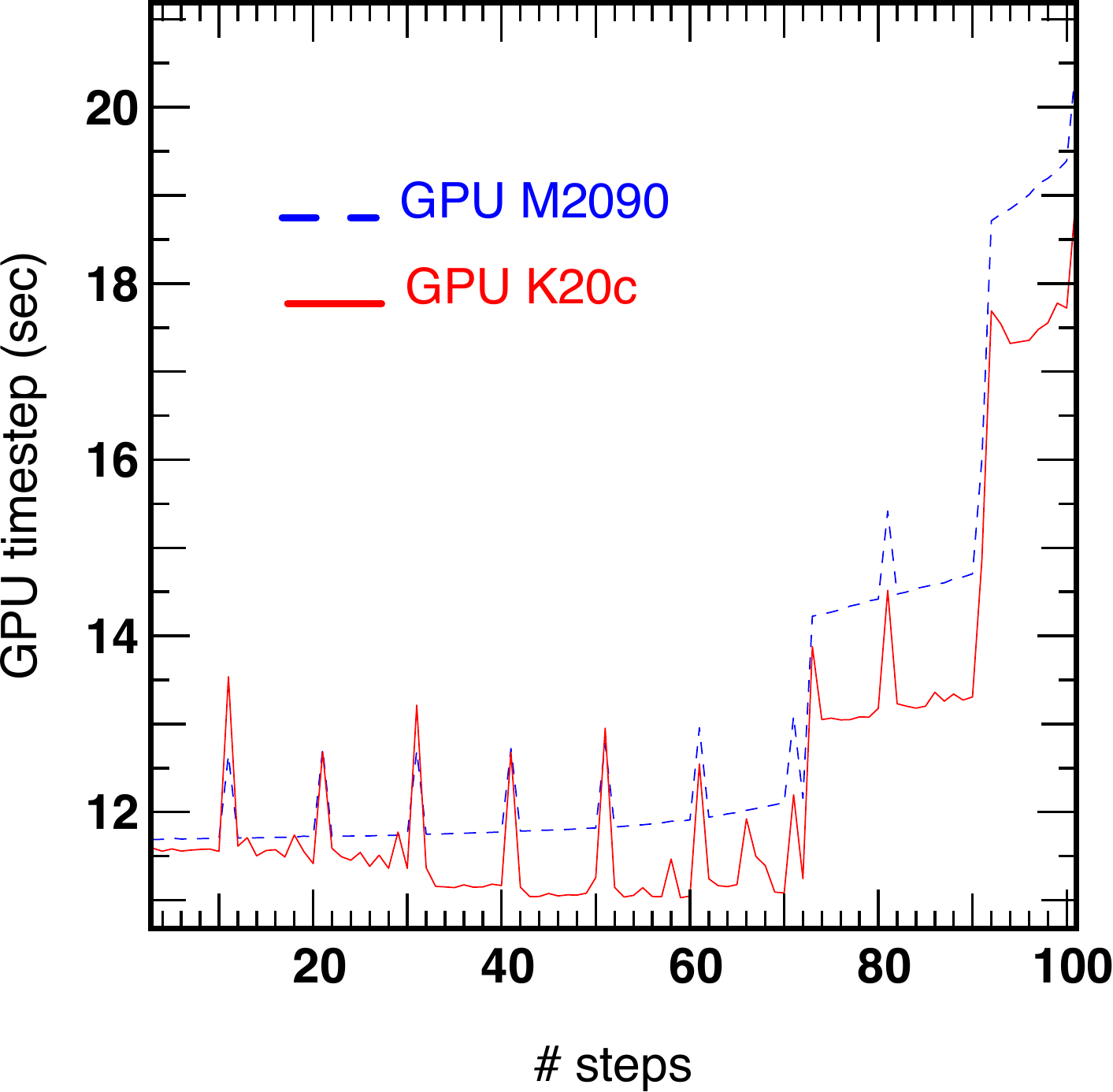}
\caption{Comparison of the time step duration on two different kind of GPU devices. Measures were taken on the first 100 steps of a 4 Mpc/h cosmological simulation with full physics with parameters similar to the runs described in Sec. \ref{s:cosmo}. The blue dashed line stands for M2090 timings, the red solid   line stand for timings on the more recent K20. The spikes seen in both curve are due to input/outputs operations. }
\label{f:K20}
\end{figure}

\dom{We ran a 4 Mpc/h cosmological simulation on a single GPU with full physics over 100 time steps, with the same settings as the one chosen in Sec. \ref{s:cosmo} and \ref{s:para}. Fig. \ref{f:K20} compares the duration of the time steps obtained from two simulations made on these two kind of devices. In both case, the timings show the same global evolution with spikes due to outputs of data and large jumps due to AMR refinement. K20 performances are marginally better than the M2090 ones, at the 10\% level, despite their greater computing power. It is expected since, \texttt{EMMA} calculations on GPU are already dominated on M2090 devices by gather/scatter  and host to device transfer operations . Future devices with greater bandwidth could improve the situation, but in its current state \texttt{EMMA} does not really benefit from the greater computing power of more recent hardware.}

%
%

\bibliographystyle{mn2e}
\bibliography{emma}

\end{document}